# Projective Limits of State Spaces
# IV. Fractal Label Sets

**Suzanne Lanéry[1,2] and Thomas Thiemann[1]**

[1] Institute for Quantum Gravity, Friedrich-Alexander University Erlangen-Nürnberg, Germany

[2] Mathematics and Theoretical Physics Laboratory, François-Rabelais University of Tours, France

**Abstract**

Instead of formulating the state space of a quantum field theory over one big Hilbert space, it has been proposed by Kijowski [6] to represent quantum states as projective families of density matrices over a collection of smaller, simpler Hilbert spaces. One can thus bypass the need to select a vacuum state for the theory, and still be provided with an explicit and constructive description of the quantum state space, at least as long as the label set indexing the projective structure is countable. Because uncountable label sets are much less practical in this context, we develop in the present article a general procedure to trim an originally uncountable label set down to countable cardinality. In particular, we investigate how to perform this tightening of the label set in a way that preserves both the physical content of the algebra of observables and its symmetries.

This work is notably motivated by applications to the holonomy-flux algebra underlying Loop Quantum Gravity. Building on earlier work by Okołów [15], a projective state space was introduced for this algebra in [7]. However, the non-trivial structure of the holonomy-flux algebra prevents the construction of satisfactory semi-classical states [11]. Implementing the general procedure just mentioned in the case of a one-dimensional version of this algebra, we show how a discrete subalgebra can be extracted without destroying universality nor diffeomorphism invariance. On this subalgebra, states can then be constructed whose semi-classicality is enforced step by step, starting from collective, macroscopic degrees of freedom and going down progressively toward smaller and smaller scales.

# Contents

# 1 Introduction

An important step toward the canonical quantization of a classical field theory is to prepare a suitable kinematical quantum state space: typically, one selects a *vacuum* for the theory, and write quantum states as *discrete excitations* around this vacuum. Such states however only span a particular *sector* of the kinematical theory. Ensuring that this sector will contain enough *physical* states (aka. solutions of the field equations) requires to understand beforehand the dynamics of the quantum theory, and to encapsulate this understanding in the choice of the right vacuum state. Like the previous articles in this series [8, 9, 10], the present work is set up in the context of a *projective formalism*, first introduced by Jerzy Kijowski in the late '70s [6] and further developed by Andrzej Okołów more recently [14, 15, 16], which provides an alternative way of building the kinematical quantum state space: instead of describing states as density matrices over a single big Hilbert space, one constructs them as projective families of partial density matrices over small 'building block' Hilbert spaces [9, subsection 2.1]. The projections are given as *partial traces*, so that each small Hilbert space can be physically interpreted as selecting out finitely many degrees of freedom. This approach tends to yields bigger quantum state spaces, as it allows to bypass the specialization to a specific sector.

Building on previous work by Okołów in [15, section 5] and [16], a projective state space of this kind was obtained in [7] for the holonomy-flux algebra underlying in Loop Quantum Gravity (LQG). This projective state space was shown to form an *extension* of the Ashtekar-Lewandowski (AL) state space used in LQG (in the case of a *compact* gauge group, where the latter can be defined). The motivation to go beyond the AL sector was to facilitate the construction of *semi-classical states*. Yet, the non-existence of such states was proved in [11] in the case of real-valued connections (with gauge group $G = \mathbb{R}$), hinting that fundamental *obstructions*, which could, to some extent, affect arbitrary gauge groups, arise from the algebra of observables itself. As discussed at the end of [11], the analysis performed in this reference also raises the question of whether the projective state space for the holonomy-flux algebra on a *non-compact* gauge group may turn out to be *empty* (note that emptiness concerns are averted in the compact group case, where this projective state space not only contains all AL states, but even additional ones, as proved in [7, prop. 3.22]).

The negative result of [11, prop. 2.14] can be traced back to the fact that the holonomy-flux algebra along analytical (or semi-analytical) edges and surfaces is generated by a *continuum* of elementary observables, which suggests the development of the present article: we will investigate how a *dense* but *discrete* subalgebra of observables could be extracted, while keeping intact the *physical content* of the theory. Specifically, the goal would be to restrict the *uncountable label set* $\mathcal{L}_{\text{HF}}$ (defined in [7, def. 2.14] to index a projective structure supporting the holonomy-flux algebra) to a carefully chosen *countable subset*. Such a restriction allows the *systematic construction* of projective quantum states, in particular semi-classical ones (subsection 2.1).

However, we have to be cautious of the dangerous side-effects such an endeavor could have:

- we do not want to introduce any objectionable arbitrariness in the theory: as stressed above, the projective approach is intended to *improve the universality* of the kinematical quantum state space (compared to the choice of a particular representation of the algebra of observables);
- in addition, going over to discrete structures carries a serious risk of breaking *diffeomorphism invariance* [19], something we want to avoid at any cost in view of applications to *background independent* quantum gravity [20, 21].

In subsection 2.2, we will therefore spell out, *in a general setting*, the properties that a label subset should satisfy to ensure that restricting the projective system to this subset will preserve suitable notions of both *universality* and diffeomorphism *invariance* (or, more generally, invariance under whatever the group of *symmetries* is for the particular theory under consideration). This general framework will be put to practice on a simple toy model in section 3, which can be seen as a (slightly simplified) *one-dimensional* version of the holonomy-flux algebra, while the generalization in dimension $d > 1$ (especially the physically relevant $d = 3$ case) is currently under progress.

Note that the kind of result we are aiming at should not be confused with various results in the context of LQG displaying how countable cardinality or universality can be obtained or restored *after* quotienting out the diffeomorphisms [1, 3]: since we do not yet fully understand how this quotienting should be done in the projective formalism (see the discussion at the end of [7]), we want to simplify the algebra of observables already at the diffeomorphism-*covariant* level, rather than at the diffeomorphism-*invariant* one (where those designations refer to the individual quantum states, not to the invariance of the overall state space). In fact, such an upfront simplification of the algebra could make it easier to implement in practice the strategy proposed in [8, section 3] (for dealing with constraints in the projective formalism): it could thus actually help *solving* the diffeomorphism constraints.

Once the label set is trimmed down to countable cardinality, a corresponding *inductive limit* Hilbert space (constructed from the choice of a vacuum, along the lines of [9, theorem 2.9]) will automatically be *separable* (assuming all 'building block' Hilbert spaces are), rather that non-separable like eg. the Ashtekar-Lewandowski Hilbert space (which results from such an inductive limit construction on the holonomy-flux algebra). Hence, constructing states will get easier on the inductive limit side too: besides lifting the technical issues plaguing non-separable Hilbert spaces [3], it will allow states to include all basis vectors at once (by contrast, non-separable Hilbert spaces tend, paradoxically, to be 'too small', as their orthonormal basis are uncountable while linear combinations can only be at most countable). Nevertheless, the advantages of the projective formalism over an inductive limit Hilbert space remains: as we will check in prop. 2.2, the semi-classical quantum states that one can construct within projective state spaces on countable label sets typically do *not* belong to corresponding inductive limit Hilbert space arising from vacuum states that are far from semi-classical. In other words, the argument put forward in the introduction of [7], that discrete quantum excitations cannot mask the core properties of the vacuum, still holds in the case of a countable label set.

When the label set is countable, one can also, in the spirit of [9, theorem 2.11], produce from the projective system associated *infinite tensor products* (ITP, see [27, 26]), and the states we will be considering often *do* belong to these ITP Hilbert spaces (see again prop. 2.2). In fact, the whole construction of section 2 is very closely connected with Algebraïc Quantum Gravity (AQG, see [4]): the idea of AQG is to choose an infinitely extended graph and to write a state space for quantum gravity as the infinite tensor product of the $L_2(G)$ Hilbert spaces carried by the individual edges. Like in the present development, this switch to discrete degrees of freedom in AQG is motivated by the search for better semi-classical states. We will comment where appropriate on the similarities and differences between the two approaches, and delineate some benefits of the projective formulation (namely a lesser dependence on arbitrary choices and an improved diffeomorphism invariance).

# 2 Quasi-cofinal sequences

Whenever the label set indexing the projective structure is countable, the projective formalism provides an *explicit and constructive* description of the quantum state space, suitable for concrete calculations and further exploration of the theory (subsection 2.1). In particular, there is no risk for a projective limit on a countable label set to be empty, since all projective states can then be constructed recursively in a systematic way: as the argument below shows, for any label $\eta$, and any partial density matrix $\rho_\eta$ on $\eta$, it will always be possible to find a projective state whose projection on the label $\eta$ coincides with $\rho_\eta$.

To take advantage of these simplifications we will, in subsection 2.2, lay out a general framework to extract a countable subset from an initially uncountable label set. More precisely, we will formulate the conditions which have to be satisfied so that the projective state space built on a label subset is *universal* – ie. can be shown to be independent of any arbitrary choices entering the selection of this subset (theorem 2.8) – and supports the *symmetries* of the original theory (prop. 2.9).

Note that, from a *physical* point of view, it seems in fact very reasonable to expect the *elementary* observables of a theory (in the sense discussed in [8, section 1]) to form a *countable* set: these observables are meant to be in one-to-one correspondence with the experimental protocols describing their measurement, and such protocols should indeed form a countable set (since they can be encoded eg. as finite sequences of chars). To say it differently, if there would be uncountably many elementary observables, we would not even be able to accurately *tell* which one we are measuring in a given experiment.

One could at first think that such an argument should be made at the level of the *dynamical* observables, since those are often thought of as the only 'real' ones. However, in the spirit of [8, appendix A] (viz. the extended discussions in [8, section 3]), we adopt the interpretation that the kinematical observables are not just byproducts of the construction of the final, dynamical state space, to be discarded as soon as the latter has been obtained, but that they instead play a prominent role to formulate the interface between the mathematical theory and the experimental reality: they are used to label with physical meaning the dynamical observables to which they give rise (as stressed in the discussion preceding [8, def. A.2], the redundancy of this labeling is deliberate: it reflects the predictive power of the theory). So, in this perspective, we indeed expect countability already at the level of the *kinematical* elementary observables.

## 2.1 Factorized states on cofinal sequences

As underlined in [9, prop. 2.6], restricting the label set to a cofinal part does not affect the projective quantum state space, so, rather than considering a countable label set, it is sufficient to look at a label set admitting a *countable cofinal subset*. This weaker condition is of course equivalent if the part of the label set that is below any given label is countable, like in $\mathcal{L}_{\text{HF}}$ (as the labels in $\mathcal{L}_{\text{HF}}$ are finite collections of edges and faces, they actually have only finitely many sublabels, see prop. 2.3). But, for example, in the label set considered in [10, prop. 3.3], which consists of finite dimensional vector subspaces, most labels (namely the vector subspaces of dimension greater than 2) are above uncountably many others (while the label set $\mathcal{L}$ of [10, prop. 3.3] itself does not admit a countable cofinal subset, one could easily construct an uncountable part of $\mathcal{L}$ that does).

If we do have a countable cofinal part, then we can construct recursively an *increasing* cofinal sequence from it, and, along this sequence, we can use the fact that the partial traces $\mathrm{Tr}_{\eta_{n+1}\to\eta_n}$ are surjective to construct a projective quantum state, by recursively choosing a density matrix $\rho_{\eta_{n+1}}$ in the preimage of $\rho_{\eta_n}$. Clearly, all projective states can be constructed in this way. The 'factorized pure states', satisfying $\rho_{\eta_{n+1}} \approx |\psi_{n+1}\,\rangle\langle\,\psi_{n+1}| \otimes \rho_{\eta_n}$ for some vector $\psi_{n+1} \in \mathcal{H}_{\eta_{n+1}\to\eta_n}$, are particularly simple, and their convex closure is dense in the projective state space (with respect to a topology defined like in [7, prop. 3.21]).

**Proposition 2.1** Let $(\mathcal{L}, \mathcal{H}, \Phi)^\otimes$ be a projective system of quantum state spaces [9, def. 2.1] and suppose that $\mathcal{L}$ admits a *countable* cofinal subset $\widetilde{\mathcal{L}}_{\mathrm{seq}}$. Then, there exists an increasing sequence $(\eta_n)_{n\in\mathbb{N}}$ such that $\{\eta_n \mid n \in \mathbb{N}\}$ is cofinal in $\widetilde{\mathcal{L}}_{\mathrm{seq}}$, hence in $\mathcal{L}$. We *choose* such an increasing sequence, and we define $\mathcal{L}_{\mathrm{seq}} := \{\eta_n \mid n \in \mathbb{N}\}$ as well as:

$$\mathcal{J}_o := \mathcal{H}_{\eta_o} \quad \& \quad \forall n > 0,\ \mathcal{J}_n := \mathcal{H}_{\eta_n\to\eta_{n-1}}\,.$$

Then, for any sequence $\psi = (\psi_n)_{n\in\mathbb{N}}$ such that:

$$\forall n \in \mathbb{N},\quad \psi_n \in \mathcal{J}_n \quad \& \quad \|\psi_n\|_{\mathcal{J}_n} = 1\,, \tag{2.1.1}$$

there exists a *unique* state $\rho[\psi] \in \mathcal{S}^\otimes_{(\mathcal{L},\mathcal{H},\Phi)}$ such that:

$$\rho_{\eta_o}[\psi] = |\psi_o\,\rangle\langle\,\psi_o| \quad \& \quad \forall n > 0,\ \rho_{\eta_n}[\psi] = \Phi^{-1}_{\eta_n\to\eta_{n-1}} \circ \Big(\,|\psi_n\,\rangle\langle\,\psi_n| \otimes \rho_{\eta_{n-1}}[\psi]\Big) \circ \Phi_{\eta_n\to\eta_{n-1}}\,. \tag{2.1.2}$$

**Proof** *Auxiliary projective system on* $\mathbb{N}$. Since $\widetilde{\mathcal{L}}_{\mathrm{seq}}$ is countable, there exists a sequence $\left(\widetilde{\eta}_n\right)_{n\in\mathbb{N}}$ such that $\widetilde{\mathcal{L}}_{\mathrm{seq}} = \left\{\widetilde{\eta}_n \middle| n \in \mathbb{N}\right\}$ (if $\widetilde{\mathcal{L}}_{\mathrm{seq}}$ happens to be finite, we can simply choose the sequence to be eventually constant). Next, $\mathcal{L}$ being directed and $\widetilde{\mathcal{L}}_{\mathrm{seq}}$ being cofinal in $\mathcal{L}$, there exists, for any $n, n' \in \mathbb{N}$, $N \in \mathbb{N}$ such that:

$$\widetilde{\eta}_n\,,\, \widetilde{\eta}_{n'} \preccurlyeq \widetilde{\eta}_N\,.$$

Hence, we can define recursively a sequence $(N_n)_{n\in\mathbb{N}}$ via:

$$N_o = 0 \quad \& \quad \forall n > 0,\ N_n := \min\left\{N \in \mathbb{N} \mid \widetilde{\eta}_n\,,\, \widetilde{\eta}_{N_{n-1}} \preccurlyeq \widetilde{\eta}_N\right\}.$$

By construction, the sequence $\left(\eta_n := \widetilde{\eta}_{N_n}\right)_{n\in\mathbb{N}}$ is increasing and $\mathcal{L}_{\mathrm{seq}} := \{\eta_n \mid n \in \mathbb{N}\}$ is cofinal in $\widetilde{\mathcal{L}}_{\mathrm{seq}}$.

We define recursively a family of unitary isomorphisms $\left(\widetilde{\Phi}_n : \mathcal{H}_{\eta_n} \to \mathcal{J}_n \otimes \ldots \otimes \mathcal{J}_o\right)_{n\in\mathbb{N}}$ via:

$$\widetilde{\Phi}_o := \mathrm{id}_{\mathcal{J}_o} \quad \& \quad \forall n > 0,\ \widetilde{\Phi}_n := \left(\mathrm{id}_{\mathcal{J}_n} \otimes \widetilde{\Phi}_{n-1}\right) \circ \Phi_{\eta_n\to\eta_{n-1}}\,.$$

Next, for any $n \in \mathbb{N}$, we define recursively a family of unitary isomorphisms $\Big(\widetilde{\Phi}_{n'\to n} : \mathcal{H}_{\eta_{n'}\to\eta_n} \to \mathcal{J}_{n'} \otimes \ldots \otimes \mathcal{J}_{n+1}\Big)_{n'>n}$ via:

$$\widetilde{\Phi}_{(n+1)\to n} := \mathrm{id}_{\mathcal{J}_{n+1}} \quad \& \quad \forall n' > n+1,\, \widetilde{\Phi}_{n'\to n} := \left(\mathrm{id}_{\mathcal{J}_{n'}} \otimes \widetilde{\Phi}_{(n'-1)\to n}\right) \circ \Phi_{\eta_{n'}\to\eta_{n'-1}\to\eta_n} .$$

Thus, we get, for any $n < n' \in \mathbb{N}$:

$$\left(\widetilde{\Phi}_{n'\to n} \otimes \widetilde{\Phi}_n\right) \circ \Phi_{\eta_{n'}\to\eta_n} \circ \widetilde{\Phi}_{n'}^{-1} = \mathrm{id}_{\mathcal{J}_{n'}\otimes\ldots\otimes\mathcal{J}_{n+1}\otimes\mathcal{J}_n\otimes\ldots\otimes\mathcal{J}_o} , \tag{2.1.3}$$

as can be shown by recursion over $n' - n$, using the definitions above together with [9, eq. (2.1.1)].

Let $\sigma_{\mathcal{L}\to\mathcal{L}_{\text{seq}}} : \overline{\mathcal{S}}^{\otimes}_{\mathcal{L},\mathcal{H},\Phi} \to \overline{\mathcal{S}}^{\otimes}_{(\mathcal{L}_{\text{seq}},\mathcal{H},\Phi)}$ and $\alpha_{\mathcal{L}\leftarrow\mathcal{L}_{\text{seq}}} : \overline{\mathcal{A}}^{\otimes}_{\mathcal{L}_{\text{seq}},\mathcal{H},\Phi} \to \overline{\mathcal{A}}^{\otimes}_{(\mathcal{L},\mathcal{H},\Phi)}$ be the bijective maps constructed as in [9, prop. 2.6]. We define, for any $n \in \mathbb{N}$:

$$\mathcal{K}_n := \mathcal{J}_n \otimes \ldots \otimes \mathcal{J}_o \quad \& \quad \mathcal{K}_{n\to n} := \mathbb{C} ,$$

and for any $n < n' \in \mathbb{N}$:

$$\mathcal{K}_{n'\to n} := \mathcal{J}_{n'} \otimes \ldots \otimes \mathcal{J}_{n+1} .$$

Moreover, we define, for any $n \leqslant n' \in \mathbb{N}$, $\Psi_{n'\to n}$ to be the natural identification $\mathcal{K}_{n'} \approx \mathcal{K}_{n'\to n} \otimes \mathcal{K}_n$ and, for any $n \leqslant n' \leqslant n'' \in \mathbb{N}$, $\Psi_{n''\to n'\to n}$ to be the natural identification $\mathcal{K}_{n''\to n} \approx \mathcal{K}_{n''\to n'} \otimes \mathcal{K}_{n'\to n}$. Thus, $(\mathbb{N}, \mathcal{K}, \Psi)^{\otimes}$ is a projective system of quantum state spaces. Now, we define:

$$\begin{array}{rccc} \sigma_{\mathcal{L}_{\text{seq}}\to\mathbb{N}} : & \overline{\mathcal{S}}^{\otimes}_{(\mathcal{L}_{\text{seq}},\mathcal{H},\Phi)} & \to & \overline{\mathcal{S}}^{\otimes}_{(\mathbb{N},\mathcal{K},\Psi)} \\ & \left(\rho_\eta\right)_{\eta\in\mathcal{L}_{\text{seq}}} & \mapsto & \left(\widetilde{\Phi}_n \circ \rho_{\eta_n} \circ \widetilde{\Phi}_n^{-1}\right)_{n\in\mathbb{N}} \end{array} .$$

Let $\rho \in \overline{\mathcal{S}}^{\otimes}_{(\mathcal{L}_{\text{seq}},\mathcal{H},\Phi)}$. For any $n \in \mathbb{N}$, we have:

$$\mathrm{Tr}_{n\to n} \widetilde{\Phi}_n \circ \rho_{\eta_n} \circ \widetilde{\Phi}_n^{-1} = \widetilde{\Phi}_n \circ \rho_{\eta_n} \circ \widetilde{\Phi}_n^{-1} ,$$

and for any $n < n' \in \mathbb{N}$, eq. (2.1.3) yields:

$$\begin{aligned} \mathrm{Tr}_{n'\to n} \widetilde{\Phi}_{n'} \circ \rho_{\eta_{n'}} \circ \widetilde{\Phi}_{n'}^{-1} &= \widetilde{\Phi}_n \circ \left[\mathrm{Tr}_{\mathcal{H}_{\eta_{n'}\to\eta_n}} \Phi_{\eta_{n'}\to\eta_n} \circ \rho_{\eta_{n'}} \circ \Phi^{-1}_{\eta_{n'}\to\eta_n}\right] \circ \widetilde{\Phi}_n^{-1} \\ &= \widetilde{\Phi}_n \circ \rho_{\eta_n} \circ \widetilde{\Phi}_n^{-1} . \end{aligned}$$

Therefore, $\sigma_{\mathcal{L}_{\text{seq}}\to\mathbb{N}}$ is well-defined as a map $\overline{\mathcal{S}}^{\otimes}_{(\mathcal{L}_{\text{seq}},\mathcal{H},\Phi)} \to \overline{\mathcal{S}}^{\otimes}_{(\mathbb{N},\mathcal{K},\Psi)}$. In addition, $\sigma_{\mathcal{L}_{\text{seq}}\to\mathbb{N}}$ is injective, since $\mathcal{L}_{\text{seq}} = \{\eta_n \mid n \in \mathbb{N}\}$.

We now want to prove that $\sigma_{\mathcal{L}_{\text{seq}}\to\mathbb{N}}$ is surjective as well. Let $\widetilde{\rho} \in \overline{\mathcal{S}}^{\otimes}_{(\mathbb{N},\mathcal{K},\Psi)}$. Let $n, n' \in \mathbb{N}$ such that $\eta_n \preccurlyeq \eta_{n'}$. If $n < n'$, we have, in a way similar to above:

$$\mathrm{Tr}_{\eta_{n'}\to\eta_n} \widetilde{\Phi}_{n'}^{-1} \circ \widetilde{\rho}_{n'} \circ \widetilde{\Phi}_{n'} = \widetilde{\Phi}_n^{-1} \circ \widetilde{\rho}_n \circ \widetilde{\Phi}_n . \tag{2.1.4}$$

Clearly, eq. (2.1.4) also holds if $n = n'$. Finally, if $n > n'$, $\eta_n \succcurlyeq \eta_{n'} \succcurlyeq \eta_n$, hence applying [9, eq. (2.1.1)] implies $\mathrm{Tr}_{\eta_{n'}\to\eta_n} = \left(\mathrm{Tr}_{\eta_n\to\eta_{n'}}\right)^{-1}$. Making use of the first case for $n', n$ then yields eq. (2.1.4) in this case too. In particular, if $\eta_n = \eta_{n'}$, we get:

$$\widetilde{\Phi}_{n'}^{-1} \circ \widetilde{\rho}_{n'} \circ \widetilde{\Phi}_{n'} = \widetilde{\Phi}_n^{-1} \circ \widetilde{\rho}_n \circ \widetilde{\Phi}_n .$$

Therefore, there exists $\rho = \left(\rho_\eta\right)_{\eta\in\mathcal{L}_{\text{seq}}} \in \overline{\mathcal{S}}^{\otimes}_{(\mathcal{L}_{\text{seq}},\mathcal{H},\Phi)}$ such that $\forall n \in \mathbb{N}$, $\rho_{\eta_n} = \widetilde{\Phi}_n^{-1} \circ \widetilde{\rho}_n \circ \widetilde{\Phi}_n$, ie. $\sigma_{\mathcal{L}_{\text{seq}}\to\mathbb{N}}(\rho) = \widetilde{\rho}$.

Next, for any $n, n' \in \mathbb{N}$, and any $A_{\eta_n} \in \mathcal{A}_{\eta_n}$, $A_{\eta_{n'}} \in \mathcal{A}_{\eta_{n'}}$, we have:

$$\Big(\exists\, \eta'' \in \mathcal{L}_{\text{seq}} \,/\, \eta'' \succcurlyeq \eta_n \,,\, \eta_{n'} \quad \&$$

$$\Phi^{-1}_{\eta''\to\eta_n} \circ (\mathrm{id}_{\mathcal{H}_{\eta''\to\eta_n}} \otimes A_{\eta_n}) \circ \Phi_{\eta''\to\eta_n} = \Phi^{-1}_{\eta''\to\eta_{n'}} \circ (\mathrm{id}_{\mathcal{H}_{\eta''\to\eta_{n'}}} \otimes A_{\eta_{n'}}) \circ \Phi_{\eta''\to\eta_{n'}}\Big) \Leftrightarrow$$

$$\Leftrightarrow \Big(\exists\, n'' \geqslant n, n' \,/\, \mathrm{id}_{\mathcal{K}_{n''\to n}} \otimes (\widetilde{\Phi}_n \circ A_{\eta_n} \circ \widetilde{\Phi}_n^{-1}) = \mathrm{id}_{\mathcal{K}_{n''\to n'}} \otimes (\widetilde{\Phi}_{n'} \circ A_{\eta_{n'}} \circ \widetilde{\Phi}_{n'}^{-1})\Big)$$

(the direction '$\Rightarrow$' can be shown by choosing $\widetilde{n}$ such that $\eta'' = \eta_{\widetilde{n}}$ and $n'' > n, n', \widetilde{n}$). Thus, we can define the algebra isomorphism:

$$\begin{array}{rccl} \alpha_{\mathcal{L}_{\text{seq}}\leftarrow\mathbb{N}} : & \mathcal{A}^{\otimes}_{\mathbb{N},\mathcal{K},\Psi} & \to & \mathcal{A}^{\otimes}_{\mathcal{L}_{\text{seq}},\mathcal{H},\Phi} \\ & \Big[A_n\Big]_{\sim} & \mapsto & \Big[\widetilde{\Phi}_n^{-1} \circ A_n \circ \widetilde{\Phi}_n\Big]_{\sim} \end{array},$$

and extends it by continuity into a $C^*$-algebra isomorphism $\alpha_{\mathcal{L}_{\text{seq}}\leftarrow\mathbb{N}} : \overline{\mathcal{A}}^{\otimes}_{\mathbb{N},\mathcal{K},\Psi} \to \overline{\mathcal{A}}^{\otimes}_{\mathcal{L}_{\text{seq}},\mathcal{H},\Phi}$. Finally, we define the bijective maps $\sigma := \sigma_{\mathcal{L}_{\text{seq}}\to\mathbb{N}} \circ \sigma_{\mathcal{L}\to\mathcal{L}_{\text{seq}}} : \overline{\mathcal{S}}^{\otimes}_{(\mathcal{L},\mathcal{H},\Phi)} \to \overline{\mathcal{S}}^{\otimes}_{(\mathbb{N},\mathcal{K},\Psi)}$ and $\alpha := \alpha_{\mathcal{L}\leftarrow\mathcal{L}_{\text{seq}}} \circ \alpha_{\mathcal{L}_{\text{seq}}\leftarrow\mathbb{N}} : \overline{\mathcal{A}}^{\otimes}_{(\mathbb{N},\mathcal{K},\Psi)} \to \overline{\mathcal{A}}^{\otimes}_{(\mathcal{L},\mathcal{H},\Phi)}$.

*Existence and uniqueness of $\rho[\psi]$.* Let $\psi = (\psi_n)_{n\in\mathbb{N}}$ be a sequence satisfying eq. (2.1.1). We define $\widetilde{\rho}[\psi] \in \mathcal{S}^{\otimes}_{(\mathbb{N},\mathcal{K},\Psi)}$ via:

$$\forall n \in \mathbb{N},\ \widetilde{\rho}_n := |\psi_n \otimes \ldots \otimes \psi_o\,\rangle\langle\, \psi_n \otimes \ldots \otimes \psi_o| .$$

Then, $\rho[\psi] := \sigma^{-1}\big(\widetilde{\rho}[\psi]\big) \in \mathcal{S}^{\otimes}_{(\mathcal{L},\mathcal{H},\Phi)}$ fulfills eq. (2.1.2). Reciprocally, if $\rho \in \mathcal{S}^{\otimes}_{(\mathcal{L},\mathcal{H},\Phi)}$ fulfills eq. (2.1.2), then $\sigma(\rho) = \widetilde{\rho}[\psi]$ (this can be checked recursively on $n$), hence $\rho = \rho[\psi]$. □

Supposing that the projective system of quantum state spaces under consideration has been obtained through the quantization of a factorizing system of symplectic manifolds (eg. along the lines of [9, section 3]), and that the latter forms a rendering [8, def. 2.6] of some classical, continuum phase space $\mathcal{M}_\infty$, we can use this technique to construct a semi-classical state centered on a classical point $x_\infty \in \mathcal{M}_\infty$. To this intend, the vector $\psi_{n+1} \in \mathcal{H}_{\eta_{n+1}\to\eta_n}$ should be chosen as a semi-classical state centered around the point $x_{\eta_{n+1}\to\eta_n} \in \mathcal{M}_{\eta_{n+1}\to\eta_n}$, computed from $x_\infty$ via $\big(x_{\eta_{n+1}\to\eta_n}\,,\, x_{\eta_n}\big) := \varphi_{\eta_{n+1}\to\eta_n}\big(x_{\eta_{n+1}}\big) := \varphi_{\eta_{n+1}\to\eta_n} \circ \pi_{\infty\to\eta_{n+1}}(x_\infty)$.

For small $n$, we can think of the coarse labels $\eta_n$ as describing some *collective*, *macroscopic* degrees of freedom, so that the prescription above offers a concrete implementation of the approach advocated in [17]: namely, we start by forming states having good peaking properties at macroscopic scales, and, going down step by step toward smaller and smaller scales, we impose, at each step, as much semi-classicality as the Heisenberg uncertainty relations will allow (taking heed of the already fixed behavior at larger scales). This is readily achieved here because the largest part of the work was done beforehand while setting up the factorizing system, by identifying the degrees of freedom in $\eta_{n+1}$ that commute with the ones from $\eta_n$ (recall the discussion before [8, prop. 2.10]): those are precisely the variables on which semi-classicality can be imposed independently of the already chosen state on $\eta_n$.

We can then ask whether a semi-classical state constructed this way would belong to the inductive limit Hilbert space arising from a choice of vacuum state [9, prop. 2.8]. Assuming this vacuum is itself a factorized pure state, the characterization given in [9, theorem 2.9] can be reformulated into the condition 2.2.3 below. In particular, if the vacuum state is a momentum eigenstate, like the Ashtekar-Lewandowski vacuum, a factorized semi-classical state could only be made an element of the corresponding inductive limit by deteriorating the semi-classicality of $\psi_n$: eg. if $\psi_n$ is taken as a coherent state, controlled by a semi-classicality parameter that determines the repartition of the quantum uncertainties between position and momentum variables, this parameter will have to be shifted fast enough, as $n$ grows, toward maximally peaked momenta and maximally spread positions. By contrast, if the vacuum state is itself a coherent state, like the Fock vacuum, the condition 2.2.3 can be interpreted as delimiting a domain in the classical projective limit [8, def. 2.3] such that, for $x_\infty$ belonging to this domain, the factorized semi-classical state centered around $x_\infty$ will belong to the corresponding inductive limit: assuming the vacuum is centered around 0, this requires that $x_{\eta_{n+1}\rightarrow\eta_n}$ tends to 0 fast enough. The question will then be whether the image of $\mathcal{M}_\infty$ in the projective limit [8, prop. 2.7] happens to be contained in this admissible domain.

Finally, we also notice that the tensor product factors $\mathcal{H}_{\eta_{n+1}\rightarrow\eta_n}$ can be arranged into an infinite tensor product (ITP, see [27, 26] and [9, theorem 2.11]), and, not surprisingly, all factorized states do belong to this ITP. Still, as we will argue below, working with a projective state space instead of an ITP Hilbert space allows to overcome certain limitations of the ITP construction, in particular with respect to universality (prop. 3.4).

To comment on the relation with the Algebraic Quantum Gravity framework (AQG, see [4] and the brief explanation at the end of the introduction), note that, while a primary motivation for introducing an ITP Hilbert space in AQG was the availability of factorized coherent states very similar to the one discussed above, an important difference lies in the type of tensor product factors we are using: the building blocs of the ITP in AQG describe individual, *microscopic* degrees of freedom, meant to represent the smallest atoms of a quantum geometry (presumably at Plank scale), instead of holding complementary degrees of freedom added step by step as we refine our description from macroscopic to microscopic scales.

**Proposition 2.2** We consider the same objects as in prop. 2.1. We denote by $\mathcal{H}_{\text{seq}}$ the infinite tensor product of $(\mathcal{J}_n)_{n\in\mathbb{N}}$ (see [27] and [9, theorem 2.11]). There exist a map $\sigma_{\text{seq}} : \overline{\mathcal{S}}_{\text{seq}} \rightarrow \overline{\mathcal{S}}^{\otimes}_{(\mathcal{L},\mathcal{H},\Phi)}$ and an algebra morphism $\alpha_{\text{seq}} : \overline{\mathcal{A}}^{\otimes}_{(\mathcal{L},\mathcal{H},\Phi)} \rightarrow \mathcal{A}_{\text{seq}}$ ($\overline{\mathcal{S}}_{\text{seq}}$, resp. $\mathcal{A}_{\text{seq}}$, being the space of non-negative traceclass operators, resp. the algebra of bounded operators, on $\mathcal{H}_{\text{seq}}$) such that:

$$\forall \rho \in \overline{\mathcal{S}}_{\text{seq}}\,,\ \forall A \in \overline{\mathcal{A}}^{\otimes}_{(\mathcal{L},\mathcal{H},\Phi)}\,,\quad \text{Tr}_{\mathcal{H}_{\text{seq}}}\, \rho\, \alpha_{\text{seq}}(A) = \text{Tr}\, \sigma_{\text{seq}}(\rho)\, A\,.$$

Similarly, for any sequence $\psi$ satisfying eq. (2.1.1), we denote by $\mathcal{H}_{[\psi]}$ the GNS representation of $\overline{\mathcal{A}}^{\otimes}_{(\mathcal{L},\mathcal{H},\Phi)}$ arising from the state $\rho[\psi]$ (see [9, props. 2.4 and 2.8]). There exist an *injective* map $\sigma_{[\psi]} : \overline{\mathcal{S}}_{[\psi]} \rightarrow \overline{\mathcal{S}}^{\otimes}_{(\mathcal{L},\mathcal{H},\Phi)}$ and an algebra morphism $\alpha_{[\psi]} : \overline{\mathcal{A}}^{\otimes}_{(\mathcal{L},\mathcal{H},\Phi)} \rightarrow \mathcal{A}_{[\psi]}$ ($\overline{\mathcal{S}}_{[\psi]}$, resp. $\mathcal{A}_{[\psi]}$, being the space of non-negative traceclass operators, resp. the algebra of bounded operators, on $\mathcal{H}_{[\psi]}$) such that:

$$\forall \rho \in \overline{\mathcal{S}}_{[\psi]}\,,\ \forall A \in \overline{\mathcal{A}}^{\otimes}_{(\mathcal{L},\mathcal{H},\Phi)}\,,\quad \text{Tr}_{\mathcal{H}_{[\psi]}}\, \rho\, \alpha_{[\psi]}(A) = \text{Tr}\, \sigma_{[\psi]}(\rho)\, A\,.$$

Moreover, $\rho[\psi] \in \sigma_{[\psi]}\left\langle \mathcal{S}_{[\psi]} \right\rangle$ and $\sigma_{[\psi]}\left\langle \mathcal{S}_{[\psi]} \right\rangle \subset \sigma_{\text{seq}}\left\langle \mathcal{S}_{\text{seq}} \right\rangle$ (with $\mathcal{S}_{[\psi]}$, resp. $\mathcal{S}_{\text{seq}}$, the space of density matrices on $\mathcal{H}_{[\psi]}$, resp. $\mathcal{H}_{\text{seq}}$).

Let $\psi, \psi'$ be two sequences satisfying eq. (2.1.1). The following statements are equivalents:

1. $\rho[\psi'] \in \sigma_{[\psi]}\left\langle \mathcal{S}_{[\psi]} \right\rangle$;

2. $\sigma_{[\psi']} \left\langle \mathcal{S}_{[\psi']} \right\rangle = \sigma_{[\psi]} \left\langle \mathcal{S}_{[\psi]} \right\rangle$ ;

3. $\displaystyle\sum_{n=0}^{\infty} \Big(1 - \big| \langle \psi_n \mid \psi'_n \rangle_{\mathcal{J}_n} \big|\Big) < \infty$ .

**Proof** *Construction of $\mathcal{H}_{[\psi]}$, $\sigma_{[\psi]}$ and $\alpha_{[\psi]}$.* We define:

$$\mathcal{Z}^{\otimes}_{(\mathbb{N},\mathcal{J})} := \left\{ (\psi_n)_{n\in\mathbb{N}} \,\middle|\, \forall n \in \mathbb{N},\ \psi_n \in \mathcal{J}_n \quad \& \quad \|\psi_n\|_{\mathcal{J}_n} = 1 \right\}.$$

Let $\psi \in \mathcal{Z}^{\otimes}_{(\mathbb{N},\mathcal{J})}$ and let $\widetilde{\rho}[\psi] \in \mathcal{S}^{\otimes}_{(\mathbb{N},\mathcal{K},\Psi)}$ be defined as in the proof of prop. 2.1. Let $\mathcal{H}_{[\psi]}$ be the GNS representation of $\overline{\mathcal{A}}^{\otimes}_{(\mathbb{N},\mathcal{K},\Psi)}$ arising from the state $\widetilde{\rho}[\psi]$. From [9, prop. 2.8 and theorem 2.9], there exist an injective map $\widetilde{\sigma}_{[\psi]} : \overline{\mathcal{S}}_{[\psi]} \to \overline{\mathcal{S}}^{\otimes}_{(\mathbb{N},\mathcal{K},\Psi)}$ and a $C^*$-algebra morphism $\widetilde{\alpha}_{[\psi]} : \overline{\mathcal{A}}^{\otimes}_{(\mathbb{N},\mathcal{K},\Psi)} \to \mathcal{A}_{[\psi]}$, such that $\mathrm{Tr}_{\mathcal{H}_{[\psi]}} \left( \,\cdot\, \widetilde{\alpha}_{[\psi]}(\,\cdot\,)\right) = \mathrm{Tr}\left(\widetilde{\sigma}_{[\psi]}(\,\cdot\,)\,\cdot\,\right)$. Moreover, we have:

$$\widetilde{\sigma}_{[\psi]} \left\langle \mathcal{S}_{[\psi]} \right\rangle := \left\{ \left(\widetilde{\rho}_n\right)_{n\in\mathbb{N}} \in \mathcal{S}^{\otimes}_{(\mathbb{N},\mathcal{K},\Psi)} \,\middle|\, \sup_{n\in\mathbb{N}} \inf_{n'>n} \left\langle \zeta_{n'\to n} \middle| \left(\mathrm{Tr}_{\mathcal{K}_n} \widetilde{\rho}_{n'}\right) \zeta_{n'\to n} \right\rangle = \mathrm{Tr}\,\widetilde{\rho} = 1 \right\},$$

where $\forall n < n' \in \mathbb{N}$, $\zeta_{n'\to n} := \psi_{n'} \otimes \ldots \otimes \psi_{n+1} \in \mathcal{K}_{n'\to n}$ .

Using the Tr-intertwined bijective maps $\sigma$ and $\alpha$ defined in the proof of prop. 2.1, we can identify $\mathcal{H}_{[\psi]}$ with the GNS representation of $\overline{\mathcal{A}}^{\otimes}_{(\mathcal{L},\mathcal{H},\Phi)}$ arising from the state $\rho[\psi]$, and define the injective map $\sigma_{[\psi]} := \sigma^{-1} \circ \widetilde{\sigma}_{[\psi]} : \overline{\mathcal{S}}_{[\psi]} \to \overline{\mathcal{S}}^{\otimes}_{(\mathcal{L},\mathcal{H},\Phi)}$, as well as the algebra morphism $\alpha_{[\psi]} := \widetilde{\alpha}_{[\psi]} \circ \alpha^{-1} : \overline{\mathcal{A}}^{\otimes}_{(\mathcal{L},\mathcal{H},\Phi)} \to \mathcal{A}_{[\psi]}$. We then have $\mathrm{Tr}_{\mathcal{H}_{[\psi]}} \left( \,\cdot\, \alpha_{[\psi]}(\,\cdot\,)\right) = \mathrm{Tr}\left(\sigma_{[\psi]}(\,\cdot\,)\,\cdot\,\right)$, and:

$$\sigma_{[\psi]} \left\langle \mathcal{S}_{[\psi]} \right\rangle := \left\{ \left(\rho_n\right)_{\eta\in\mathcal{L}} \in \mathcal{S}^{\otimes}_{(\mathcal{L},\mathcal{H},\Phi)} \,\middle|\, \sup_{n\in\mathbb{N}} \inf_{n'>n} \left\langle \zeta_{n'\to n} \middle| \left(\mathrm{Tr}_{\mathcal{K}_n} \widetilde{\Phi}_{n'} \circ \rho_{\eta_{n'}} \circ \widetilde{\Phi}^{-1}_{n'}\right) \zeta_{n'\to n} \right\rangle = \mathrm{Tr}\,\rho = 1 \right\}. \tag{2.2.1}$$

In particular, $\rho[\psi] \in \sigma_{[\psi]} \left\langle \mathcal{S}_{[\psi]} \right\rangle$.

*Comparing $\sigma_{[\psi]} \left\langle \mathcal{S}_{[\psi]} \right\rangle$ and $\sigma_{[\psi']} \left\langle \mathcal{S}_{[\psi']} \right\rangle$.* Let $\psi, \psi' \in \mathcal{Z}^{\otimes}_{(\mathbb{N},\mathcal{J})}$. Since $\rho[\psi'] \in \sigma_{[\psi']} \left\langle \mathcal{S}_{[\psi']} \right\rangle$, statement 2.2.2 implies statement 2.2.1. We suppose that statement 2.2.1 holds. Then, the characterization above implies:

$$\begin{aligned} 1 &= \sup_{n\in\mathbb{N}} \inf_{n'>n} \left| \left\langle \psi_{n'} \otimes \ldots \otimes \psi_{n+1} \middle| \psi'_{n'} \otimes \ldots \otimes \psi'_{n+1} \right\rangle \right|^2 \\ &= \sup_{n\in\mathbb{N}} \inf_{n'>n} \prod_{k=n+1}^{n'} \left| \langle \psi_k \mid \psi'_k \rangle_{\mathcal{J}_k} \right|^2 . \end{aligned}$$

This can only holds if there exists $N \in \mathbb{N}$ such that $\forall k \geqslant N$, $\langle \psi_k \mid \psi'_k \rangle_{\mathcal{J}_k} \neq 0$. Then, for any $k \geqslant N$, $0 < \left| \langle \psi_k \mid \psi'_k \rangle_{\mathcal{J}_k} \right| \leqslant \|\psi_k\|_{\mathcal{J}_k} \, \|\psi'_k\|_{\mathcal{J}_k} = 1$, so that $0 \leqslant -\log \left| \langle \psi_k \mid \psi'_k \rangle_{\mathcal{J}_k} \right| < \infty$. Hence, we get:

$$0 = \inf_{n\geqslant N} \sum_{k=n+1}^{\infty} \left( -\log \left| \langle \psi_k \mid \psi'_k \rangle_{\mathcal{J}_k} \right| \right),$$

in other words $\displaystyle\sum_{n=N+1}^{\infty} \left( -\log \left| \langle \psi_n \mid \psi'_n \rangle_{\mathcal{J}_n} \right| \right)$ converges. Then, using that $\log x \leqslant x - 1$ for $x \in$

$]0, \infty[$, this implies that:

$$\sum_{n=N+1}^{\infty} \left(1 - \left| \langle \psi_n \mid \psi'_n \rangle_{\mathcal{J}_n} \right|\right)$$

converges, hence statement 2.2.3 holds.

Reciprocally, we now suppose that 2.2.3 holds. Let $\rho \in \sigma_{[\psi]} \left\langle \mathcal{S}_{[\psi]} \right\rangle$ and let $\epsilon > 0$. Then, there exists $N \in \mathbb{N}$ such that:

$$\forall n' > N, \;\; 1 - \epsilon \leqslant \left\langle \psi_{n'} \otimes \ldots \otimes \psi_{N+1} \middle| \left(\mathrm{Tr}_{\mathcal{K}_N} \widetilde{\Phi}_{n'} \circ \rho_{\eta_{n'}} \circ \widetilde{\Phi}_{n'}^{-1}\right) \psi_{n'} \otimes \ldots \otimes \psi_{N+1} \right\rangle .$$

Thus, for any $n \geqslant N$ and any $n' > n$, we have:

$$\begin{aligned}
1 - \epsilon &\leqslant \left\langle \psi_{n'} \otimes \ldots \otimes \psi_{N+1} \middle| \left(\mathrm{Tr}_{\mathcal{K}_N} \widetilde{\Phi}_{n'} \circ \rho_{\eta_{n'}} \circ \widetilde{\Phi}_{n'}^{-1}\right) \psi_{n'} \otimes \ldots \otimes \psi_{N+1} \right\rangle \\
&= \left\langle \psi_{n'} \otimes \ldots \otimes \psi_{n+1} \otimes \ldots \otimes \psi_{N+1} \middle| \left(\mathrm{Tr}_{\mathcal{K}_N} \widetilde{\Phi}_{n'} \circ \rho_{\eta_{n'}} \circ \widetilde{\Phi}_{n'}^{-1}\right) \psi_{n'} \otimes \ldots \otimes \psi_{n+1} \otimes \ldots \otimes \psi_{N+1} \right\rangle \\
&\leqslant \left\langle \psi_{n'} \otimes \ldots \otimes \psi_{n+1} \middle| \mathrm{Tr}_{\mathcal{K}_{n \to N}} \left(\mathrm{Tr}_{\mathcal{K}_N} \widetilde{\Phi}_{n'} \circ \rho_{\eta_{n'}} \circ \widetilde{\Phi}_{n'}^{-1}\right) \psi_{n'} \otimes \ldots \otimes \psi_{n+1} \right\rangle \\
&= \left\langle \psi_{n'} \otimes \ldots \otimes \psi_{n+1} \middle| \left(\mathrm{Tr}_{\mathcal{K}_n} \widetilde{\Phi}_{n'} \circ \rho_{\eta_{n'}} \circ \widetilde{\Phi}_{n'}^{-1}\right) \psi_{n'} \otimes \ldots \otimes \psi_{n+1} \right\rangle .
\end{aligned}$$

Let $n \geqslant N$, $n' > n$, $\zeta_{n' \to n} := \psi_{n'} \otimes \ldots \otimes \psi_{n+1}$, $\zeta'_{n' \to n} := \psi'_{n'} \otimes \ldots \otimes \psi'_{n+1}$ and $\rho_{n' \to n} := \mathrm{Tr}_{\mathcal{K}_n} \left(\widetilde{\Phi}_{n'} \circ \rho_{\eta_{n'}} \circ \widetilde{\Phi}_{n'}^{-1}\right)$. We have:

$$\begin{aligned}
1 - &\left\langle \zeta'_{n' \to n} \middle| \rho_{n' \to n} \, \zeta'_{n' \to n} \right\rangle = \\
&= 1 - \left\langle \zeta_{n' \to n} \middle| \rho_{n' \to n} \, \zeta_{n' \to n} \right\rangle + \mathrm{Tr}_{\mathcal{K}_{n' \to n}} \rho_{n' \to n} \left( \left|\zeta_{n' \to n}\right\rangle\!\left\langle \zeta_{n' \to n}\right| - \left|\zeta'_{n' \to n}\right\rangle\!\left\langle \zeta'_{n' \to n}\right| \right) \\
&\leqslant \epsilon + \left\| \left|\zeta_{n' \to n}\right\rangle\!\left\langle \zeta_{n' \to n}\right| - \left|\zeta'_{n' \to n}\right\rangle\!\left\langle \zeta'_{n' \to n}\right| \right\|_{\mathcal{A}_{n' \to n}} ,
\end{aligned} \tag{2.2.2}$$

where $\|\cdot\|_{\mathcal{A}_{n' \to n}}$ denotes the operator norm on $\mathcal{K}_{n' \to n}$ and we have used that $\rho_{n' \to n}$ is a density matrix on $\mathcal{K}_{n' \to n}$ (as $\rho \in \sigma_{[\psi]} \left\langle \mathcal{S}_{[\psi]} \right\rangle \subset \mathcal{S}^{\otimes}_{(\mathcal{L}, \mathcal{H}, \Phi)}$). Now, $\|\zeta_{n' \to n}\|_{\mathcal{K}_{n' \to n}} = \|\zeta'_{n' \to n}\|_{\mathcal{K}_{n' \to n}} = 1$, so:

$$\begin{aligned}
\left\| \left|\zeta_{n' \to n}\right\rangle\!\left\langle \zeta_{n' \to n}\right| - \left|\zeta'_{n' \to n}\right\rangle\!\left\langle \zeta'_{n' \to n}\right| \right\|_{\mathcal{A}_{n' \to n}} &= \inf_{\theta \in [0, 2\pi[} \left\| \left|\zeta_{n' \to n}\right\rangle\!\left\langle \zeta_{n' \to n}\right| - \left|e^{i\theta} \zeta'_{n' \to n}\right\rangle\!\left\langle e^{i\theta} \zeta'_{n' \to n}\right| \right\|_{\mathcal{A}_{n' \to n}} \\
&\leqslant 2 \inf_{\theta \in [0, 2\pi[} \left\| \zeta_{n' \to n} - e^{i\theta} \zeta'_{n' \to n} \right\|_{\mathcal{K}_{n' \to n}} \\
&= 2\sqrt{2} \sqrt{1 - \left| \langle \zeta_{n' \to n} \mid \zeta'_{n' \to n} \rangle_{\mathcal{K}_{n' \to n}} \right|} .
\end{aligned} \tag{2.2.3}$$

Let $\epsilon_1 := \min(\epsilon, 2) > 0$ and $\epsilon_2 := -\log\left(1 - \frac{\epsilon_1^2}{8}\right) > 0$. Making use of statement 2.2.3, let $N' \geqslant N$ such that:

$$\sum_{k=} N' + 1^{\infty} \left(1 - \left| \langle \psi_k \mid \psi'_k \rangle_{\mathcal{J}_k} \right|\right) \leqslant \frac{\epsilon_2}{2} .$$

In particular, this implies:

$$\forall k > N', \; \left| \langle \psi_k \mid \psi'_k \rangle_{\mathcal{J}_k} \right| \in \left[1 - \frac{\log 2}{2}, 1\right] \subset \left[\frac{1}{2}, 1\right] .$$

Using that $\log' x = 1/x \leqslant 2$ for any $x \in [1/2, 1]$, we thus get:

$$\forall k > N',\, -\log \left| \langle \psi_k \mid \psi'_k \rangle_{\mathcal{J}_k} \right| \leqslant 2 \left( 1 - \left| \langle \psi_k \mid \psi'_k \rangle_{\mathcal{J}_k} \right| \right).$$

Therefore, we have, for any $n' > N'$:

$$-\log \left| \langle \zeta_{n'\to N'} \mid \zeta'_{n'\to N'} \rangle_{\mathcal{K}_{n'\to N'}} \right| \leqslant -\sum_{k=} N' + 1^{\infty} \log \left| \langle \psi_k \mid \psi'_k \rangle_{\mathcal{J}_k} \right| \leqslant \epsilon_2 \,,$$

and, using the definition of $\epsilon_2$:

$$1 - \left| \langle \zeta_{n'\to N'} \mid \zeta'_{n'\to N'} \rangle_{\mathcal{K}_{n'\to N'}} \right| \leqslant \frac{\epsilon^2}{8} \,. \tag{2.2.4}$$

Finally, for $n' > N'$, combining eqs. (2.2.2), (2.2.3) and (2.2.4) yields:

$$1 - \left\langle \zeta'_{n'\to N'} \middle| \rho_{n'\to N'}\, \zeta'_{n'\to N'} \right\rangle \leqslant \epsilon + 2\sqrt{2} \sqrt{1 - \left| \langle \zeta_{n'\to N'} \mid \zeta'_{n'\to N'} \rangle_{\mathcal{K}_{n'\to N'}} \right|} \leqslant 2\,\epsilon \,.$$

Hence, we get:

$$\sup_{n\in\mathbb{N}} \inf_{n'>n} \left\langle \zeta'_{n'\to n} \middle| \rho_{n'\to n}\, \zeta'_{n'\to n} \right\rangle \geqslant 1 - 2\,\epsilon \,.$$

Since this holds for any $\epsilon > 0$ and the right hand side is bounded above by 1 (for $\mathrm{Tr}\,\rho = 1$), $\rho \in \sigma_{[\psi']} \left\langle \mathcal{S}_{[\psi']} \right\rangle$. Thus, we have proved that $\sigma_{[\psi]} \left\langle \mathcal{S}_{[\psi]} \right\rangle \subset \sigma_{[\psi']} \left\langle \mathcal{S}_{[\psi']} \right\rangle$, and, statement 2.2.3 being symmetric in $\psi$, $\psi'$, we can prove as well $\sigma_{[\psi']} \left\langle \mathcal{S}_{[\psi']} \right\rangle \subset \sigma_{[\psi]} \left\langle \mathcal{S}_{[\psi]} \right\rangle$, ie. statement 2.2.2 holds.

*Construction of* $\mathcal{H}_{\mathrm{seq}}$*,* $\sigma_{\mathrm{seq}}$ *and* $\alpha_{\mathrm{seq}}$*.* Like in the proof of [9, theorem 2.11], the ITP $\mathcal{H}_{\mathrm{seq}}$ of $(\mathcal{J}_n)_{n\in\mathbb{N}}$ can be written as:

$$\mathcal{H}_{\mathrm{seq}} = \overline{\bigoplus_{[|\psi|]} \mathcal{H}_{[|\psi|]}} \,,$$

where the $[|\psi|]$ are the equivalence classes in $\mathcal{Z}^{\otimes}_{(\mathbb{N},\mathcal{J})}$ for the equivalence relation:

$$(\psi_n)_{n\in\mathbb{N}} \simeq (\psi'_n)_{n\in\mathbb{N}} \Leftrightarrow \sum_{n\in\mathbb{N}} \left| 1 - \langle \psi'_n \,,\, \psi_n \rangle_{\mathcal{J}_n} \right| < \infty \,,$$

and the Hilbert space $\mathcal{H}_{[|\psi|]}$ can be identified with $\mathcal{H}_{[\psi]}$ for some representative $\psi$ of $[|\psi|]$ (to check that the inductive limit mentioned in the proof of [9, theorem 2.11] coincides with the one defining $\mathcal{H}_{[\psi]}$, as described in [9, prop. 2.8], we notice that the subsets of $\mathbb{N}$ of the form $\{0, \ldots, n\}$ for some $n \in \mathbb{N}$ constitute a cofinal part, with respect to the inclusion order, in the set of all *finite* subsets of $\mathbb{N}$). We choose such a representative $\psi$ for each equivalence class $[|\psi|]$ and we define:

$$\begin{array}{rccl} \sigma_{\mathrm{seq}} : & \overline{\mathcal{S}}_{\mathrm{seq}} & \to & \overline{\mathcal{S}}^{\otimes}_{(\mathcal{L},\mathcal{H},\Phi)} \\ & \rho & \mapsto & \sum_{[|\psi|]} \sigma_{[\psi]} \left( \Pi_{[|\psi|]}\, \rho\, \Pi_{[|\psi|]} \right) \end{array} ,$$

where, for any equivalence class $[|\psi|]$, $\Pi_{[|\psi|]}$ denotes the orthogonal projection on $\mathcal{H}_{[|\psi|]} \approx \mathcal{H}_{[\psi]}$. Note that the sum over $[|\psi|]$ is absolutely convergent in trace-norm, since, for each $[|\psi|]$, $\Pi_{[|\psi|]}\, \rho\, \Pi_{[|\psi|]}$ is a non-negative traceclass operator and we have $\sum_{[|\psi|]} \mathrm{Tr}_{\mathcal{H}_{[\psi]}} \left( \Pi_{[|\psi|]}\, \rho\, \Pi_{[|\psi|]} \right) = \mathrm{Tr}_{\mathcal{H}_{\mathrm{seq}}}\, \rho < \infty$. We also define:

$$\begin{array}{rcl} \alpha_{\text{seq}} \;:\; \overline{\mathcal{A}}^{\otimes}_{(\mathcal{L},\mathcal{H},\Phi)} & \to & \mathcal{A}_{\text{seq}} \\ A & \mapsto & \sum_{[\|\psi\|]} \Pi_{[\|\psi\|]}\, \alpha_{[\psi]}\big(A\big)\, \Pi_{[\|\psi\|]} \end{array} .$$

Again, the sum involved converges, because the projections $\Pi_{[\|\psi\|]}$ are mutually orthogonal. We have, for any $\rho \in \overline{\mathcal{S}}_{\text{seq}}$ and any $A \in \overline{\mathcal{A}}^{\otimes}_{(\mathcal{L},\mathcal{H},\Phi)}$ :

$$\mathrm{Tr}_{\mathcal{H}_{\text{seq}}}\big(\rho\, \alpha_{\text{seq}}(A)\big) = \mathrm{Tr}\,\big(\sigma_{\text{seq}}(\rho)\, A\big)\,,$$

as follows from the corresponding property fulfilled by each pair $\sigma_{[\psi]}$ , $\alpha_{[\psi]}$ .

Now, let $\psi' \in \mathcal{Z}^{\otimes}_{(\mathbb{N},\mathcal{J})}$ and let $\psi$ be the representative chosen in $[\|\psi'\|]$. The definition of $\simeq$ implies that statement 2.2.3 holds for $\psi$, $\psi'$ , hence:

$$\sigma_{[\psi']}\left\langle \mathcal{S}_{[\psi']} \right\rangle = \sigma_{[\psi]}\left\langle \mathcal{S}_{[\psi]} \right\rangle \subset \sigma_{\text{seq}}\left\langle \mathcal{S}_{\text{seq}} \right\rangle .$$

*Note.* The detailed description of the mapping from the state space of the ITP into the projective state space obtained for this proof, in particular the characterization of which super-selection sectors of the ITP are sent onto identical images (owing to the equivalence relation from statement 2.2.3 being strictly coarser than the relation $\simeq$, as the latter is sensible to the relative *phase* of the factors $\psi_n$ , while the former is not), could be easily generalized to the situation considered in [9, theorem 2.11] (with possibly uncountably many tensor product factors). □

The non-existence of narrow states in the $G = \mathbb{R}$ case [11, prop. 2.14] indirectly proves that the label set $\mathcal{L}_{\text{HF}}$ defined in [7, def. 2.14] does not admit a countable cofinal subset (otherwise, such states could be constructed as described above): indeed, this can be checked directly.

**Proposition 2.3** Let $\mathcal{L}_{\text{HF}}$ be the directed label set defined in [7, def. 2.14]. $\mathcal{L}_{\text{HF}}$ is *uncountable* and for any $\eta' \in \mathcal{L}_{\text{HF}}$:

$$\mathcal{L}_{\text{HF}}[\eta'] := \{\eta \in \mathcal{L}_{\text{HF}} \mid \eta \preccurlyeq \eta'\}$$

is *finite*. Hence, $\mathcal{L}_{\text{HF}}$ does not admit any countable cofinal subset.

**Proof** *$\mathcal{L}_{\text{HF}}$ is uncountable.* Let $\Psi : U \to V$ be an analytical coordinate patch on $\Sigma$, with $U$ an open neighborhood of 0 in $\mathbb{R}^d$, $\epsilon > 0$ such that $B^{(d)}_{\epsilon} \subset U$, and, for any $\tau \neq \tau' \in [0, 2]$, $\breve{e}_{\tau,\tau'}$ be defined as in the proof of [11, prop. 2.14]. For any $\tau \in \,]0, 1[$, let $e^-_\tau$ , resp. $e^+_\tau$ , be the edge corresponding to $\breve{e}_{\tau,0}$ , resp. $\breve{e}_{\tau,1}$ , and $S_\tau$ be the surface corresponding to $\breve{S}_\tau \equiv \breve{e}_{\tau,2}$ . We have, for any $\tau \in \,]0, 1[$ :

$$r(e^-_\tau) \cap r(e^+_\tau) = \left\{\Psi(\tfrac{\epsilon\tau}{2}, 0)\right\} = \{b(e^-_\tau)\} = \{b(e^+_\tau)\}\ ,\quad e^-_\tau \downarrow S_\tau \quad \& \quad e^+_\tau \uparrow S_\tau\,,$$

hence $\eta_\tau := (\gamma_\tau\,,\,\lambda_\tau) \in \mathcal{L}_{\text{HF}}$, with:

$$\gamma_\tau := \{e^-_\tau\,,\, e^+_\tau\} \in \mathcal{L}_{\text{graphs}} \quad \& \quad \lambda_\tau := \big[\,\{S_\tau\}\,\big]_\sim \in \mathcal{L}_{\text{profls}}\,.$$

Moreover, for any $(\tau, s)$, $(\tau', s') \in \,]0, 1[ \times \{0, 1\}$ , we have:

$$\breve{e}_{\tau,s} \sim \breve{e}_{\tau',s'} \Rightarrow r(\breve{e}_{\tau,s}) = r(\breve{e}_{\tau',s'}) \Rightarrow \left[\tfrac{\epsilon\tau}{2}, \tfrac{\epsilon s}{2}\right] = \left[\tfrac{\epsilon\tau'}{2}, \tfrac{\epsilon s'}{2}\right] \Rightarrow (\tau, s) = (\tau', s')\,,$$

hence $\eta_\tau = \eta_{\tau'} \Leftrightarrow \tau = \tau'$. Since $]0, 1[$ is uncountable, so is $\mathcal{L}_{\text{HF}}$.

*The part below any label $\eta'$ is finite.* Let $\gamma' \in \mathcal{L}_{\text{graphs}}$ and let $N = \#\gamma'$. For any $\gamma \preccurlyeq \gamma'$, we define:

$$\mathsf{A}_{\gamma'}(\gamma) := \left\{ \Big( \big(a_{\gamma'\to e}(1), \epsilon_{\gamma'\to e}(1)\big), \ldots, \big(a_{\gamma'\to e}(n_{\gamma'\to e}), \epsilon_{\gamma'\to e}(n_{\gamma'\to e})\big) \Big) \,\Big|\, e \in \gamma \right\} \subset \bigcup_{1\leqslant n\leqslant N} \big(\gamma' \times \{\pm 1\}\big)^n,$$

where, for any $e \in \gamma$, $a_{\gamma'\to e}$, $\epsilon_{\gamma'\to e}$ and $n_{\gamma'\to e}$ have been defined in [7, prop. 3.2], and $n_{\gamma'\to e} \leqslant N$ for $a_{\gamma'\to e}$ is injective $\{1, \ldots, n_{\gamma'\to e}\} \to \gamma'$. We have:

$$\gamma = \left\{ (e'_n)_n^{\epsilon'} \circ \ldots \circ (e'_1)_1^{\epsilon'} \,\Big|\, \big((e'_1, \epsilon'_1), \ldots, (e'_n, \epsilon'_n)\big) \in \mathsf{A}_{\gamma'}(\gamma) \right\},$$

hence $\mathsf{A}_{\gamma'}$ is injective from $\mathcal{L}_{\text{graphs}}[\gamma'] := \{\gamma \in \mathcal{L}_{\text{graphs}} \mid \gamma \preccurlyeq \gamma'\}$ into the set of parts of $\bigcup_{1\leqslant n\leqslant N}(\gamma' \times \{\pm 1\})^n$, so:

$$\#\mathcal{L}_{\text{graphs}}[\gamma'] \leqslant 2^{\frac{(2N)^{N+1}-2N}{2N-1}} < \infty\,.$$

Next, let $\lambda' \in \mathcal{L}_{\text{profls}}$ and let $N' = \#\mathcal{F}(\lambda')$. For any $\lambda \preccurlyeq \lambda'$, we define:

$$\mathsf{H}_{\lambda'}(\lambda) := \left\{ H_{\lambda'\to\overline{F}} \,\middle|\, \overline{F} = F^{\perp} \circ F,\ F \in \mathcal{F}(\lambda) \right\} \subset \mathcal{P}(\mathcal{F}(\lambda'))\,,$$

where, for any $F \in \mathcal{F}(\lambda)$, $F^{\perp}$, $\overline{F}$ and $H_{\lambda'\to\overline{F}}$ have been defined in [7, prop. 3.3], and $\mathcal{P}(\mathcal{F}(\lambda'))$ denotes the set of parts of $\mathcal{F}(\lambda')$. Let $F \in \mathcal{F}(\lambda)$. Since $\lambda \preccurlyeq \lambda'$, there exist $F'_1, \ldots, F'_m \in \mathcal{F}(\lambda')$ ($m \geqslant 1$) such that:

$$F = F_{\urcorner}(\lambda) \circ \bigcup_{i=1}^{m} F'_i\,,$$

and for any $i \in \{1, \ldots, m\}$, $F'_i \in H_{\lambda'\to\overline{F}}$ (see the proof of [7, prop. 3.3]). Moreover, for any $F' \in H_{\lambda'\to\overline{F}} \subset \mathcal{F}(\lambda')$, there exists, by definition of $\mathcal{F}(\lambda')$ [7, prop. 2.11], $e \in F'$, and there exists, by definition of $H_{\lambda'\to\overline{F}}$, $i \in \{1, \ldots, m\}$ such that $e = e'' \circ e'$ with $e' \in F'_i$. Thus, using [7, props. 2.8.6 and 2.11.1], $F'_i = F'$. Hence, $\{F'_1, \ldots, F'_m\} = H_{\lambda'\to\overline{F}}$, in other words:

$$F = F_{\urcorner}(\lambda) \circ \bigcup_{F' \in H_{\lambda'\to\overline{F}}} F'. \tag{2.3.1}$$

Now, using [7, props. 2.11.2 and 2.11.3], we get:

$$F_{\urcorner}(\lambda) = \left( F_{\urcorner}(\lambda) \circ \bigcup_{H \in \mathsf{H}_{\lambda'}(\lambda)} \bigcup_{F' \in H} F' \right)^{\perp} = \left( \bigcup_{H \in \mathsf{H}_{\lambda'}(\lambda)} \bigcup_{F' \in H} F' \right)^{\perp}.$$

Together with eq. (2.3.1) and [7, def. 2.12], this ensures that $\mathsf{H}_{\lambda'}$ is injective from $\mathsf{L}_{\text{profls}}[\lambda'] := \{\lambda \in \mathcal{L}_{\text{profls}} \mid \lambda \preccurlyeq \lambda'\}$ into the set of parts of $\mathcal{P}(\mathcal{F}(\lambda'))$, hence:

$$\#\mathcal{L}_{\text{profls}}[\lambda'] \leqslant 2^{2^{N'}} < \infty\,.$$

Finally, for any $\eta' = (\gamma', \lambda') \in \mathcal{L}_{\text{HF}}$, $\mathcal{L}_{\text{HF}}[\eta'] \subset \mathcal{L}_{\text{graphs}}[\gamma'] \times \mathcal{L}_{\text{profls}}[\lambda']$, so:

$$\#\mathcal{L}_{\text{HF}}[\eta'] \leqslant 2^{2^N + \frac{(2N)^{N+1}-2N}{2N-1}} < \infty,$$

where $N := \#\gamma' = \#\mathcal{F}(\lambda)$. In particular, if $\mathcal{L}'$ is a cofinal subset of $\mathcal{L}_{\text{HF}}$, we have:

$$\mathcal{L}_{\mathsf{HF}} = \bigcup_{\eta' \in \mathcal{L}'} \mathcal{L}_{\mathsf{HF}}[\eta'],$$

so $\mathcal{L}'$ must be uncountable. □

The problem with the label set $\mathcal{L}_{\mathsf{HF}}$ is basically that the slightest deformation or displacement of an edge (or surface) yields an observable, which, according to the structure set up in [7, subsection 3.1], is completely independent of the original one. As argued at the beginning of the present section, this is physically not justifiable and the task of the next subsection will therefore be to formalize the idea that, whenever two edges (or surfaces) are related to each other by an infinitesimal deformation, they should be considered indistinguishable. This will allow us to cut down the algebra to a countable cardinal, while preserving both universality and diffeomorphism-invariance.

## 2.2 Quasi-cofinal sequences: definition and properties

This subsection intends to clarify, in a general setting, which requirements should an increasing sequence of labels satisfy to ensure that it captures the whole algebra of observables, up to small deformations. To give a precise meaning to this notion of 'small deformations', closeness of observables will be defined with respect to a (topological) group a transformations acting on the algebra. Our definition for the action of a group on a projective system (def. 2.5) is inspired by [14, section 3.5].

We call such sequences *quasi-cofinal*, to underline their affinity with cofinal sequences, which, as recalled in the previous subsection, capture the whole algebra *exactly*. Indeed, as we will show below, a rather innocent-looking condition (def. 2.7.3), which can be understood as 'cofinality up to small deformations', is sufficient to prove two strong results:

- the projective system of quantum state spaces obtained by restricting the label set to a quasi-cofinal sequence is universal: it only depends on the original projective system and on the action of the group of transformations (theorem 2.8);
- and any transformation in this group can be approximated by a transformation acting on the restricted projective system (prop. 2.9).

Since the initial label set will eventually be restricted to a part admitting an *increasing cofinal sequence* (and thus automatically directed, for $\mathbb{N}$, $\leqslant$ is directed), we can afford to start from a very large, 'extended' label set $\mathcal{L}^{(\mathsf{ext})}$, which will not even be required to be directed: there can sometimes be a tension between ensuring the pivotal three-spaces consistency condition [9, eq. (2.1.1) and fig. 2.1], and preserving the directedness of the label set, so that it might prove convenient to initially relax this requirement (we will come back to how this added flexibility could be exploited in the outlook).

To make the abstract construction of the present subsection clearer, it will be sufficient for now to imagine $\mathcal{L}^{(\mathsf{ext})}$ to be the semi-analytical version of $\mathcal{L}_{\mathsf{HF}}$ and the group of transformations $\mathcal{T}$ to consist of all semi-analytic diffeomorphisms (see [25, section IV.20], as well as the beginning of [7, subsection 3.2]). In contrast to *fully* analytic diffeomorphisms, semi-analytic ones can be local, so are usable as small deformations, and while the group of semi-analytic diffeomorphisms *do* act on the algebra generated by $\mathcal{L}_{\text{HF}}$ (since, as underlined many times in [7], this algebra is *identical* to the one generated by semi-analytical labels), its action is easier to write down if we use semi-analytical labels: in particular, it can then be put in the convenient form described in def. 2.5.

**Definition 2.4** A projective pre-system of quantum state spaces is a quintuple:

$$\left(\mathcal{L}^{(\text{ext})}, \left(\mathcal{H}_\eta\right)_{\eta\in\mathcal{L}^{(\text{ext})}}, \left(\mathcal{H}_{\eta'\to\eta}\right)_{\eta\preccurlyeq\eta'}, \left(\Phi_{\eta'\to\eta}\right)_{\eta\preccurlyeq\eta'}, \left(\Phi_{\eta''\to\eta'\to\eta}\right)_{\eta\preccurlyeq\eta'\preccurlyeq\eta''}\right)$$

where $\mathcal{L}^{(\text{ext})}$ is a pre-ordered set (*not* necessarily directed) and the rest of [9, def. 2.1] holds. Whenever possible, we will use the shortened notation $/\mathcal{L}^{(\text{ext})}, \mathcal{H}, \Phi/^{\otimes}$ instead of $\Big(\mathcal{L}^{(\text{ext})}, \left(\mathcal{H}_\eta\right)_{\eta\in\mathcal{L}^{(\text{ext})}}, \left(\mathcal{H}_{\eta'\to\eta}\right)_{\eta\preccurlyeq\eta'}, \left(\Phi_{\eta'\to\eta}\right)_{\eta\preccurlyeq\eta'}, \left(\Phi_{\eta''\to\eta'\to\eta}\right)_{\eta\preccurlyeq\eta'\preccurlyeq\eta''}\Big)$.

For any $\eta \preccurlyeq \eta' \in \mathcal{L}^{(\text{ext})}$, we define $\text{Tr}_{\eta'\to\eta} : \overline{\mathcal{S}}_{\eta'} \to \overline{\mathcal{S}}_\eta$ and $\iota_{\eta'\leftarrow\eta} : \mathcal{A}_\eta \to \mathcal{A}_{\eta'}$ as in [9, defs. 2.2 and 2.3]. From [9, eq. (2.1.1)], we have, for any $\eta \preccurlyeq \eta' \preccurlyeq \eta''$:

$$\text{Tr}_{\eta''\to\eta} = \text{Tr}_{\eta'\to\eta} \circ \text{Tr}_{\eta''\to\eta'} \quad \& \quad \iota_{\eta''\leftarrow\eta} = \iota_{\eta''\leftarrow\eta'} \circ \iota_{\eta'\leftarrow\eta}\,.$$

**Definition 2.5** Let $/\mathcal{L}^{(\text{ext})}, \mathcal{H}, \Phi/^{\otimes}$ be a projective pre-system of quantum state spaces and let $\mathcal{T}$ be a group. An action of $\mathcal{T}$ on $/\mathcal{L}^{(\text{ext})}, \mathcal{H}, \Phi/^{\otimes}$ is an action $T, \eta \mapsto T\eta$ of $\mathcal{T}$ on $\mathcal{L}^{(\text{ext})}$ together with families $\left(\text{U}_\eta\right)_{\eta\in\mathcal{L}^{(\text{ext})}}$ and $\left(\text{U}_{\eta'\to\eta}\right)_{\eta\preccurlyeq\eta'}$ such that:

1. $\forall T \in \mathcal{T}$, $\forall \eta \in \mathcal{L}^{(\text{ext})}$, $\text{U}_\eta(T)$ is an isomorphism of Hilbert spaces $\mathcal{H}_\eta \to \mathcal{H}_{T\eta}$;
2. $\forall T \in \mathcal{T}$, $\forall \eta \preccurlyeq \eta' \in \mathcal{L}^{(\text{ext})}$, $T\eta \preccurlyeq T\eta'$ and $\text{U}_{\eta'\to\eta}(T)$ is an isomorphism of Hilbert spaces $\mathcal{H}_{\eta'\to\eta} \to \mathcal{H}_{T\eta'\to T\eta}$, such that:
$$\Phi_{T\eta'\to T\eta} \circ \text{U}_{\eta'}(T) = \left(\text{U}_{\eta'\to\eta}(T) \otimes \text{U}_\eta(T)\right) \circ \Phi_{\eta'\to\eta}\,;$$
3. for any $\eta \in \mathcal{L}^{(\text{ext})}$,
$$\forall T, T' \in \mathcal{T},\ \ \text{U}_{T\eta}(T^{-1}) = \text{U}_\eta(T)^{-1} \quad \& \quad \text{U}_{T\eta}(T') \circ \text{U}_\eta(T) = \text{U}_\eta(T'.T)\,,$$
and for any $\eta \preccurlyeq \eta' \in \mathcal{L}^{(\text{ext})}$:
$$\forall T, T' \in \mathcal{T},\ \ \text{U}_{T\eta'\to T\eta}(T^{-1}) = \text{U}_{\eta'\to\eta}(T)^{-1} \quad \& \quad \text{U}_{T\eta'\to T\eta}(T') \circ \text{U}_{\eta'\to\eta}(T) = \text{U}_{\eta'\to\eta}(T'.T)\,.$$

In particular, for any $\eta \in \mathcal{L}^{(\text{ext})}$, $\text{U}_\eta(\mathbf{1}) = \text{id}_{\mathcal{H}_\eta}$ and, for any $\eta \preccurlyeq \eta' \in \mathcal{L}^{(\text{ext})}$, $\text{U}_{\eta'\to\eta}(\mathbf{1}) = \text{id}_{\mathcal{H}_{\eta'\to\eta}}$.

For any $\eta \in \mathcal{L}^{(\text{ext})}$ and any $T \in \mathcal{T}$, we define $T\triangleright : \overline{\mathcal{S}}_\eta \to \overline{\mathcal{S}}_{T\eta}$, resp. $\mathcal{A}_\eta \to \mathcal{A}_{T\eta}$, via:

$$\forall \rho_\eta \in \overline{\mathcal{S}}_\eta\,,\ T\triangleright\rho_\eta := \text{U}_\eta(T)\,\rho_\eta\,\text{U}_\eta(T)^{-1}\,, \quad \text{resp. } \forall A_\eta \in \mathcal{A}_\eta\,,\ T\triangleright A_\eta := \text{U}_\eta(T)\,A_\eta\,\text{U}_\eta(T)^{-1}\,.$$

From assumption 2.5.3, $\triangleright$ is a group action of $\mathcal{T}$ on $\displaystyle\bigsqcup_{\eta\in\mathcal{L}^{(\text{ext})}} \mathcal{A}_\eta$, and, from assumption 2.5.2, we have, for any $\eta \preccurlyeq \eta' \in \mathcal{L}^{(\text{ext})}$ and any $T \in \mathcal{T}$:

$$\mathrm{Tr}_{T\eta' \to T\eta}\big(T \triangleright \cdot\ \big) = T \triangleright \big(\mathrm{Tr}_{\eta' \to \eta}(\,\cdot\,)\big) \quad \& \quad \iota_{T\eta' \leftarrow T\eta}\big(T \triangleright \cdot\ \big) = T \triangleright \big(\iota_{\eta' \leftarrow \eta}(\,\cdot\,)\big)\,. \tag{2.5.1}$$

**Proposition 2.6** Let $\left/\mathcal{L}^{(\text{ext})}, \mathcal{H}, \Phi\right/^{\otimes}$ be a projective pre-system of quantum state spaces and let $\mathcal{T}$ be a *topological* group acting on $\left/\mathcal{L}^{(\text{ext})}, \mathcal{H}, \Phi\right/^{\otimes}$. We denote by $\mathcal{A}^{\sqcup}_{/\mathcal{L}^{(\text{ext})}, \mathcal{H}, \Phi/}$ the set of all subsets in $\bigsqcup_{\eta \in \mathcal{L}^{(\text{ext})}} \mathcal{A}_\eta$, and we define, for any open neighborhood $V$ of $\mathbf{1}$ in $\mathcal{T}$:

$$\mathcal{U}_V := \left\{ (Y,\, Y') \in \mathcal{A}^{\sqcup}_{/\mathcal{L}^{(\text{ext})}, \mathcal{H}, \Phi/} \times \mathcal{A}^{\sqcup}_{/\mathcal{L}^{(\text{ext})}, \mathcal{H}, \Phi/} \,\middle|\, Y' \subset V \triangleright Y \;\;\&\;\; Y \subset V^{-1} \triangleright Y' \right\},$$

where $V^{-1} := \left\{ T^{-1} \,\middle|\, T \in V \right\}$ and, for any $Y \in \mathcal{A}^{\sqcup}_{/\mathcal{L}^{(\text{ext})}, \mathcal{H}, \Phi/}$, $V \triangleright Y := \left\{ T \triangleright A \,\middle|\, T \in V,\, A \in Y \right\}$.

The set $\left\{ \mathcal{U} \,\middle|\, \exists\, V \text{ open neighborhood of } \mathbf{1} \text{ in } \mathcal{T} \,/\, \mathcal{U}_V \subset \mathcal{U} \right\}$ is a uniform structure on $\sqcup_{\eta \in \mathcal{L}^{(\text{ext})}} \mathcal{A}_\eta$ [2, def. IX.11.1]. For any $Y, Y' \in \mathcal{A}^{\sqcup}_{/\mathcal{L}^{(\text{ext})}, \mathcal{H}, \Phi/}$, we say that $Y'$ is $V$-close to $Y$ if $(Y, Y') \in \mathcal{U}_V$.

**Proof** To prove that $\left\{ \mathcal{U} \,\middle|\, \exists\, V \text{ open neighborhood of } \mathbf{1} \text{ in } \mathcal{T} \,/\, \mathcal{U}_V \subset \mathcal{U} \right\}$ is a uniform structure, we need to prove that:

1. for any open neighborhood $V$ of $\mathbf{1}$ in $\mathcal{T}$, $\left\{ (Y, Y) \,\middle|\, Y \in \mathcal{A}^{\sqcup}_{/\mathcal{L}^{(\text{ext})}, \mathcal{H}, \Phi/} \right\} \subset \mathcal{U}_V$;
2. for any open neighborhoods $V_1$, $V_2$ of $\mathbf{1}$ in $\mathcal{T}$, there exists an open neighborhood $W$ of $\mathbf{1}$ in $\mathcal{T}$ such that $\mathcal{U}_W \subset \mathcal{U}_{V_1} \cap \mathcal{U}_{V_2}$;
3. for any open neighborhood $V$ of $\mathbf{1}$ in $\mathcal{T}$, there exists an open neighborhood $W$ of $\mathbf{1}$ in $\mathcal{T}$ such that:

$$\left\{ (Y,\, Y') \in \mathcal{A}^{\sqcup}_{/\mathcal{L}^{(\text{ext})}, \mathcal{H}, \Phi/} \times \mathcal{A}^{\sqcup}_{/\mathcal{L}^{(\text{ext})}, \mathcal{H}, \Phi/} \,\middle|\, \exists Y'' \in \mathcal{A}^{\sqcup}_{/\mathcal{L}^{(\text{ext})}, \mathcal{H}, \Phi/} \,/\, (Y,\, Y''),\, (Y',\, Y'') \in \mathcal{U}_W \right\} \subset \mathcal{U}_V\,,$$

which can be rewritten as:

$$\left\{ (Y,\, Y') \in \mathcal{A}^{\sqcup}_{/\mathcal{L}^{(\text{ext})}, \mathcal{H}, \Phi/} \times \mathcal{A}^{\sqcup}_{/\mathcal{L}^{(\text{ext})}, \mathcal{H}, \Phi/} \,\middle|\, (Y \cup Y') \subset W^{-1} \triangleright (W \triangleright Y \,\cap\, W \triangleright Y') \right\} \subset \mathcal{U}_V\,.$$

Def. 2.5.3 ensures that for any $A \in \bigsqcup_{\eta \in \mathcal{L}^{(\text{ext})}} \mathcal{A}_\eta$, $\mathbf{1} \triangleright A = A$, so statement 2.6.1 holds. Next, for any open neighborhoods $V_1$, $V_2$ of $\mathbf{1}$ in $\mathcal{T}$, $W := V_1 \cap V_2$ is an open neighborhood of $\mathbf{1}$ in $\mathcal{T}$ and we have $W^{-1} = V_1^{-1} \cap V_2^{-1}$, so $\mathcal{U}_W \subset \mathcal{U}_{V_1} \cap \mathcal{U}_{V_2}$, hence statement 2.6.2 holds.

Let $V$ be an open neighborhood of $\mathbf{1}$ in $\mathcal{T}$. Since $V$ is a topological group $T, T' \mapsto (T')^{-1} . T$ is continuous, hence:

$$\widetilde{W} := \left\{ (T, T') \in \mathcal{T} \times \mathcal{T} \,\middle|\, (T')^{-1} . T \in V \right\}$$

is an open neighborhood of $(\mathbf{1}, \mathbf{1})$ in $\mathcal{T} \times \mathcal{T}$. Then, there exists open neighborhoods $W'$, $W''$ of $\mathbf{1}$ in $\mathcal{T}$ such that $W' \times W'' \subset \widetilde{W}$. Defining $W := W' \cap W''$, $W$ is also an open neighborhood of $\mathbf{1}$ in $\mathcal{T}$, and we have:

$$\forall Y \in \mathcal{A}^{\sqcup}_{/\mathcal{L}^{(\text{ext})}, \mathcal{H}, \Phi/}\,, \quad W^{-1} \triangleright (W \triangleright Y) \subset (V \cap V^{-1}) \triangleright Y\,.$$

Thus, for any $Y,\, Y' \in \mathcal{A}^{\sqcup}_{/\mathcal{L}^{(\text{ext})}, \mathcal{H}, \Phi/}$, we get:

$$W^{-1} \triangleright (W \triangleright Y \,\cap\, W \triangleright Y') \subset \big(W^{-1} \triangleright (W \triangleright Y)\big) \cap \big(W^{-1} \triangleright (W \triangleright Y')\big) \subset (V \triangleright Y) \cap (V^{-1} \triangleright Y')\,,$$

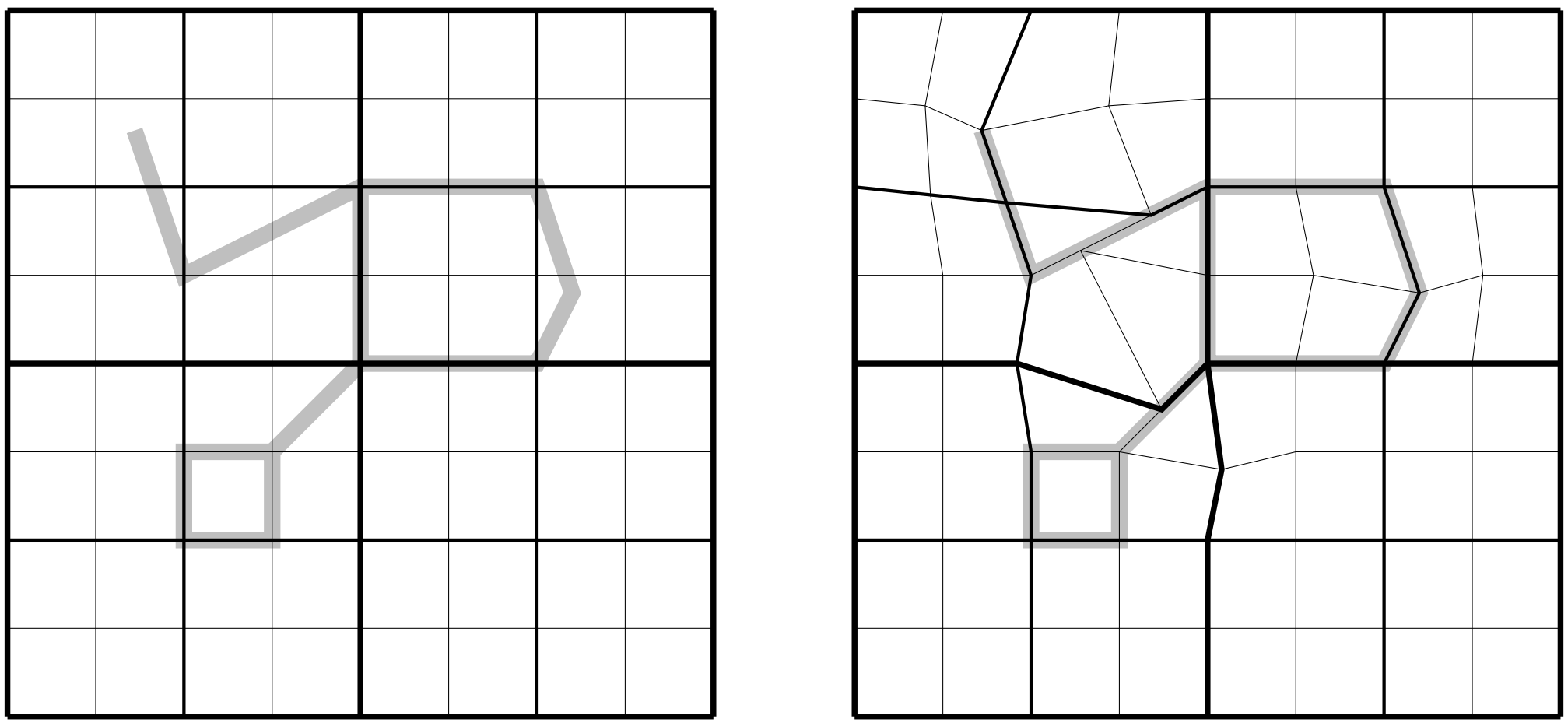

We symbolically represent the quasi-cofinal sequence by finer and finer grids (in black)
and the label to be approximated by thick line segments (in gray)

Figure 2.1 – Deforming the quasi-cofinal sequence to adapt it to an arbitrary label, while preserving all parts that are already in place

which proves statement 2.6.3. □

A first idea to express the notion of an increasing sequence $(\kappa_n)_n$ being 'cofinal up to small deformations' would be to require that, for any label $\eta' \in \mathcal{L}^{(\text{ext})}$, an arbitrarily small deformation of the sequence $(\kappa_n)_n$ should be sufficient to make $\eta'$ a sublabel of some sufficiently fine $\kappa_n$ . However, it turns out that a slighly stronger 'quasi-cofinality' condition (def. 2.7.3) can be much more powerful, leading to the advertised results regarding universality of the restricted projective system and approximation of the transformations in $\mathcal{T}$. The key adjustment, that will be crucial to prove these results, is to require that, whenever $\eta'$ have some parts that are *already adapted* to the quasi-cofinal sequence, the small deformation mentioned above should *leave these parts untouched* (fig. 2.1).

**Definition 2.7** Let $\left/\mathcal{L}^{(\text{ext})}, \mathcal{H}, \Phi\right/^{\otimes}$ be a projective pre-system of quantum state spaces and let $\mathcal{T}$ be a topological group acting on $\left/\mathcal{L}^{(\text{ext})}, \mathcal{H}, \Phi\right/^{\otimes}$. A quasi-cofinal sequence in $\mathcal{L}^{(\text{ext})}$ with respect to this action is a sequence $(\kappa_n)_{n\in\mathbb{N}}$ in $\mathcal{L}^{(\text{ext})}$ such that:

1. $\kappa_o$ is a least element in $\mathcal{L}^{(\text{ext})}$, ie. $\forall \eta \in \mathcal{L}^{(\text{ext})},\ \kappa_o \preccurlyeq \eta$ ;
2. $(\kappa_n)_{n\in\mathbb{N}}$ is increasing, ie. $\forall n \leqslant n' \in \mathbb{N},\ \kappa_n \preccurlyeq \kappa_{n'}$ ;
3. for any open neighborhood $V$ of $\mathbf{1}$ in $\mathcal{T}$, any $n \in \mathbb{N}$ and any $\eta' \succcurlyeq \eta \in \mathcal{L}^{(\text{ext})}$ such that $\eta \preccurlyeq \kappa_n$ , there exists $n' \geqslant n$ and $T \in V$ such that:

$$T\eta = \eta\,,\quad \mathrm{U}_\eta(T) = \mathrm{id}_{\mathcal{H}_\eta} \quad \& \quad \eta' \preccurlyeq T\kappa_{n'}\,.$$

For any quasi-cofinal sequence $\kappa = (\kappa_n)_{n\in\mathbb{N}}$ , we define:

$$\mathcal{L}^{[\kappa]} := \left\{\eta \in \mathcal{L}^{(\text{ext})} \;\middle|\; \exists n \in \mathbb{N} \,\middle/\, \eta \preccurlyeq \kappa_n\right\}.$$

$\mathcal{L}^{[\kappa]}$ is a directed set (for $\mathbb{N}$ is), so that $\left(\mathcal{L}^{[\kappa]}, \mathcal{H}, \Phi\right)$ is a projective system of quantum state spaces.

By construction, $\{\kappa_n \mid n \in \mathbb{N}\}$ is cofinal in $\mathcal{L}^{[\kappa]}$.

While the choice of a quasi-cofinal sequence is far from unique (if only because omitting terms will not void the requirements of def. 2.7), we now want to show that the resulting projective system does *not* depend on this choice. More precisely, the projective systems defined from two different quasi-cofinal sequences $(\kappa_n)_n$ and $(\lambda_m)_m$ can be matched through an arbitrarily small deformation. The idea of the proof is to *interlace* the two sequences, by applying small deformations to both $(\kappa_n)_n$ and $(\lambda_m)_m$: we will then be able to identify their associated projective systems using the same extension/restriction routine that we used repeatedly, eg. in [9, subsection 2.2]. Here is the reason why we insisted, in the formulation of the quasi-cofinality property 2.7.3, to protect against deformation any part of the quasi-cofinal sequence that happens to be already adapted to the target label: this allows us to *recursively* construct the required deformations of $(\kappa_n)_n$ and $(\lambda_m)_m$, by alternately adapting $(\kappa_n)_n$ to a certain $\lambda_m$, and, in the next step, $(\lambda_m)_m$ to a certain $\kappa_n$.

**Theorem 2.8** Let $\big/\mathcal{L}^{(\text{ext})}, \mathcal{H}, \Phi\big/^{\otimes}$ be a projective pre-system of quantum state spaces and let $\mathcal{T}$ be a topological group acting on $\big/\mathcal{L}^{(\text{ext})}, \mathcal{H}, \Phi\big/^{\otimes}$. Let $\kappa = (\kappa_n)_{n\in\mathbb{N}}$ and $\lambda = (\lambda_m)_{m\in\mathbb{N}}$ be two quasi-cofinal sequences in $\mathcal{L}^{(\text{ext})}$ with respect to this action. Then, there exist, for any open neighborhood $V$ of $\mathbf{1}$ in $\mathcal{T}$, a bijective map $\sigma : \overline{\mathcal{S}}^{\otimes}_{(\mathcal{L}^{[\kappa]},\mathcal{H},\Phi)} \to \overline{\mathcal{S}}^{\otimes}_{(\mathcal{L}^{[\lambda]},\mathcal{H},\Phi)}$ and a $C^*$-algebra isomorphism $\alpha : \overline{\mathcal{A}}^{\otimes}_{(\mathcal{L}^{[\lambda]},\mathcal{H},\Phi)} \to \overline{\mathcal{A}}^{\otimes}_{(\mathcal{L}^{[\kappa]},\mathcal{H},\Phi)}$ such that:

1. for any $\rho \in \overline{\mathcal{S}}^{\otimes}_{(\mathcal{L}^{[\kappa]},\mathcal{H},\Phi)}$ and any $A \in \overline{\mathcal{A}}^{\otimes}_{(\mathcal{L}^{[\lambda]},\mathcal{H},\Phi)}$, $\operatorname{Tr} \rho\, \alpha(A) = \operatorname{Tr} \sigma(\rho)\, A$;
2. $\alpha\left\langle \mathcal{A}^{\otimes}_{(\mathcal{L}^{[\lambda]},\mathcal{H},\Phi)} \right\rangle = \mathcal{A}^{\otimes}_{(\mathcal{L}^{[\kappa]},\mathcal{H},\Phi)}$ and for any $A \in \mathcal{A}^{\otimes}_{(\mathcal{L}^{[\lambda]},\mathcal{H},\Phi)}$, $\alpha(A)$ is $V$-close to $A$ (note that $\mathcal{A}^{\otimes}_{(\mathcal{L}^{[\kappa]},\mathcal{H},\Phi)}$, $\mathcal{A}^{\otimes}_{(\mathcal{L}^{[\lambda]},\mathcal{H},\Phi)} \subset \mathcal{A}^{\sqcup}_{/\mathcal{L}^{(\text{ext})},\mathcal{H},\Phi/}$).

**Proof** *Sequences of deformations* $(T_k)_{k\in\mathbb{N}}$, $(S_k)_{k\in\mathbb{N}}$. We define recursively families $(V_k)_{k\in\mathbb{N}}$, $(W_k)_{k\in\mathbb{N}}$ of open neighborhoods of $\mathbf{1}$ in $\mathcal{T}$ as follows:

3. $V_o$ and $W_o$ are chosen (in a way similar to the proof of prop. 2.6) so that:

$$\forall (T, S) \in V_o \times W_o\,,\ T^{-1} . S \in V\,;$$

4. for any $k \geqslant 1$, $V_k$ and $W_k$ are chosen (again in a similar way) so that:

$$\forall T, T' \in V_k\,,\ T . T' \in V_{k-1} \quad \& \quad \forall S, S' \in W_k\,,\ S . S' \in W_{k-1}\,.$$

Next, we define recursively families $(n_k)_{k\in\mathbb{N}}$, $(m_k)_{k\in\mathbb{N}}$ of integers, and families $(T_k)_{k\in\mathbb{N}}$, $(S_k)_{k\in\mathbb{N}}$ of elements in $\mathcal{T}$, in the following way:

5. $n_o := 0$, $m_o := 0$, $T_o := \mathbf{1}$ and $S_o := \mathbf{1}$.
6. For any $k \geqslant 1$, we have $\kappa_{n_{k-1}} \preccurlyeq \big(T_{k-1}^{-1} . S_{k-1}\big)\lambda_{m_{k-1}}$ (as follows from either point 2.8.5 if $k = 1$ or from point 2.8.7 for the step $k-1$ if $k > 1$) and $\kappa_{n_{k-1}} \preccurlyeq \kappa_{n_{k-1}+1}$ (from def. 2.7.2). Using def. 2.7.3 for the cofinal family $(\kappa_n)_{n\in\mathbb{N}}$, we choose $n_k \geqslant n_{k-1} + 1$ and $\widetilde{T}_k \in V_k$ such that:

$$\widetilde{T}_k \kappa_{n_{k-1}} = \kappa_{n_{k-1}}\,,\ \ \mathrm{U}_{\kappa_{n_{k-1}}}(\widetilde{T}_k) = \mathrm{id}_{\mathcal{H}_{\kappa_{n_{k-1}}}} \quad \& \quad \big(T_{k-1}^{-1} . S_{k-1}\big)\lambda_{m_{k-1}} \preccurlyeq \widetilde{T}_k \kappa_{n_k}\,,$$

and, defining $T_k := T_{k-1} . \widetilde{T}_k$, we have:

$$T_k \kappa_{n_{k-1}} = T_{k-1} \kappa_{n_{k-1}} , \quad \mathrm{U}_{\kappa_{n_{k-1}}}(T_k) = \mathrm{U}_{\kappa_{n_{k-1}}}(T_{k-1}) \quad \& \quad S_{k-1} \lambda_{m_{k-1}} \preccurlyeq T_k \kappa_{n_k} .$$

7. For any $k \geqslant 1$, we have $\lambda_{m_{k-1}} \preccurlyeq \left(S_{k-1}^{-1} \,.\, T_k\right) \kappa_{n_k}$ (as follows from point 2.8.6) and $\lambda_{m_{k-1}} \preccurlyeq \lambda_{m_{k-1}+1}$ (from def. 2.7.2). Using def. 2.7.3 for the cofinal family $(\lambda_m)_{m \in \mathbb{N}}$, we choose $m_k \geqslant m_{k-1} + 1$ and $\widetilde{S}_k \in W_k$ such that:

$$\widetilde{S}_k \lambda_{m_{k-1}} = \lambda_{m_{k-1}} , \quad \mathrm{U}_{\lambda_{m_{k-1}}}(\widetilde{S}_k) = \mathrm{id}_{\mathcal{H}_{\lambda_{m_{k-1}}}} \quad \& \quad \left(S_{k-1}^{-1} \,.\, T_k\right) \kappa_{n_k} \preccurlyeq \widetilde{S}_k \lambda_{m_k} ,$$

and, defining $S_k := S_{k-1} \,.\, \widetilde{S}_k$, we have:

$$S_k \lambda_{m_{k-1}} = S_{k-1} \lambda_{m_{k-1}} , \quad \mathrm{U}_{\lambda_{m_{k-1}}}(S_k) = \mathrm{U}_{\lambda_{m_{k-1}}}(S_{k-1}) \quad \& \quad T_k \kappa_{n_k} \preccurlyeq S_k \lambda_{m_k} .$$

For any $k \in \mathbb{N}$, we introduce the notations $K_k := \kappa_{n_k}$ & $L_k := \lambda_{m_k}$, so that we have:

$$T_k K_k \preccurlyeq S_k L_k \preccurlyeq T_{k+1} K_{k+1} ,$$

as well as:

$$T_{k+1} K_k = T_k K_k , \quad \mathrm{U}_{K_k}(T_{k+1}) = \mathrm{U}_{K_k}(T_k) \quad \& \quad S_{k+1} L_k = S_k L_k , \quad \mathrm{U}_{L_k}(S_{k+1}) = \mathrm{U}_{L_k}(S_k) .$$

We also define, for any $k \in \mathbb{N}$, $R_k := S_k^{-1} \,.\, T_k$.

Using the definitions of the sequences $(V_k)_{k \in \mathbb{N}}$, $(W_k)_{k \in \mathbb{N}}$ and $(T_k)_{k \in \mathbb{N}}$, $(S_k)_{k \in \mathbb{N}}$, we can prove recursively that:

$$\forall k \in \mathbb{N}, \quad \forall T \in V_k , \ T_k \,.\, T \in V_o \quad \& \quad \forall S \in W_k , \ S_k \,.\, S \in W_o .$$

Thus, for any $k \in \mathbb{N}$, $(T_k , S_k) \in V_o \times W_o$ (since $\mathbf{1} \in V_k \cap W_k$), so $R_k^{-1} \in V$ and $R_k \in V^{-1}$.

In addition, for any $k \geqslant 1$, we have $n_k > n_{k-1}$, resp. $m_k > m_{k-1}$, so the sequence $(n_k)_{k \in \mathbb{N}}$, resp. $(m_k)_{k \in \mathbb{N}}$, is strictly increasing, and $\mathcal{K} := \{K_k \mid k \in \mathbb{N}\}$, resp. $\mathcal{L} := \{L_k \mid k \in \mathbb{N}\}$, is cofinal in $\mathcal{L}^{[\kappa]}$, resp. $\mathcal{L}^{[\lambda]}$. Thus, from [9, prop. 2.6], there exist a bijective map $\sigma_{\mathcal{K}} : \overline{\mathcal{S}}^{\otimes}_{(\mathcal{L}^{[\kappa]},\mathcal{H},\Phi)} \to \overline{\mathcal{S}}^{\otimes}_{(\mathcal{K},\mathcal{H},\Phi)}$, resp. $\sigma_{\mathcal{L}} : \overline{\mathcal{S}}^{\otimes}_{(\mathcal{L}^{[\lambda]},\mathcal{H},\Phi)} \to \overline{\mathcal{S}}^{\otimes}_{(\mathcal{L},\mathcal{H},\Phi)}$, and a $C^*$-algebra isomorphism $\alpha_{\mathcal{K}} : \overline{\mathcal{A}}^{\otimes}_{(\mathcal{K},\mathcal{H},\Phi)} \to \overline{\mathcal{A}}^{\otimes}_{(\mathcal{L}^{[\kappa]},\mathcal{H},\Phi)}$, resp. $\alpha_{\mathcal{L}} : \overline{\mathcal{A}}^{\otimes}_{(\mathcal{L},\mathcal{H},\Phi)} \to \overline{\mathcal{A}}^{\otimes}_{(\mathcal{L}^{[\lambda]},\mathcal{H},\Phi)}$, such that $\sigma_{\mathcal{K}}$ and $\alpha_{\mathcal{K}}$, resp. $\sigma_{\mathcal{L}}$ and $\alpha_{\mathcal{L}}$, are Tr-intertwined.

*Mapping states.* Let $\rho \in \overline{\mathcal{S}}^{\otimes}_{(\mathcal{K},\mathcal{H},\Phi)}$. For any $k \in \mathbb{N}$, we have $S_k L_k \preccurlyeq T_{k+1} K_{k+1}$, so we can define a non-negative traceclass operator $\widetilde{\rho}_k$ on $\mathcal{H}_{L_k}$ via:

$$\widetilde{\rho}_k := S_k^{-1} \triangleright \left(\mathrm{Tr}_{T_{k+1}K_{k+1} \to S_k L_k} \left(T_{k+1} \triangleright \rho_{K_{k+1}}\right)\right) . \tag{2.8.1}$$

Using $S_k L_k = S_{k+1} L_k$ and $\mathrm{U}_{L_k}(S_k) = \mathrm{U}_{L_k}(S_{k+1})$ together with eq. (2.5.1) yields:

$$\widetilde{\rho}_k = \mathrm{Tr}_{R_{k+1}K_{k+1} \to L_k} \left(R_{k+1} \triangleright \rho_{K_{k+1}}\right) . \tag{2.8.2}$$

Moreover, from eqs. (2.5.1) and (2.8.1), we get:

$$\begin{aligned}
\mathrm{Tr}_{L_{k+1} \to L_k} \widetilde{\rho}_{k+1} &= S_k^{-1} \triangleright \left(\mathrm{Tr}_{T_{k+2}K_{k+2} \to S_k L_k} \, T_{k+2} \triangleright \rho_{K_{k+2}}\right) \\
&= S_k^{-1} \triangleright \left[\mathrm{Tr}_{T_{k+1}K_{k+1} \to S_k L_k} \left(\mathrm{Tr}_{T_{k+2}K_{k+2} \to T_{k+1}K_{k+1}} \, T_{k+2} \triangleright \rho_{K_{k+2}}\right)\right] \\
&= S_k^{-1} \triangleright \left(\mathrm{Tr}_{T_{k+1}K_{k+1} \to S_k L_k} \left(T_{k+1} \triangleright \rho_{K_{k+1}}\right)\right) = \widetilde{\rho}_k ,
\end{aligned}$$

where we have used $L_k \preccurlyeq L_{k+1}$, $S_{k+1} L_k = S_k L_k$ and $\mathrm{U}_{L_k}(S_{k+1}) = \mathrm{U}_{L_k}(S_k)$ in the first line, $S_k L_k \preccurlyeq T_{k+1} K_{k+1} = T_{k+2} K_{k+1} \preccurlyeq T_{k+2} K_{k+2}$ in the second, and $\mathrm{U}_{K_{k+1}}(T_{k+2}) = \mathrm{U}_{K_{k+1}}(T_{k+1})$ and

$\mathrm{Tr}_{K_{k+2}\to K_{k+1}}\, \rho_{K_{k+2}} = \rho_{K_{k+1}}$ in the third.

Now, for any $k \leqslant k' \in \mathbb{N}$, this allows to prove recursively:

$$\mathrm{Tr}_{L_{k'}\to L_k}\, \widetilde{\rho}_{k'} = \widetilde{\rho}_k\,, \tag{2.8.3}$$

so for any $k, k' \in \mathbb{N}$, such that $L_k \preccurlyeq L_{k'}$, either $k \leqslant k'$, in which case eq. (2.8.3) holds, or $k > k'$, in which case $L_k \preccurlyeq L_{k'} \preccurlyeq L_k$, so $\mathrm{Tr}_{L_{k'}\to L_k} = \left(\mathrm{Tr}_{L_k\to L_{k'}}\right)^{-1}$ and eq. (2.8.3) follows from the equality for $k' \leqslant k$. In particular, if $L_k = L_{k'}$, $\widetilde{\rho}_{k'} = \widetilde{\rho}_k$. Thus, the map:

$$\begin{array}{cccc} \sigma_{\mathcal{K}\to\mathcal{L}} & : & \overline{\mathcal{S}}^{\otimes}_{(\mathcal{K},\mathcal{H},\Phi)} & \to & \overline{\mathcal{S}}^{\otimes}_{(\mathcal{L},\mathcal{H},\Phi)} \\ & & \left(\rho_{K_k}\right)_{K_k\in\mathcal{K}} & \mapsto & \left(\mathrm{Tr}_{R_{k+1}K_{k+1}\to L_k}\left(R_{k+1}\triangleright\rho_{K_{k+1}}\right)\right)_{L_k\in\mathcal{L}} \end{array},$$

is well-defined as a map $\overline{\mathcal{S}}^{\otimes}_{(\mathcal{K},\mathcal{H},\Phi)} \to \overline{\mathcal{S}}^{\otimes}_{(\mathcal{L},\mathcal{H},\Phi)}$.

*Mapping observables.* Let $A_{L_k} \in \mathcal{A}_{L_k}$. Since $S_{k+1}L_k = S_kL_k \preccurlyeq T_{k+1}K_{k+1}$ and $\mathrm{U}_{L_k}(S_{k+1}) = \mathrm{U}_{L_k}(S_k)$, we have, using eq. (2.5.1):

$$\left(\widetilde{A_{L_k}}\right)_{k+1} := \iota_{K_{k+1}\leftarrow R^{-1}_{k+1}L_k}\left(R^{-1}_{k+1}\triangleright A_{L_k}\right) = T^{-1}_{k+1}\triangleright\left(\iota_{T_{k+1}K_{k+1}\leftarrow S_kL_k}\left(S_k\triangleright A_{L_k}\right)\right) \in \mathcal{A}_{K_{k+1}}\,.$$

Next, let $A_{L_{k+1}} = \iota_{L_{k+1}\leftarrow L_k}(A_{L_k}) \in \mathcal{A}_{L_{k+1}}$. In a way similar to the computation of $\mathrm{Tr}_{L_{k+1}\to L_k}\, \widetilde{\rho}_{k+1}$ above, we have:

$$\begin{aligned} \left(\widetilde{A_{L_{k+1}}}\right)_{k+2} &= T^{-1}_{k+2}\triangleright\left(\iota_{T_{k+2}K_{k+2}\leftarrow T_{k+1}K_{k+1}} \circ \iota_{T_{k+1}K_{k+1}\leftarrow S_kL_k}\left(S_k\triangleright A_{L_k}\right)\right) \\ &\sim_{\mathcal{K}} T^{-1}_{k+1}\triangleright\left(\iota_{T_{k+1}K_{k+1}\leftarrow S_kL_k}\left(S_k\triangleright A_{L_k}\right)\right) = \left(\widetilde{A_{L_k}}\right)_{k+1}, \end{aligned}$$

with $\sim_{\mathcal{K}}$ defined as in [9, eq. (2.3.2)] for the projective system $(\mathcal{K},\, \mathcal{H},\, \Phi)^{\otimes}$.

We can then prove recursively that for any $k \leqslant k'$ and any $A_{L_k} \in \mathcal{A}_{L_k}$, $A_{L_{k'}} := \iota_{L_{k'}\leftarrow L_k}(A_{L_k}) \in \mathcal{A}_{L_{k'}}$:

$$\left(\widetilde{A_{L_k}}\right)_{k+1} \sim_{\mathcal{K}} \left(\widetilde{A_{L_{k'}}}\right)_{k'+1}.$$

Now, for any $k, k' \in \mathbb{N}$ and any $A_{L_k} \in \mathcal{A}_{L_k}$, $A_{L_{k'}} \in \mathcal{A}_{L_{k'}}$ such that $A_{L_k} \sim_{\mathcal{L}} A_{L_{k'}}$ (with $\sim_{\mathcal{L}}$ defined as in [9, eq. (2.3.2)] for the projective system $(\mathcal{L},\, \mathcal{H},\, \Phi)^{\otimes}$), there exists $\widetilde{k} \in \mathbb{N}$ such that $L_{\widetilde{k}} \succcurlyeq L_k$, $L_{k'}$ and $\iota_{L_{\widetilde{k}}\leftarrow L_k}(A_{L_k}) = \iota_{L_{\widetilde{k}}\leftarrow L_{k'}}(A_{L_{k'}})$. Hence, there exists $k'' \geqslant k, k', \widetilde{k}$, such that $\iota_{L_{k''}\leftarrow L_k}(A_{L_k}) = \iota_{L_{k''}\leftarrow L_{k'}}(A_{L_{k'}}) =: A_{L_{k''}}$, and:

$$\left(\widetilde{A_{L_k}}\right)_{k+1} \sim_{\mathcal{K}} \left(\widetilde{A_{L_{k''}}}\right)_{k''+1} \sim_{\mathcal{K}} \left(\widetilde{A_{L_{k'}}}\right)_{k'+1}.$$

Thus, the map:

$$\begin{array}{cccc} \alpha_{\mathcal{L}\to\mathcal{K}} & : & \mathcal{A}^{\otimes}_{(\mathcal{L},\mathcal{H},\Phi)} & \to & \mathcal{A}^{\otimes}_{(\mathcal{K},\mathcal{H},\Phi)} \\ & & \left[A_{L_k}\right]_{\sim_{\mathcal{L}}} & \mapsto & \left[\iota_{K_{k+1}\leftarrow R^{-1}_{k+1}L_k}\left(R^{-1}_{k+1}\triangleright A_{L_k}\right)\right]_{\sim_{\mathcal{K}}} \end{array},$$

is well-defined as an isometric $*$-algebra morphism $\mathcal{A}^{\otimes}_{(\mathcal{L},\mathcal{H},\Phi)} \to \mathcal{A}^{\otimes}_{(\mathcal{K},\mathcal{H},\Phi)}$, and it can be extended by continuity into a $C^*$-algebra morphism $\overline{\mathcal{A}}^{\otimes}_{(\mathcal{L},\mathcal{H},\Phi)} \to \overline{\mathcal{A}}^{\otimes}_{(\mathcal{K},\mathcal{H},\Phi)}$. Moreover, $\alpha_{\mathcal{L}\to\mathcal{K}}$ and the previously defined map $\sigma_{\mathcal{K}\to\mathcal{L}}$ are Tr-intertwined.

*Inverse mapping.* In a similar fashion, we can define the maps:

$$\begin{array}{rcl} \sigma_{\mathcal{L}\to\mathcal{K}} : & \overline{\mathcal{S}}^{\otimes}_{(\mathcal{L},\mathcal{H},\Phi)} & \to \overline{\mathcal{S}}^{\otimes}_{(\mathcal{K},\mathcal{H},\Phi)} \\ & \left(\rho_{L_k}\right)_{L_k\in\mathcal{L}} & \mapsto \left(\mathrm{Tr}_{R_k^{-1}L_k\to K_k}\left(R_k^{-1}\triangleright\rho_{L_k}\right)\right)_{K_k\in\mathcal{K}} \end{array},$$

and:

$$\begin{array}{rcl} \alpha_{\mathcal{K}\to\mathcal{L}} : & \mathcal{A}^{\otimes}_{(\mathcal{K},\mathcal{H},\Phi)} & \to \mathcal{A}^{\otimes}_{(\mathcal{L},\mathcal{H},\Phi)} \\ & \left[A_{K_k}\right]_{\sim\mathcal{K}} & \mapsto \left[\iota_{L_k\leftarrow R_kK_k}(R_k\triangleright A_{K_k})\right]_{\sim\mathcal{L}} \end{array}.$$

We have $\sigma_{\mathcal{K}\to\mathcal{L}}\circ\sigma_{\mathcal{L}\to\mathcal{K}} = \mathrm{id}_{\overline{\mathcal{S}}^{\otimes}_{(\mathcal{L},\mathcal{H},\Phi)}}$ and $\sigma_{\mathcal{L}\to\mathcal{K}}\circ\sigma_{\mathcal{K}\to\mathcal{L}} = \mathrm{id}_{\overline{\mathcal{S}}^{\otimes}_{(\mathcal{K},\mathcal{H},\Phi)}}$, hence $\sigma_{\mathcal{K}\to\mathcal{L}}$ is bijective with $\sigma^{-1}_{\mathcal{K}\to\mathcal{L}} = \sigma_{\mathcal{L}\to\mathcal{K}}$. Similarly, $\alpha_{\mathcal{L}\to\mathcal{K}}\circ\alpha_{\mathcal{K}\to\mathcal{L}} = \mathrm{id}_{\overline{\mathcal{A}}^{\otimes}_{(\mathcal{K},\mathcal{H},\Phi)}}$ and $\alpha_{\mathcal{K}\to\mathcal{L}}\circ\alpha_{\mathcal{L}\to\mathcal{K}} = \mathrm{id}_{\overline{\mathcal{A}}^{\otimes}_{(\mathcal{L},\mathcal{H},\Phi)}}$, hence $\alpha_{\mathcal{L}\to\mathcal{K}}$ is bijective with $\alpha^{-1}_{\mathcal{L}\to\mathcal{K}} = \alpha_{\mathcal{K}\to\mathcal{L}}$.

*Closeness.* We define the bijective map $\sigma := \sigma_{\mathcal{L}}^{-1}\circ\sigma_{\mathcal{K}\to\mathcal{L}}\circ\sigma_{\mathcal{K}} : \overline{\mathcal{S}}^{\otimes}_{(\mathcal{L}^{[\kappa]},\mathcal{H},\Phi)} \to \overline{\mathcal{S}}^{\otimes}_{(\mathcal{L}^{[\lambda]},\mathcal{H},\Phi)}$, and the $C^*$-algebra isomorphism $\alpha := \alpha_{\mathcal{K}}\circ\alpha_{\mathcal{L}\to\mathcal{K}}\circ\alpha_{\mathcal{L}}^{-1} : \overline{\mathcal{A}}^{\otimes}_{(\mathcal{L}^{[\lambda]},\mathcal{H},\Phi)} \to \overline{\mathcal{A}}^{\otimes}_{(\mathcal{L}^{[\kappa]},\mathcal{H},\Phi)}$. Let $A\in\mathcal{A}^{\otimes}_{(\mathcal{L}^{[\lambda]},\mathcal{H},\Phi)}$ and $\widetilde{A} := \alpha(A)\in\overline{\mathcal{A}}^{\otimes}_{(\mathcal{L}^{[\kappa]},\mathcal{H},\Phi)}$.

Let $\eta\in\mathcal{L}^{[\lambda]}$ and $A_\eta\in\mathcal{A}_\eta$, such that $A_\eta\in A$. Let $n\in\mathbb{N}$ such that $\eta\preccurlyeq\lambda_n$ and $k\in\mathbb{N}$ such that $n\leqslant m_k$. Let $A_k := \iota_{L_k\leftarrow\eta}(A_\eta)\in\mathcal{A}_{L_k}$ and $\widetilde{A}_{k+1} := \iota_{K_{k+1}\leftarrow R^{-1}_{k+1}L_k}(R^{-1}_{k+1}\triangleright A_{L_k})\in\mathcal{A}_{K_{k+1}}$. We have $\widetilde{A} = \left[\widetilde{A}_{k+1}\right]_{\sim\mathcal{L}^{[\kappa]}}$. In particular, $\widetilde{A}\in\mathcal{A}^{\otimes}_{(\mathcal{L}^{[\kappa]},\mathcal{H},\Phi)}$. Moreover, $\widetilde{A}_{k+1} = \iota_{K_{k+1}\leftarrow R^{-1}_{k+1}\eta}(R^{-1}_{k+1}\triangleright A_\eta)$ and $R^{-1}_{k+1}\eta\preccurlyeq R^{-1}_{k+1}L_k\preccurlyeq K_{k+1}$, so $R^{-1}_{k+1}\eta\in\mathcal{L}^{[\kappa]}$ and $R^{-1}_{k+1}\triangleright A_\eta\in\widetilde{A}$. Since $R_{k+1}\in V^{-1}$, $A_\eta\in V^{-1}\triangleright\widetilde{A}$. Therefore, $A\subset V^{-1}\triangleright\widetilde{A}$.

Similarly, for any $\widetilde{A}\in\mathcal{A}^{\otimes}_{(\mathcal{L}^{[\kappa]},\mathcal{H},\Phi)}$, $A := \alpha^{-1}(\widetilde{A})\in\mathcal{A}^{\otimes}_{(\mathcal{L}^{[\lambda]},\mathcal{H},\Phi)}$ and $\widetilde{A}\subset V\triangleright A$, which proves statement 2.8.2. □

It is an immediate corollary of the just proven universality result that any transformation $T\in\mathfrak{T}$ can be approximated, at an arbitrary precision, by a transformation that *stabilizes* the restricted projective system over a quasi-cofinal sequence $(\kappa_n)_n$: indeed, $T$ maps $(\kappa_n)_n$ to a new quasi-cofinal sequence $(\lambda_m)_m$, and the projective system over $(\lambda_m)_m$ can then, by universality, be deformed back into the one over $(\kappa_n)_n$.

**Proposition 2.9** Let $\left/\mathcal{L}^{(\mathrm{ext})},\mathcal{H},\Phi\right/^{\otimes}$ be a projective pre-system of quantum state spaces and let $\mathfrak{T}$ be a topological group acting on $\left/\mathcal{L}^{(\mathrm{ext})},\mathcal{H},\Phi\right/^{\otimes}$. Let $\kappa = (\kappa_n)_{n\in\mathbb{N}}$ be a quasi-cofinal sequence in $\mathcal{L}^{(\mathrm{ext})}$ with respect to this action. Then, there exist, for any $T\in\mathfrak{T}$ and any open neighborhood $V$ of $\mathbf{1}$ in $\mathfrak{T}$, a bijective map $\sigma : \overline{\mathcal{S}}^{\otimes}_{(\mathcal{L}^{[\kappa]},\mathcal{H},\Phi)}\to\overline{\mathcal{S}}^{\otimes}_{(\mathcal{L}^{[\kappa]},\mathcal{H},\Phi)}$ and a $C^*$-algebra isomorphism $\alpha : \overline{\mathcal{A}}^{\otimes}_{(\mathcal{L}^{[\kappa]},\mathcal{H},\Phi)}\to\overline{\mathcal{A}}^{\otimes}_{(\mathcal{L}^{[\kappa]},\mathcal{H},\Phi)}$ such that:

1. for any $\rho\in\overline{\mathcal{S}}^{\otimes}_{(\mathcal{L}^{[\kappa]},\mathcal{H},\Phi)}$ and any $A\in\overline{\mathcal{A}}^{\otimes}_{(\mathcal{L}^{[\kappa]},\mathcal{H},\Phi)}$, $\mathrm{Tr}\,\rho A = \mathrm{Tr}\,\sigma(\rho)\,\alpha(A)$;
2. $\alpha\left\langle\mathcal{A}^{\otimes}_{(\mathcal{L}^{[\kappa]},\mathcal{H},\Phi)}\right\rangle = \mathcal{A}^{\otimes}_{(\mathcal{L}^{[\kappa]},\mathcal{H},\Phi)}$ and for any $A\in\mathcal{A}^{\otimes}_{(\mathcal{L}^{[\kappa]},\mathcal{H},\Phi)}$, $\alpha(A)$ is $V$-close to $T\triangleright A := \{T\triangleright A_\bullet \mid A_\bullet\in A\}$.

**Proof** Let $T\in\mathfrak{T}$ and let $V$ be an open neighborhood of $\mathbf{1}$ in $\mathfrak{T}$. For any $n\in\mathbb{N}$, we define $\underline{\kappa}_n := T\kappa_n$. For any $\underline{\eta}\in\mathcal{L}^{(\mathrm{ext})}$, $\kappa_o\preccurlyeq T^{-1}\underline{\eta}$, hence $\underline{\kappa}_o\preccurlyeq\underline{\eta}$, and, for any $n\leqslant n'\in\mathbb{N}$, $\kappa_n\preccurlyeq\kappa_{n'}$, hence $\underline{\kappa}_n \preccurlyeq \underline{\kappa}_{n'}$. Let $\underline{W}$ be an open neighborhood of $\mathbf{1}$ in $\mathcal{T}$, $n \in \mathbb{N}$ and $\underline{\eta}' \succcurlyeq \underline{\eta} \in \mathcal{L}^{(\text{ext})}$ such that $\underline{\eta} \preccurlyeq \underline{\kappa}_n$. Then, $W := \left\{ S \in \mathcal{T} \middle| T \,.\, S \,.\, T^{-1} \in \underline{W} \right\}$ is an open neighborhood of $\mathbf{1}$ in $\mathcal{T}$, $T^{-1}\underline{\eta}' \succcurlyeq T^{-1}\underline{\eta}$ and $T^{-1}\underline{\eta} \preccurlyeq \kappa_n$. Thus, there exists $n' \geqslant n$ and $S \in W$ such that:

$$S(T^{-1}\underline{\eta}) = T^{-1}\underline{\eta}\,, \quad \mathrm{U}_{T^{-1}\underline{\eta}}(S) = \mathrm{id}_{\mathcal{H}_{T^{-1}\underline{\eta}}} \quad \& \quad T^{-1}\underline{\eta}' \preccurlyeq S\kappa_{n'}\,.$$

Defining $\underline{S} := T \,.\, S \,.\, T^{-1} \in \underline{W}$, this can be rewritten as:

$$\underline{S}\underline{\eta} = \underline{\eta}\,, \quad \mathrm{U}_{\underline{\eta}}(\underline{S}) = \mathrm{id}_{\mathcal{H}_{\underline{\eta}}} \quad \& \quad \underline{\eta}' \preccurlyeq \underline{S}\,\underline{\kappa}_{n'}\,.$$

Therefore, $(\underline{\kappa}_n)_{n\in\mathbb{N}}$ is a quasi-cofinal sequence in $\mathcal{L}^{(\text{ext})}$. Moreover, $\mathcal{L}^{[\underline{\kappa}]} = \left\{ \underline{\eta} \,\middle|\, T^{-1}\underline{\eta} \in \mathcal{L}^{[\kappa]} \right\}$.

Now, we define:

$$\begin{array}{rccc} \underline{\sigma} : & \overline{\mathcal{S}}^{\otimes}_{(\mathcal{L}^{[\kappa]},\mathcal{H},\Phi)} & \to & \overline{\mathcal{S}}^{\otimes}_{(\mathcal{L}^{[\underline{\kappa}]},\mathcal{H},\Phi)} \\ & \left(\rho_\eta\right)_{\eta\in\mathcal{L}^{[\kappa]}} & \mapsto & \left(T \triangleright \rho_{T^{-1}\underline{\eta}}\right)_{\underline{\eta}\in\mathcal{L}^{[\underline{\kappa}]}} \end{array},$$

as well as:

$$\begin{array}{rccc} \underline{\alpha} : & \mathcal{A}^{\otimes}_{(\mathcal{L}^{[\underline{\kappa}]},\mathcal{H},\Phi)} & \to & \mathcal{A}^{\otimes}_{(\mathcal{L}^{[\kappa]},\mathcal{H},\Phi)} \\ & \left[\underline{A}_{\underline{\eta}}\right]_{\sim_{\mathcal{L}^{[\underline{\kappa}]}}} & \mapsto & \left[T^{-1} \triangleright \underline{A}_{\underline{\eta}}\right]_{\sim_{\mathcal{L}^{[\kappa]}}} \end{array}.$$

Def. 2.5 ensures that $\underline{\sigma}$ is well-defined as a bijective map $\overline{\mathcal{S}}^{\otimes}_{(\mathcal{L}^{[\kappa]},\mathcal{H},\Phi)} \to \overline{\mathcal{S}}^{\otimes}_{(\mathcal{L}^{[\underline{\kappa}]},\mathcal{H},\Phi)}$, that $\underline{\alpha}$ is well-defined as an isometric $*$-algebra isomorphism $\mathcal{A}^{\otimes}_{(\mathcal{L}^{[\underline{\kappa}]},\mathcal{H},\Phi)} \to \mathcal{A}^{\otimes}_{(\mathcal{L}^{[\kappa]},\mathcal{H},\Phi)}$ and can be extended by continuity into a $C^*$-algebra isomorphism $\overline{\mathcal{A}}^{\otimes}_{(\mathcal{L}^{[\underline{\kappa}]},\mathcal{H},\Phi)} \to \overline{\mathcal{A}}^{\otimes}_{(\mathcal{L}^{[\kappa]},\mathcal{H},\Phi)}$, that $\underline{\sigma}$ and $\underline{\alpha}$ are Tr-intertwined, and that, for any $\underline{A} \in \mathcal{A}^{\otimes}_{(\mathcal{L}^{[\underline{\kappa}]},\mathcal{H},\Phi)}$, $\underline{\alpha}(\underline{A}) = T^{-1} \triangleright \underline{A}$.

Next, $V^{-1}$ is an open neighborhood of $\mathbf{1}$ in $\mathcal{T}$ and, from theorem 2.8, there exists a bijective map $\widetilde{\sigma} : \overline{\mathcal{S}}^{\otimes}_{(\mathcal{L}^{[\underline{\kappa}]},\mathcal{H},\Phi)} \to \overline{\mathcal{S}}^{\otimes}_{(\mathcal{L}^{[\kappa]},\mathcal{H},\Phi)}$ and a $C^*$-algebra isomorphism $\widetilde{\alpha} : \overline{\mathcal{A}}^{\otimes}_{(\mathcal{L}^{[\kappa]},\mathcal{H},\Phi)} \to \overline{\mathcal{A}}^{\otimes}_{(\mathcal{L}^{[\underline{\kappa}]},\mathcal{H},\Phi)}$, such that $\widetilde{\sigma}$ and $\widetilde{\alpha}$ are Tr-intertwined, $\widetilde{\alpha}\left\langle \mathcal{A}^{\otimes}_{(\mathcal{L}^{[\kappa]},\mathcal{H},\Phi)} \right\rangle = \mathcal{A}^{\otimes}_{(\mathcal{L}^{[\underline{\kappa}]},\mathcal{H},\Phi)}$ and:

$$\forall A \in \mathcal{A}^{\otimes}_{(\mathcal{L}^{[\kappa]},\mathcal{H},\Phi)}\,, \quad \widetilde{\alpha}(A) \text{ is } V^{-1}\text{-close to } A\,.$$

Thus, $\sigma := \widetilde{\sigma} \circ \underline{\sigma}$ is a bijective map $\overline{\mathcal{S}}^{\otimes}_{(\mathcal{L}^{[\kappa]},\mathcal{H},\Phi)} \to \overline{\mathcal{S}}^{\otimes}_{(\mathcal{L}^{[\kappa]},\mathcal{H},\Phi)}$, $\alpha := \widetilde{\alpha}^{-1} \circ \underline{\alpha}^{-1}$ is a $C^*$-algebra isomorphism $\overline{\mathcal{A}}^{\otimes}_{(\mathcal{L}^{[\kappa]},\mathcal{H},\Phi)} \to \overline{\mathcal{A}}^{\otimes}_{(\mathcal{L}^{[\kappa]},\mathcal{H},\Phi)}$ and statements 2.9.1 and 2.9.2 are fulfilled. □

Taking $\mathcal{T}$ to be the group of diffeomorphisms, and assuming the *existence* of a quasi-cofinal sequence for the projective system under consideration (see the next section for a $d = 1$ example; the existence proof in higher dimensions is currently under study), the previous result would allow to define a *discretized* theory, while preserving a notion of *diffeomorphism invariance*: such a theory would only have *countably* many observables, instead of a *continuum* thereof, but it would have *enough* automorphisms to approximate the full group of diffeomorphisms.

Restoring diffeomorphism invariance is indeed a serious concern when discretizing a background-independent theory [13, 19]: for example, a *fixed* lattice does *not* have enough automorphisms to appropriately account for diffeomorphism invariance. It would be tempting to bypass this issue altogether, by declaring such a lattice to be 'non-embedded', in the hope that one would thus quotient out any coordinate dependency. This strategy is however known to give the wrong answer, as it fails to remove enough degrees of freedom from the theory. The intuitive reason for this failure is that the lattice *itself* effectively provides a coordinate system: the disposition of the fields with respect to the lattice should therefore *also* be quotiented out when going diffeomorphism-invariant. In the context of AQG, these difficulties are in particular the reason why diffeomorphisms have to be treated through the so-called 'Extended Master Constraint' approach [24], which can accommodate the absence of an action of the diffeomorphism group on the ITP Hilbert space of AQG.

# 3 One-dimensional toy-model

To illustrate the abstract framework laid in the last subsection, we now want to work out a concrete example. The projective system we are considering here can be though as a one-dimensional version of the label set $\mathcal{L}_{\text{HF}}$ defined in [7, def. 2.14]. To further simplify the argument below, we take $\Sigma$ to be the line segment $]0, 1]$, and for each surface (which, in dimension $d = 1$, is pointlike), we only keep its downward face (ie. the face oriented toward 0): these additional simplifications are purely for convenience, and could be easily lifted. The resulting projective system is precisely the one we set up in [11, subsection 2.2] on the label set $\mathcal{L}^{(\text{aux})}$ (see in particular [11, prop. 2.10] and the proof of [11, prop. 2.14]), except that we will, in the following, keep the gauge group $G$ arbitrary. As group of transformations $\mathcal{T}$ we will use the homeomorphisms of $]0, 1]$, which act on this projective system in a transparent way, mapping a label (which is a finite set of points in $]0, 1]$) to its image, and identifying the associated Hilbert spaces accordingly (this is similar to [14, section 3.5]).

In this section, $G$ will denote a finite-dimensional Lie group and $\mathfrak{g}$ its Lie algebra.

**Proposition 3.1** Let $\mathcal{L}^{(\text{aux})}$ be defined as in [11, prop. 2.10], $G$ be a finite-dimensional Lie-group and $\mu$ be a right-invariant Haar measure on $G$. For any $\kappa \in \mathcal{L}^{(\text{aux})} = (e_1, \ldots, e_n)$ with $0 < e_1 < \ldots < e_n \leqslant 1$, we define $\mathcal{C}_\kappa := \{h : \kappa \to G\}$ and:

$$\begin{array}{rcl} E_\kappa : \mathcal{C}_\kappa & \to & G^{\#\kappa} \\ h & \mapsto & \left(h_{(-1)}(e_k)^{-1}.h(e_k)\right)_{k\in\{1,\ldots,n\}} \end{array},$$

where for any $h \in \mathcal{C}_\kappa$, $h_{(-1)}$ is given by:

$$h_{(-1)}(e_1) = \mathbf{1} \quad \& \quad \forall k \in \{2, \ldots, n\}\,,\ h_{(-1)}(e_k) = h(e_{k-1})\,.$$

$E_\kappa$ is a diffeomorphism $\mathcal{C}_\kappa \to G^{\#\kappa}$ and we equip $\mathcal{C}_\kappa$ with the push-forward measure $\mu_\kappa := E^{-1}_{\kappa,*}\,\mu^n$. Next, for any $\kappa \subset \kappa'$, we define $\mathcal{C}_{\kappa'\to\kappa} := \mathcal{C}_{\kappa'\setminus\kappa}$ as well as:

$$\begin{array}{rcl} \varphi_{\kappa'\to\kappa} & : & \mathcal{C}_{\kappa'} \ \to \ \mathcal{C}_{\kappa'\to\kappa} \times \mathcal{C}_{\kappa} \\ & & h' \ \mapsto \ \left((h'_{(-1)})^{-1}.h'\right)\Big|_{\kappa'\setminus\kappa} \ , \ h'\big|_{\kappa} \end{array} ,$$

Then, these objects can be completed into a factorizing system of measured manifolds $\left(\mathcal{L}^{(\mathrm{aux})}, (\mathcal{C}, \mu), \varphi\right)^{\times}$ [9, def. 3.1] and we denote by $\left(\mathcal{L}^{(\mathrm{aux})}, \mathcal{H}, \Phi\right)^{\otimes}$ the corresponding projective system of quantum state spaces [9, prop. 3.3].

Let $\mathcal{T}$ be the group of homeomorphism $]0, 1] \to ]0, 1]$ equipped with the metric:

$$\forall T, T' \in \mathcal{T}, \ \ d(T, T') := \sup_{x\in]0,1]} |T(x) - T'(x)| \,.$$

For any $T \in \mathcal{T}$ and any $\kappa \in \mathcal{L}^{(\mathrm{aux})}$, we define $T\kappa := T\langle\kappa\rangle$ and:

$$\begin{array}{rcl} \mathrm{U}_{\kappa}(T) & : & \mathcal{H}_{\kappa} \ \to \ \mathcal{H}_{T\kappa} \\ & & \psi \ \mapsto \ \psi\left( \ \cdot \circ T\big|_{\kappa} \right) \end{array} .$$

Then, these objects can be completed into an action of $\mathcal{T}$ on $\left(\mathcal{L}^{(\mathrm{aux})}, \mathcal{H}, \Phi\right)^{\otimes}$ (which, being a projective system of quantum state spaces, is *a fortiori* a projective pre-system of quantum state spaces).

**Proof** Let $\kappa \subset \kappa' \in \mathcal{L}^{(\mathrm{aux})}$ with $\kappa' = (e'_1, \ldots, e'_{n'})$ $(0 < e'_1 < \ldots < e'_{n'} \leqslant 1)$. $G$ being a Lie group, $\varphi_{\kappa'\to\kappa}$ is smooth $\mathcal{C}_{\kappa'} \to \mathcal{C}_{\kappa'\to\kappa} \times \mathcal{C}_{\kappa}$. Next, for any $(j, h) \in \mathcal{C}_{\kappa'\to\kappa} \times \mathcal{C}_{\kappa}$, we define $\widetilde{h}' \in \mathcal{C}_{\kappa'}$ recursively via:

$$\widetilde{h}'(e'_1) = \begin{cases} h(e'_1) & \text{if } e'_1 \in \kappa \\ j(e'_1) & \text{if } e'_1 \notin \kappa \end{cases} \quad \& \quad \forall k \in \{2, \ldots, n'\} \ , \ \widetilde{h}'(e'_k) = \begin{cases} h(e'_k) & \text{if } e'_k \in \kappa \\ \widetilde{h}'(e'_{k-1}).j(e'_k) & \text{if } e'_k \notin \kappa \end{cases} ,$$

and we define $\widetilde{\varphi}_{\kappa'\to\kappa} : \mathcal{C}_{\kappa'\to\kappa} \times \mathcal{C}_{\kappa} \to \mathcal{C}_{\kappa'}$, $(j, h) \to \widetilde{h}'$. $\widetilde{\varphi}_{\kappa'\to\kappa}$ is smooth and we have $\widetilde{\varphi}_{\kappa'\to\kappa} \circ \varphi_{\kappa'\to\kappa} = \mathrm{id}_{\mathcal{C}_{\kappa'}}$ as well as $\varphi_{\kappa'\to\kappa} \circ \widetilde{\varphi}_{\kappa'\to\kappa} = \mathrm{id}_{\mathcal{C}_{\kappa'\to\kappa}} \times \mathrm{id}_{\mathcal{C}_{\kappa}}$, so $\varphi_{\kappa'\to\kappa}$ is a diffeomorphism. In particular, for any $\kappa \in \mathcal{L}^{(\mathrm{aux})}$, $E_{\kappa} = \varphi_{\kappa\to\varnothing} : \mathcal{C}_{\kappa} \to \mathcal{C}_{\kappa'\to\varnothing} \times \mathcal{C}_{\varnothing} \approx G^{\#\kappa}$ is a diffeomorphism.

Next, for any $\kappa \subset \kappa' \in \mathcal{L}^{(\mathrm{aux})}$ with $\kappa = (e_1, \ldots, e_n)$ $(0 < e_1 < \ldots < e_n \leqslant 1)$ and $\kappa' = (e'_1, \ldots, e'_{n'})$ $(0 < e'_1 < \ldots < e'_{n'} \leqslant 1)$, we define an action of $G^n$ (equipped with a Lie group structure using pointwise operations) on $\mathcal{C}_{\kappa'}$ as:

$$\forall g \in G^n, \forall h' \in \mathcal{C}_{\kappa'} \, , \ \ R^{(\kappa,g)}_{\kappa'}(h') := E^{-1}_{\kappa'}\left(E_{\kappa'}(h') \,.\, g'\right),$$

where:

$$\forall g \in G^n, \forall l \in \{1, \ldots, n'\} \ , \ \ g'_l := \begin{cases} g_k & \text{if } e'_l = e_k \\ \mathbf{1} & \text{if } e'_l \notin \kappa \end{cases} .$$

Also, for any $\kappa' \subset \kappa''$, with $\kappa' = (e'_1, \ldots, e'_{n'})$ $(0 < e'_1 < \ldots < e'_{n'} \leqslant 1)$ and $\kappa'' = (e''_1, \ldots, e''_{n''})$ $(0 < e''_1 < \ldots < e''_{n''} \leqslant 1)$, we have:

$$\left(\mathrm{id}_{\mathcal{C}_{\kappa''\to\kappa'}} \times E_{\kappa'}\right) \circ \varphi_{\kappa''\to\kappa'} \circ E^{-1}_{\kappa''}(j'') = \left(j''_m\right)_{m\in\{1,\ldots,n''\}\setminus\{m_l \,|\, 1\leqslant l\leqslant n'\}} , \left(j''_{m_{l-1}+1} \cdot \ldots \cdot j''_{m_l}\right)_{l\in\{1,\ldots,n'\}} , \tag{3.1.1}$$

where $(m_l)_{l\in\{0,\ldots,n'\}}$ is defined via:

$$m_o := 0 \quad \& \quad \forall l \in \{1, \dots, n'\}\,,\; e''_{m_l} = e'_l\,.$$

and we have identified $\mathcal{C}_{\kappa''\to\kappa'} \approx G^{n''-n'}$. Hence, for any $\kappa \subset \kappa' \subset \kappa''$, we obtain:

$$\forall g \in G^{\#\kappa},\; \varphi_{\kappa''\to\kappa'} \circ R^{(\kappa,g)}_{\kappa''} = (\mathrm{id}_{\mathcal{C}_{\kappa''\to\kappa'}} \times R^{(\kappa,g)}_{\kappa'}) \circ \varphi_{\kappa''\to\kappa'}\,. \tag{3.1.2}$$

Applying eq. (3.1.2) repeatedly yields, for any $\kappa \subset \kappa' \subset \kappa''$ and any $g \in G^{\#\kappa}$:

$$\begin{aligned}\big(\mathrm{id}_{\mathcal{C}_{\kappa''\to\kappa'}} \times \varphi_{\kappa'\to\kappa}\big) \circ \varphi_{\kappa''\to\kappa'} \circ \varphi^{-1}_{\kappa''\to\kappa} \circ \big(\mathrm{id}_{\mathcal{C}_{\kappa''\to\kappa}} \times R^{(\kappa,g)}_{\kappa}\big) = \\ = \big(\mathrm{id}_{\mathcal{C}_{\kappa''\to\kappa'}} \times \mathrm{id}_{\mathcal{C}_{\kappa'\to\kappa}} \times R^{(\kappa,g)}_{\kappa}\big) \circ \big(\mathrm{id}_{\mathcal{C}_{\kappa''\to\kappa'}} \times \varphi_{\kappa'\to\kappa}\big) \circ \varphi_{\kappa''\to\kappa'} \circ \varphi^{-1}_{\kappa''\to\kappa}\,.\end{aligned}$$

Now, there exists a map $\varphi_{\kappa''\to\kappa'\to\kappa} : \mathcal{C}_{\kappa''\to\kappa} \to \mathcal{C}_{\kappa''\to\kappa'} \times \mathcal{C}_{\kappa'\to\kappa}$ such that:

$$\forall j'' \in \mathcal{C}_{\kappa''\to\kappa}\,,\quad \big(\mathrm{id}_{\mathcal{C}_{\kappa''\to\kappa'}} \times \varphi_{\kappa'\to\kappa}\big) \circ \varphi_{\kappa''\to\kappa'} \circ \varphi^{-1}_{\kappa''\to\kappa}(j'', \mathbf{1}_\kappa) = \big(\varphi_{\kappa''\to\kappa'\to\kappa}(j''), \mathbf{1}_\kappa\big)\,,$$

with $\mathbf{1}_\kappa : \kappa \to G,\; e \mapsto \mathbf{1}$, so using:

$$\forall h \in \mathcal{C}_\kappa\,,\quad h = R^{(\kappa, E_\kappa(h))}_{\kappa}(\mathbf{1}_\kappa)\,,$$

we get:

$$\big(\mathrm{id}_{\mathcal{C}_{\kappa''\to\kappa'}} \times \varphi_{\kappa'\to\kappa}\big) \circ \varphi_{\kappa''\to\kappa'} = (\varphi_{\kappa''\to\kappa'\to\kappa} \times \mathrm{id}_{\mathcal{C}_\kappa}) \circ \varphi_{\kappa''\to\kappa}\,. \tag{3.1.3}$$

In particular, this requires that $\varphi_{\kappa''\to\kappa'\to\kappa}$ is a diffeomorphism $\mathcal{C}_{\kappa''\to\kappa} \to \mathcal{C}_{\kappa''\to\kappa'} \times \mathcal{C}_{\kappa'\to\kappa}$. Therefore, $\big(\mathcal{L}^{(\mathrm{aux})}, \mathcal{C}, \varphi\big)$ is a factorizing system of smooth, finite dimensional manifolds.

Moreover, for any $\kappa \subset \kappa' \in \mathcal{L}^{(\mathrm{aux})}$ and any $g \in G^{\#\kappa}$, we have:

$$R^{(\kappa,g)}_{\kappa',*}\,\mu_{\kappa'} = \mu_{\kappa'}\,,$$

for $\mu$ is invariant under right-translations on $G$. Eq. (3.1.2) then yields, for any $g \in G^{\#\kappa}$:

$$\varphi_{\kappa'\to\kappa,*}\,\mu_{\kappa'} = (\mathrm{id}_{\mathcal{C}_{\kappa'\to\kappa}} \times R^{(\kappa,g)}_{\kappa})_*\,[\varphi_{\kappa'\to\kappa,*}\,\mu_{\kappa'}]$$

Using the uniqueness up to a global positive factor of the right-invariant Haar measure on $G^{\#\kappa}$, we conclude that there exists a smooth measure $\mu_{\kappa'\to\kappa}$ on $\mathcal{C}_{\kappa'\to\kappa}$ such that:

$$\varphi_{\kappa'\to\kappa,*}\mu_{\kappa'} = \mu_{\kappa'\to\kappa} \times \mu_\kappa\,.$$

Finally, from eq. (3.1.3), this also implies:

$$\varphi_{\kappa''\to\kappa'\to\kappa,*}\mu_{\kappa''\to\kappa} = \mu_{\kappa''\to\kappa'} \times \mu_{\kappa'\to\kappa}\,.$$

*Action of* $\mathcal{T}$. Let $T$ be an homeomorphism $]0,1] \to ]0,1]$. From the intermediate value theorem, $T$ is strictly monotonous. If $T$ would be decreasing, we would have $T\left\langle\, ]0,1] \right\rangle \subset [T(1), 1]$ with $T(1) > 0$, which would contradict the surjectivity of $T$, so $T$ has to be strictly increasing. $T$ can then be extended into an homeomorphism $\widetilde{T} : [0,1] \to [0,1]$ with $\widetilde{T}(0) := 0$, and, in particular, $T$ is uniformly continuous. Therefore, for any $\epsilon > 0$ there exists $\tau > 0$ such that:

$$\forall x, y \in ]0,1]\,,\quad |x - y| \leqslant \tau \Rightarrow |T(x) - T(y)| \leqslant \epsilon/2\,,$$

so, for any $T' \in \mathcal{T}$:

$$\begin{aligned}&\forall S, S' \in \mathcal{T},\; \big(d(S,T) \leqslant \epsilon/2 \quad \& \quad d(S',T') \leqslant \tau\big) \\ &\qquad \Rightarrow\; d(S \circ S', T \circ T') \leqslant d(S \circ S', T \circ S') + d(T \circ S', T \circ T') \leqslant \epsilon\,.\end{aligned}$$

Similarly, there exists $\tau' > 0$ such that:

$$\forall S \in \mathfrak{T},\, d(S, T) \leqslant \tau' \;\Rightarrow\; d\left(T^{-1} \circ S, \mathrm{id}_{]0,1]}\right) = d\left(T^{-1}, S^{-1}\right) \leqslant \epsilon\,.$$

Thus, the metric $d$ makes $\mathfrak{T}$ into a topological group.

For any $\kappa \in \mathcal{L}^{(\mathrm{aux})}$, we have $\mathrm{id}_{]0,1]}\left\langle \kappa \right\rangle = \kappa$, $\forall T,\, T\left\langle \kappa \right\rangle \in \mathcal{L}^{(\mathrm{aux})}$ and $\forall T, T' \in \mathfrak{T}$, $(T \circ T')\left\langle \kappa \right\rangle = T\big\langle T'\left\langle \kappa \right\rangle \big\rangle$, hence $\kappa \mapsto T\kappa$ is a group action of $\mathfrak{T}$ on $\mathcal{L}^{(\mathrm{aux})}$. Next, for any $T \in \mathfrak{T}$, $T$ is strictly increasing, as mentioned above, so:

$$\forall h \in \mathcal{C}_{T\kappa}\,,\quad E_\kappa\left(h \circ T\big|_\kappa\right) = E_{T\kappa}(h)\,.$$

Therefore, $h \mapsto h \circ T\big|_\kappa$ is a volume-preserving diffeomorphism $(\mathcal{C}_{T\kappa}\,, \mu_{T\kappa}) \to (\mathcal{C}_\kappa\,, \mu_\kappa)$, so $\mathrm{U}_\kappa(T)$ is a unitary isomorphism from $\mathcal{H}_\kappa = L_2\left(\mathcal{C}_\kappa\,, d\mu_\kappa\right)$ into $\mathcal{H}_{T\kappa} = L_2\left(\mathcal{C}_{T\kappa}\,, d\mu_{T\kappa}\right)$. Moreover, we have:

$$\forall h \in \mathcal{C}_\kappa\,,\, h \circ \mathrm{id}_{]0,1]}\big|_\kappa = h \quad \& \quad \forall T, T' \in \mathfrak{T},\, \forall h \in \mathcal{C}_{(T' \circ T)\kappa}\,,\, h \circ (T' \circ T)\big|_\kappa = \left(h \circ T'\big|_{T\kappa}\right) \circ T\big|_\kappa\,,$$

so that:

$$\mathrm{U}_\kappa\left(\mathrm{id}_{]0,1]}\right) = \mathrm{id}_{\mathcal{H}_\kappa} \quad \& \quad \forall T, T' \in \mathfrak{T},\, \mathrm{U}_{T\kappa}(T') \circ \mathrm{U}_\kappa(T) = \mathrm{U}_\kappa(T' \circ T)\,, \tag{3.1.4}$$

which, in addition, implies:

$$\forall T \in \mathfrak{T},\, \mathrm{U}_{T\kappa}(T^{-1}) = \mathrm{U}_\kappa(T)^{-1}\,. \tag{3.1.5}$$

Let $\kappa \subset \kappa' \in \mathcal{L}^{(\mathrm{aux})}$ and $T \in \mathfrak{T}$. We then have $T\left\langle \kappa \right\rangle \subset T\left\langle \kappa' \right\rangle$ as well as $T\left\langle \kappa' \setminus \kappa \right\rangle = T\left\langle \kappa' \right\rangle \setminus T\left\langle \kappa \right\rangle$ (for $T$ is bijective), and, $T$ being strictly increasing:

$$\forall h' \in \mathcal{C}_{T\kappa'}\,,\quad \left(h' \circ T\big|_{\kappa'}\right)_{(-1)} = h'_{(-1)} \circ T\big|_{\kappa'}\,.$$

Thus, defining:

$$\begin{array}{rccl} \mathrm{U}_{\kappa' \to \kappa}(T) : & \mathcal{H}_{\kappa' \to \kappa} & \to & \mathcal{H}_{T\kappa' \to T\kappa} \\ & \psi & \mapsto & \psi\left(\,\cdot \circ T\big|_{\kappa' \setminus \kappa}\right) \end{array},$$

we get:

$$\mathrm{U}_{\kappa'}(T) \circ \Phi^{-1}_{\kappa' \to \kappa} = \Phi^{-1}_{T\kappa' \to T\kappa} \circ \left(\mathrm{U}_{\kappa' \to \kappa}(T) \otimes \mathrm{U}_\kappa(T)\right). \tag{3.1.6}$$

Therefore, $\mathrm{U}_{\kappa' \to \kappa}(T)$ is a unitary isomorphism $\mathcal{H}_{\kappa' \to \kappa} \to \mathcal{H}_{T\kappa' \to T\kappa}$ satisfying def. 2.5.2.

Finally, let $\kappa \subset \kappa' \in \mathcal{L}^{(\mathrm{aux})}$ and let $T, T' \in \mathfrak{T}$. Using eqs. (3.1.5) and (3.1.6), we have:

$$\begin{aligned} \mathrm{U}_{T\kappa' \to T\kappa}(T^{-1}) \otimes \mathrm{U}_{T\kappa}(T^{-1}) &= \Phi_{\kappa' \to \kappa} \circ \mathrm{U}_{T\kappa'}(T^{-1}) \circ \Phi^{-1}_{T\kappa' \to T\kappa} \\ &= \left[\Phi_{T\kappa' \to T\kappa} \circ \mathrm{U}_{\kappa'}(T) \circ \Phi^{-1}_{\kappa' \to \kappa}\right]^{-1} \\ &= \mathrm{U}_{\kappa' \to \kappa}(T)^{-1} \otimes \mathrm{U}_\kappa(T)^{-1}\,, \end{aligned}$$

hence $\mathrm{U}_{T\kappa' \to T\kappa}(T^{-1}) = \mathrm{U}_{\kappa' \to \kappa}(T)^{-1}$, and, similarly, using eqs. (3.1.4) and (3.1.6), $\mathrm{U}_{T\kappa' \to T\kappa}(T') \circ \mathrm{U}_{\kappa' \to \kappa}(T) = \mathrm{U}_{\kappa' \to \kappa}(T' \circ T)$. □

A simple quasi-cofinal sequence for this system is obtained by taking the set $K$ of all points of the form $k/2^n$ (as will become clear in the course of the present section, it in fact does not really matter how these points are layered on finer and finer levels of the quasi-cofinal sequence). To prove that the quasi-cofinality property 2.7.3 holds, we will consider a set of points $\kappa'$ (to be approximated), of which a subset $\kappa$ already belong to $K$. For any two *successive* points $e$, $e'$ in $\kappa$, we simply need to approximate the points in $\kappa' \cap \left]e, e'\right[$ by points in $K \cap \left]e, e'\right[$, and to find a deformation of $\left]e, e'\right[$ mapping one set of points into the other: we can, for example, use the corresponding piecewise-linear mapping.

**Proposition 3.2** We consider the same objects as in prop. 3.1. Then, there exists a quasi-cofinal sequence in $\mathcal{L}^{(\text{aux})}$ with respect to the action of $\mathcal{T}$ on $\left(\mathcal{L}^{(\text{aux})}, \mathcal{H}, \Phi\right)^{\otimes}$.

**Proof** We define $\kappa_o := \varnothing \in \mathcal{L}^{(\text{aux})}$ and, for any $n \geqslant 1$:

$$\kappa_n := \left\{ \frac{k}{2^n} \,\middle|\, k \in \{1, \dots, 2^n\} \right\} \in \mathcal{L}^{(\text{aux})}.$$

We have, for any $\kappa \in \mathcal{L}^{(\text{aux})}$, $\kappa_o \subset \kappa$, and, for any $n \leqslant n'$, $\kappa_n \subset \kappa_{n'}$.

Let $V$ be an open neighborhood of $\mathbf{1}$ in $\mathcal{T}$, $n \in \mathbb{N}$ and $\kappa \subset \kappa' \in \mathcal{L}^{(\text{aux})}$ with $\kappa \subset \kappa_n$. Let $\epsilon > 0$ such that $B_\epsilon \subset V$ (where $B_\epsilon$ denotes the closed ball of radius $\epsilon$ and center $\text{id}_{]0,1]}$ in $\mathcal{T}$) and let $\epsilon' > 0$ be given by:

$$\epsilon' = \min_{\substack{e,e' \in \kappa' \\ e \neq e'}} |e - e'|.$$

Let $n' \geqslant n$ such that $\frac{1}{2^{n'}} < \frac{1}{2}\min(\epsilon, \epsilon')$. For any $k' \in \left\{1, \dots, 2^{n'} - 1\right\}$, we define $I_{k'} \subset \left]0, 1\right]$ via:

$$I_1 := \left]0, \frac{3}{2^{n'+1}}\right], \quad I_{2^{n'}-1} := \left]1 - \frac{3}{2^{n'+1}}, 1\right[$$
$$\& \quad \forall k' \in \left\{2, \dots, 2^{n'} - 2\right\},\ I_{k'} := \left]\frac{2k'-1}{2^{n'+1}}, \frac{2k'+1}{2^{n'+1}}\right].$$

The family $(I_{k'})_{k' \in \left\{1, \dots, 2^{n'}-1\right\}}$ is then a partition of $\left]0, 1\right[$, and, for each $k' \in \left\{1, \dots, 2^{n'} - 1\right\}$, $I_{k'}$ contains at most one element of $\kappa'$. For $k' \in \left\{0, \dots, 2^{n'}\right\}$, we define $e'_{k'}$ via:

$$e'_o := 0, \;\; e'_{2^{n'}} := 1 \quad \& \quad \forall k' \in \left\{1, \dots, 2^{n'} - 1\right\},\ e'_{k'} = \begin{cases} e' & \text{if } I_{k'} \cap \kappa' = \{e'\} \\ \frac{k'}{2^{n'}} & \text{else} \end{cases}.$$

For any $k' \in \left\{1, \dots, 2^{n'} - 1\right\}$, $e'_{k'} \in I_{k'}$, thus the family $(e_{k'})_{k' \in \left\{0, \dots, 2^{n'}\right\}}$ is strictly increasing. Next, we define a piecewise linear homeomorphism $T : \left]0, 1\right] \to \left]0, 1\right]$, via:

$$\forall k' \in \left\{1, \dots, 2^{n'}\right\},\ \forall x \in \left]\frac{k'-1}{2^{n'}}, \frac{k'}{2^{n'}}\right], \quad T(x) := (k' - 2^{n'}x)\, e_{k'-1} + (1 - k' + 2^{n'}x)\, e_{k'}.$$

For any $x \in \left]0, 1\right]$, we have $|T(x) - x| \leqslant \frac{2}{2^{n'}} < \epsilon$, so $T \in V$.

Let $e' \in \kappa'$. If $e' = 1$, then $e' = T(1) = T(\frac{2^{n'}}{2^{n'}})$. Otherwise, $k' \in \left]0, 1\right[$, hence there exists $k' \in \left\{1, \dots, 2^{n'} - 1\right\}$ such that $e' \in I_{k'}$, so $e' = e'_{k'} = T(\frac{k'}{2^{n'}})$. Therefore, $\kappa' \subset T\kappa_{n'}$. Finally, for any $k' \in \left\{1, \dots, 2^{n'}\right\}$ such that $\frac{k'}{2^{n'}} \in \kappa'$, we have $e'_{k'} = \frac{k'}{2^{n'}}$, so $T(\frac{k'}{2^{n'}}) = \frac{k'}{2^{n'}}$. Thus, $T|_{\kappa' \cap \kappa_{n'}} = \mathrm{id}_{\kappa' \cap \kappa_{n'}}$ and, in particular, $T|_{\kappa} = \mathrm{id}_{\kappa}$ (for $\kappa \subset \kappa' \cap \kappa_n \subset \kappa' \cap \kappa_{n'}$), yielding $T\kappa = \kappa$ and $\mathrm{U}_{\kappa}(T) = \mathrm{id}_{\mathcal{H}_{\kappa}}$. □

In theorem 2.8, we stated that, given two quasi-cofinal sequences $(\kappa_n)_n$ and $(\lambda_m)_m$, the restricted projective system over $(\kappa_n)_n$ can be deformed into the one over $(\lambda_m)_m$. However, the deformation maps, acting at the level of the quantum states and observables, that we constructed when proving this result *a priori* do *not* arise from an element of $\mathcal{T}$: instead, while their action on any *given* label $\kappa_n$ coincides with the action of an element $T_n \in \mathcal{T}$, this element could be $n$-dependent. In prop. 3.3 below, we show that *in the particular example* we are now considering, the deformation actually *does arise* from an homeomorphism of $]0, 1]$, in other words that $T_n$ can be made independent of $n$. An important ingredient of the proof is to realize that a bijection from $]0, 1]$ into itself is an homeomorphism if, and only if, it is strictly increasing (basically because the topology of $\mathbb{R}$ is closely related to its order).

In particular, this result allows to derive a stronger version of prop. 2.9: not only can any element in $\mathcal{T}$ be approximated by an automorphism of the restricted projective system, but the set of automorphisms that arise in this way moreover forms a *group* (in fact, it is a subgroup of $\mathcal{T}$). This property could be relevant when turning to the imposition of the 'diffeomorphism' constraints (aka. the homeomorphism invariance in the present context).

**Proposition 3.3** We consider the same objects as in prop. 3.1 and we assume that $G$ is non-trivial (ie. it is not reduced to $\{\mathbf{1}\}$). Let $\kappa$ and $\lambda$ be quasi-cofinal sequences in $\mathcal{L}^{(\mathrm{aux})}$ and let $\epsilon > 0$. For any $C^*$-algebra isomorphism $\alpha : \overline{\mathcal{A}}^{\otimes}_{(\mathcal{L}^{[\lambda]}, \mathcal{H}, \Phi)} \to \overline{\mathcal{A}}^{\otimes}_{(\mathcal{L}^{[\kappa]}, \mathcal{H}, \Phi)}$ fulfilling statement 2.8.2 with respect to $V := B_\epsilon$ (the closed ball of radius $\epsilon$ and center $\mathrm{id}_{]0,1]}$ in $\mathcal{T}$) and any bijective map $\sigma : \overline{\mathcal{S}}^{\otimes}_{(\mathcal{L}^{[\kappa]}, \mathcal{H}, \Phi)} \to \overline{\mathcal{S}}^{\otimes}_{(\mathcal{L}^{[\lambda]}, \mathcal{H}, \Phi)}$ such that $\sigma$, $\alpha$ are Tr-intertwined, there exists $T \in B_\epsilon$ such that $\mathcal{L}^{[\lambda]} = \left\{T\eta \mid \eta \in \mathcal{L}^{[\kappa]}\right\}$ and:

$$\forall A \in \mathcal{A}^{\otimes}_{(\mathcal{L}^{[\lambda]}, \mathcal{H}, \Phi)},\ \alpha(A) = T^{-1} \triangleright A,$$

$$\forall \rho \in \overline{\mathcal{S}}^{\otimes}_{(\mathcal{L}^{[\kappa]}, \mathcal{H}, \Phi)},\ \sigma(\rho) = \left(T \triangleright \rho_{T^{-1}\underline{\eta}}\right)_{\underline{\eta} \in \mathcal{L}^{[\lambda]}}.$$

In particular, for any quasi-cofinal sequence $\kappa = (\kappa_n)_{n \in \mathbb{N}}$, the group:

$$\mathcal{T}^{[\kappa]} := \{T \in \mathcal{T} \mid T\langle K \rangle = K\}, \text{ with } K := \bigcup_{n \in \mathbb{N}} \kappa_n,$$

acts on the projective system $\left(\mathcal{L}^{[\kappa]}, \mathcal{H}, \Phi\right)^{\otimes}$ and is dense in $\mathcal{T}$.

**Proof** *Determination of $T \in B_\epsilon$.* Let $\kappa = (\kappa_n)_{n \in \mathbb{N}}$, $\lambda = (\lambda_m)_{m \in \mathbb{N}}$ be two quasi-cofinal sequences in $\mathcal{L}^{(\mathrm{aux})}$ with respect to the action of $\mathcal{T}$ on $\left(\mathcal{L}^{(\mathrm{aux})}, \mathcal{H}, \Phi\right)^{\otimes}$, and let $V := B_\epsilon$ for some $\epsilon > 0$. Let $\alpha : \overline{\mathcal{A}}^{\otimes}_{(\mathcal{L}^{[\lambda]}, \mathcal{H}, \Phi)} \to \overline{\mathcal{A}}^{\otimes}_{(\mathcal{L}^{[\kappa]}, \mathcal{H}, \Phi)}$ be a $C^*$-algebra isomorphism fulfilling statement 2.8.2 with respect to $V$ and let $\sigma : \overline{\mathcal{S}}^{\otimes}_{(\mathcal{L}^{[\kappa]}, \mathcal{H}, \Phi)} \to \overline{\mathcal{S}}^{\otimes}_{(\mathcal{L}^{[\lambda]}, \mathcal{H}, \Phi)}$ be a bijective map such that $\sigma$, $\alpha$ are Tr-intertwined.

For any bounded measurable function $m : G \to \mathbb{C}$, any $e \in ]0, 1]$ and any $\eta \in \mathcal{L}^{(\mathrm{aux})}$ such that $e \in \eta$, we define an operator $\widehat{\mathrm{h}^{(e,m)}_{\eta}} \in \mathcal{A}_\eta$ via:

$$\forall \psi \in \mathcal{H}_\eta\,, \forall h \in \mathcal{C}_\eta\,, \quad \left[\widehat{\mathrm{h}_\eta^{(e,m)}}\, \psi\right](h) := m\big(h(e)\big)\, \psi(h)\,.$$

For any $\eta, \eta' \in \mathcal{L}^{(\mathrm{aux})}$ such that $e \in \eta \cap \eta'$, we have, from the definition of $\varphi_{\eta' \to \eta}$ in prop. 3.1:

$$\widehat{\mathrm{h}_\eta^{(e,m)}} \sim \widehat{\mathrm{h}_{\eta'}^{(e,m)}}\,.$$

We denote by $\widehat{\mathrm{h}^{(e,m)}}$ the corresponding element of $\mathcal{A}^{\otimes}_{(\mathcal{L}^{(\mathrm{aux})}, \mathcal{H}, \Phi)}$. Next, for any $e \in L := \bigcup_{m\in\mathbb{N}} \lambda_m$, we define $\widehat{\mathrm{h}_{[\lambda]}^{(e,m)}} := \widehat{\mathrm{h}^{(e,m)}} \cap \bigsqcup_{\eta \in \mathcal{L}^{[\lambda]}} \mathcal{A}_\eta \neq \varnothing$. Since $\mathcal{L}^{[\lambda]}$ is cofinal in $\mathcal{L}^{(\mathrm{aux})}$, $\widehat{\mathrm{h}_{[\lambda]}^{(e,m)}} \in \mathcal{A}^{\otimes}_{(\mathcal{L}^{[\lambda]}, \mathcal{H}, \Phi)}$. By assumption, $\alpha\big(\widehat{\mathrm{h}_{[\lambda]}^{(e,m)}}\big)$ is then $V$-close to $\widehat{\mathrm{h}_{[\lambda]}^{(e,m)}}$, so there exists $S_{(e,m)} \in V$ such that:

$$S^{-1}_{(e,m)} \triangleright \widehat{\mathrm{h}_{\{e\}}^{(e,m)}} \in \alpha\big(\widehat{\mathrm{h}_{[\lambda]}^{(e,m)}}\big)\,.$$

This implies $S^{-1}_{(e,m)}\ \{e\} = \big\{S^{-1}_{(e,m)}(e)\big\} \in \mathcal{L}^{[\kappa]}$, ie. $S^{-1}_{(e,m)}(e) \in K := \bigcup_{n\in\mathbb{N}} \kappa_n$, and:

$$\alpha\big(\widehat{\mathrm{h}_{[\lambda]}^{(e,m)}}\big) = \Big[S^{-1}_{(e,m)} \triangleright \widehat{\mathrm{h}_{\{e\}}^{(e,m)}}\Big]_{\sim_{\mathcal{L}^{[\kappa]}}} = \Big[\widehat{\mathrm{h}_{\{S^{-1}_{(e,m)}(e)\}}^{\left(S^{-1}_{(e,m)}(e),m\right)}}\Big]_{\sim_{\mathcal{L}^{[\kappa]}}} = \widehat{\mathrm{h}_{[\kappa]}^{\left(S^{-1}_{(e,m)}(e),m\right)}}\,,$$

where the second equality comes from the definition of the action of $\mathcal{T}$ on $\big(\mathcal{L}^{(\mathrm{aux})}, \mathcal{H}, \Phi\big)$ (prop. 3.1). We *choose* a *non-constant*, smooth, compactly supported map $m_o : G \to \mathbb{C}$ (thanks to $G$ being non-trivial), and for any $e \in L$, we define $\tilde{t}(e) := S^{-1}_{(e,m_o)}(e)$. Since $S_{(e,m_o)} \in V = B_\epsilon$, we have $\big|\tilde{t}(e) - e\big| \leqslant \epsilon$.

Next, we consider $e, e' \in L$ such that $e < e'$. From the assumptions on $\alpha$, we get that:

$$\alpha\big(\widehat{\mathrm{h}_{[\lambda]}^{(e,m_o)}} - \widehat{\mathrm{h}_{[\lambda]}^{(e',m_o)}}\big) = \widehat{\mathrm{h}_{[\kappa]}^{(\tilde{t}(e),m_o)}} - \widehat{\mathrm{h}_{[\kappa]}^{(\tilde{t}(e'),m_o)}}$$

is $V$-close to $\widehat{\mathrm{h}_{[\lambda]}^{(e,m_o)}} - \widehat{\mathrm{h}_{[\lambda]}^{(e',m_o)}}$. Hence, there exists $S \in V$ such that:

$$S^{-1} \triangleright \big(\widehat{\mathrm{h}_{\{e,e'\}}^{(e,m_o)}} - \widehat{\mathrm{h}_{\{e,e'\}}^{(e',m_o)}}\big) \in \widehat{\mathrm{h}_{[\kappa]}^{(\tilde{t}(e),m_o)}} - \widehat{\mathrm{h}_{[\kappa]}^{(\tilde{t}(e'),m_o)}}\,,$$

which can be rewritten, in a way similar to above, as:

$$\widehat{\mathrm{h}_{\eta'}^{(S^{-1}(e),m_o)}} - \widehat{\mathrm{h}_{\eta'}^{(S^{-1}(e'),m_o)}} \sim \widehat{\mathrm{h}_{\eta}^{(\tilde{t}(e),m_o)}} - \widehat{\mathrm{h}_{\eta}^{(\tilde{t}(e'),m_o)}}\,,$$

where $\eta := \big\{\tilde{t}(e),\, \tilde{t}(e')\big\}$ and $\eta' := \big\{S^{-1}(e),\, S^{-1}(e')\big\}$. Let $\eta'' := \eta \cup \eta'$. For any $\psi \in \mathcal{H}_{\eta''}$, we then have:

$$\forall h \in \mathcal{C}_{\eta''}\,, \quad \Big[m_o \circ h\big(S^{-1}(e)\big) - m_o \circ h\big(S^{-1}(e')\big)\Big]\psi(h) = \Big[m_o \circ h\big(\tilde{t}(e)\big) - m_o \circ h\big(\tilde{t}(e')\big)\Big]\psi(h)\,.$$

Since this holds for any $\psi$, this implies:

$$\forall h \in \mathcal{C}_{\eta''}\,, \quad m_o \circ h\big(S^{-1}(e)\big) - m_o \circ h\big(S^{-1}(e')\big) = m_o \circ h\big(\tilde{t}(e)\big) - m_o \circ h\big(\tilde{t}(e')\big)\,,$$

and therefore, $m_o$ being non-constant, $S^{-1}(e) = \tilde{t}(e)$ and $S^{-1}(e') = \tilde{t}(e')$ (this can be check by distinguishing cases and plugging specific values of $h$ in the equality above). Now, $S^{-1} \in \mathcal{T}$, ie. it is an homeomorphism $]0,\, 1] \to ]0,\, 1]$, so, by the intermediate value theorem, it is *strictly increasing*. Thus, $\tilde{t}(e) < \tilde{t}(e')$.

To summarize, we have proved that there exists a strictly increasing map $\tilde{t} : L \to K$ such that:

$$\forall e \in L, \quad \alpha\left(\widehat{\mathsf{h}^{(e,m_o)}_{[\lambda]}}\right) = \widehat{\mathsf{h}^{(\tilde{t}(e),m_o)}_{[\kappa]}} \quad \& \quad \left|\tilde{t}(e) - e\right| \leqslant \epsilon \,.$$

Applying the same reasoning to the $C^*$-algebra isomorphism $\alpha^{-1} : \overline{\mathcal{A}}^{\otimes}_{(\mathcal{L}^{[\kappa]},\mathcal{H},\Phi)} \to \overline{\mathcal{A}}^{\otimes}_{(\mathcal{L}^{[\lambda]},\mathcal{H},\Phi)}$, which satisfies statement 2.8.2 with respect to $V^{-1} = B_\epsilon = V$, yields that $\tilde{t}$ is actually *bijective* $L \to K$ (for, $m_o$ being non-constant, we have $\widehat{\mathsf{h}^{(e,m_o)}} = \widehat{\mathsf{h}^{(e',m_o)}} \Rightarrow e = e'$). Now, let $x \in \,]0, 1]$ and let $\epsilon' > 0$. We have $\varnothing \subset \{x\} \in \mathcal{L}^{(\text{aux})}$ and $\varnothing \subset \lambda_o$, so, from 2.7.3, there exists $S \in B_{\epsilon'}$ and $m \in \mathbb{N}$ such that $\{x\} \subset S\lambda_m$, ie. there exists $y := S^{-1}(x) \in L$ such that $|x - y| \leqslant \epsilon'$. In other words, $L$ is dense in $]0, 1]$, and, similarly, so is $K$. Then the topology on $L$ (as a subspace of $]0, 1]$), coincides with its order topology, ie. it is generated by the base:

$$\left\{ \,]e, e'[\, \cap L \,\middle|\, e, e' \in L \right\} \cup \left\{ \,]0, e[\, \cap L \,\middle|\, e \in L \right\} \cup \left\{ \,]e, 1] \cap L \,\middle|\, e \in L \right\}.$$

The same holds for $K$ and $\tilde{t}$, being bijective and strictly increasing, is thus an homeomorphism $L \to K$. Then, we can extend $\tilde{t}$ into a continuous function $\widetilde{T} : \,]0, 1] \to [0, 1]$. Let $x \in \,]0, 1]$. $L$ being dense, there exists $e \in \,]0, x[\, \cap L$, and, for any $e' \in \,]e, x] \cap L$, $\widetilde{T}(e') = \tilde{t}(e') > \tilde{t}(e)$. By continuity, this implies $\widetilde{T}(x) \geqslant \tilde{t}(e) > 0$. So $\widetilde{T}$ is actually valued in $]0, 1]$. Similarly, $\tilde{t}^{-1}$ can be extended into a continuous function $T : \,]0, 1] \to [0, 1]$, and we have $T \circ \widetilde{T}\Big|_{L \to L} = \mathrm{id}_L$, as well as $\widetilde{T} \circ T\Big|_{K \to K} = \mathrm{id}_K$. Therefore, $T$ is an homeomorphism $]0, 1] \to \,]0, 1]$ with $T^{-1} = \widetilde{T}$. Moreover, for any $e \in K$, $|T(e) - e| = \left|\tilde{t}^{-1}(e) - \tilde{t}\left(\tilde{t}^{-1}(e)\right)\right| \leqslant \epsilon$, so by continuity $T \in B_\epsilon$.

Since $\mathcal{L}^{(\text{aux})}$ consists of *finite* subsets of $]0, 1]$ and the sequence $(\lambda_m)_{m\in\mathbb{N}}$ is increasing, we have:

$$\mathcal{L}^{[\lambda]} := \left\{\eta \in \mathcal{L}^{(\text{aux})} \,\middle|\, \exists m \in \mathbb{N} \,/\, \eta \subset \lambda_m \right\} = \left\{\eta \in \mathcal{L}^{(\text{aux})} \,\middle|\, \eta \subset L \right\},$$

and, similarly, $\mathcal{L}^{[\kappa]} = \left\{\eta \in \mathcal{L}^{(\text{aux})} \,\middle|\, \eta \subset K \right\}$. Thus, $T\langle K\rangle = L$ yields $\mathcal{L}^{[\lambda]} = \left\{T\eta \,\middle|\, \eta \in \mathcal{L}^{[\kappa]}\right\}$. Like in the proof of prop. 2.9, we can then define an isometric $*$-algebra isomorphism $\widetilde{\alpha}$ via:

$$\begin{array}{rcl} \widetilde{\alpha} \;:\; \mathcal{A}^{\otimes}_{(\mathcal{L}^{[\lambda]},\mathcal{H},\Phi)} & \to & \mathcal{A}^{\otimes}_{(\mathcal{L}^{[\kappa]},\mathcal{H},\Phi)} \\ A & \mapsto & T^{-1} \triangleright A \end{array},$$

and extend it into a $C^*$-algebra isomorphism $\widetilde{\alpha} : \overline{\mathcal{A}}^{\otimes}_{(\mathcal{L}^{[\lambda]},\mathcal{H},\Phi)} \to \overline{\mathcal{A}}^{\otimes}_{(\mathcal{L}^{[\kappa]},\mathcal{H},\Phi)}$. We now want to prove that $\alpha = \widetilde{\alpha}$.

*$\alpha$ and $\sigma$ come from $T$.* Let $e \in L$ and let $m : G \to \mathbb{C}$ be a bounded measurable function. Like above, there exists $S \in V$ such that:

$$S^{-1} \triangleright \left(\widehat{\mathsf{h}^{(e,m)}_{\{e\}}} - \widehat{\mathsf{h}^{(e,m_o)}_{\{e\}}}\right) \in \widehat{\mathsf{h}^{(S^{-1}_{(e,m)}(e),m_o)}_{[\kappa]}} - \widehat{\mathsf{h}^{(T^{-1}(e),m_o)}_{[\kappa]}},$$

and, defining $\eta' := \left\{S^{-1}(e),\, S^{-1}_{(e,m)}(e),\, T^{-1}(e)\right\}$, we get:

$$\forall h \in \mathcal{C}_{\eta'}, \quad m \circ h\left(S^{-1}(e)\right) - m_o \circ h\left(S^{-1}(e)\right) = m \circ h\left(S^{-1}_{(e,m)}(e)\right) - m_o \circ h\left(T^{-1}(e)\right).$$

This requires either $S^{-1}_{(e,m)}(e) = T^{-1}(e)$, or $m$ being constant. In the latter case, $\widehat{\mathsf{h}^{(e',m)}_{[\kappa]}} = m(\mathbf{1}) \left[\mathrm{id}_{\mathcal{H}_\varnothing}\right]_{\sim_{\mathcal{L}^{[\kappa]}}}$ for any $e' \in K$, so in both cases:

$$\alpha\big(\widehat{\mathsf{h}^{(e,m)}_{[\lambda]}}\big) = \widehat{\mathsf{h}^{(T^{-1}(e),m)}_{[\kappa]}} = \widetilde{\alpha}\big(\widehat{\mathsf{h}^{(e,m)}_{[\lambda]}}\big)\,.$$

Next, let $e \in L$ and $A_{\{e\}} \in \mathcal{A}_{\{e\}}$. Again, there exists $S \in V$ such that:

$$S^{-1} \triangleright A_{\{e\}} \in \alpha\left(\left[A_{\{e\}}\right]_{\mathcal{L}^{[\lambda]}}\right).$$

Defining $\eta := \left\{S^{-1}(e)\right\}$ and $\eta' := \left\{T^{-1}(e),\, S^{-1}(e)\right\}$, we get, for any bounded measurable function $m : G \to \mathbb{C}$:

$$\left[\iota_{\eta'\leftarrow\eta}\big(S^{-1} \triangleright A_{\{e\}}\big),\, \widehat{\mathsf{h}^{(T^{-1}(e),m)}_{\eta'}}\right] \in \alpha\left(\left[A^{(e,m)}_{\{e\}}\right]_{\mathcal{L}^{[\lambda]}}\right),$$

where $A^{(e,m)}_{\{e\}} := \left[A_{\{e\}}\,,\, \widehat{\mathsf{h}^{(e,m)}_{\{e\}}}\right]$.

If we suppose $T^{-1}(e) < S^{-1}(e)$, we have, from the definition of $\varphi_{\eta'\to\left\{S^{-1}(e)\right\}}$:

$$\widehat{\mathsf{h}^{(T^{-1}(e),m)}_{\eta'}} = \Phi^{-1}_{\eta'\to\eta} \circ \left(\widehat{\mathsf{h}^{(T^{-1}(e),m)}_{\eta'\to\eta}} \otimes \mathrm{id}_{\mathcal{H}_\eta}\right) \circ \Phi_{\eta'\to\eta}\,,$$

where:

$$\forall \psi \in \mathcal{H}_{\eta'\to\eta}\,,\ \forall j \in \mathcal{C}_{\eta'\to\eta}\,,\ \left[\widehat{\mathsf{h}^{(T^{-1}(e),m)}_{\eta'\to\eta}}\,\psi\right](j) = m \circ j\big(T^{-1}(e)\big)\,\psi(j)\,.$$

Thus, we get $\alpha\left(\left[A^{(e,m)}_{\{e\}}\right]_{\mathcal{L}^{[\lambda]}}\right) = 0$. $\alpha$ being a $C^*$-algebra isomorphism, this implies that, for any bounded measurable function $m$, $A_{\{e\}}$ commutes with $\widehat{\mathsf{h}^{(e,m)}_{\{e\}}}$. Now, let $\psi_o$ be a smooth, nowhere-vanishing, square-integrable function on $\mathcal{C}_{\{e\}}$ (eg. using a partition of unity [12, lemma 2.16 and theorem 2.18] with suitable dumping factors). For any smooth, compactly supported function $\psi$ on $\mathcal{C}_{\{e\}}$, $m_\psi := {}^{\psi}\!/\!_{\psi_o}$ is a bounded measurable function on $\mathcal{C}_{\{e\}} \approx G$ and we have:

$$\widehat{\mathsf{h}^{(e,m_\psi)}_{\{e\}}}\,\psi_o = \psi\,.$$

We then get:

$$\forall h \in \mathcal{C}_{\{e\}}\,,\ \ \left[A_{\{e\}}\,\psi\right](h) = \left[\widehat{\mathsf{h}^{(e,m_\psi)}_{\{e\}}}\,A_{\{e\}}\,\psi_o\right](h) = \frac{\left[A_{\{e\}}\,\psi_o\right](h)}{\psi_o(h)}\,\psi(h)\,.$$

Since this holds for any smooth compactly supported $\psi$, $m := \left[A_{\{e\}}\,\psi_o\right]/\psi_o$ is almost everywhere bounded by the operator norm $\left\|A_{\{e\}}\right\|_{\mathcal{A}_{\{e\}}}$ of $A_{\{e\}}$, and, by density, $A_{\{e\}} = \widehat{\mathsf{h}^{(e,m)}_{\{e\}}}$.

Therefore, we either have $\alpha\left(\left[A_{\{e\}}\right]_{\mathcal{L}^{[\lambda]}}\right) = \widetilde{\alpha}\left(\left[A_{\{e\}}\right]_{\mathcal{L}^{[\lambda]}}\right)$ or $S^{-1}(e) \leqslant T^{-1}(e)$. In the latter case, the same reasoning applied to $\alpha^{-1}$ yields $S^{-1}(e) = T^{-1}(e)$, thus $S^{-1}\big|_{\{e\}} = T^{-1}\big|_{\{e\}}$, so that, in this case too, $\alpha\left(\left[A_{\{e\}}\right]_{\mathcal{L}^{[\lambda]}}\right) = \widetilde{\alpha}\left(\left[A_{\{e\}}\right]_{\mathcal{L}^{[\lambda]}}\right)$.

We want to prove:

$$\forall \eta \in \mathcal{L}^{[\lambda]},\ \forall A_\eta \in \mathcal{A}_\eta\,,\ \alpha\left([A_\eta]_{\mathcal{L}^{[\lambda]}}\right) = \widetilde{\alpha}\left([A_\eta]_{\mathcal{L}^{[\lambda]}}\right). \tag{3.3.1}$$

We proceed by recursion on $\#\eta$. The case $\#\eta = 1$ has been treated above and it implies the case $\#\eta = 0$. We now suppose that eq. (3.3.1) holds up to $\#\eta = N \geqslant 1$. Let $\eta = (e_1\,, \ldots, e_{N+1}) \in \mathcal{L}^{[\lambda]}$ (with $0 < e_1 < \ldots < e_{N+1} \leqslant 1$). Let $\eta_1 := \{e_1\}$ and $\eta_2 := \eta \setminus \{e_1\}$. Using eq. (3.1.1), we have:

$$\forall j \in G^{N+1},\ \left(\mathrm{id}_{\mathcal{C}_{\eta\to\eta_1}} \times E_{\eta_1}\right) \circ \varphi_{\eta\to\eta_1} \circ E_\eta^{-1}(j) = \left(j_2\,,\,\dots\,,\,j_{N+1}\ ;\ j_1\right),$$

$$\&\quad \left(\mathrm{id}_{\mathcal{C}_{\eta\to\eta_2}} \times E_{\eta_2}\right) \circ \varphi_{\eta\to\eta_2} \circ E_\eta^{-1}(j) = \left(j_1\ ;\ (j_1\,.\,j_2)\,,\,j_3\,,\,\dots\,,\,j_{N+1}\right).$$

We define, for any $\eta' \in \mathcal{L}^{(\mathrm{aux})}$, the unitary isomorphism $\Phi_{\eta'} : \mathcal{H}_{\eta'} \to \mathcal{H}^{\otimes \#\eta'}$, $\psi \mapsto \psi \circ E_\eta^{-1}$, where $\mathcal{H} := L_2(G,\, d\mu)$. We then have, for any bounded operator $A_1$ on $\mathcal{H}$:

$$\forall \psi \in \mathcal{H}_\eta\,,\ \forall j \in G^{N+1},\ \left[\Phi_\eta \circ \iota_{\eta\leftarrow\eta_1}\left(\Phi_{\eta_1}^{-1} A_1\, \Phi_{\eta_1}\right)(\psi)\right](j) =$$
$$= \left[A_1\left(\Phi_\eta(\psi)\left(\,\cdot\,,\,j_2\,,\,\dots\,,\,j_{N+1}\right)\right)\right](j_1) = \left(A_1 \otimes \mathrm{id}_{\mathcal{H}}^{\otimes N}\right)\left(\Phi_\eta(\psi)\right)(j)\,,$$

and, for any bounded operator $A_2$ on $\mathcal{H}^{\otimes N}$:

$$\forall \psi \in \mathcal{H}_\eta\,,\ \forall j \in G^{N+1},\ \left[\Phi_\eta \circ \iota_{\eta\leftarrow\eta_2}\left(\Phi_{\eta_2}^{-1} A_2\, \Phi_{\eta_2}\right)(\psi)\right](j) =$$
$$= \left[A_2\left(\Phi_\eta(\psi)\left(j_1\,,\,(j_1^{-1}\,.\,\cdot\,),\,\cdot\,,\,\dots\,,\,\cdot\,\right)\right)\right]\left((j_1\,.\,j_2),\,j_3\,,\,\dots\,,\,j_{N+1}\right)$$
$$= \left(\mathrm{id}_{\mathcal{H}} \otimes \left(\mathsf{L}_{(j_1)}^{-1} A_2\, \mathsf{L}_{(j_1)}\right)\right)\left(\Phi_\eta(\psi)\right)(j)\,,$$

where, for any $u \in G$:

$$\begin{array}{rccl} \mathsf{L}_{(u)} : & \mathcal{H}^{\otimes N} & \to & \mathcal{H}^{\otimes N} \\ & \psi & \mapsto & \left[(j_2\,,\,\dots\,,\,j_{N+1}) \mapsto \psi\left((u^{-1}\,.\,j_2),\,j_3\,,\,\dots\,,\,j_{N+1}\right)\right] \end{array}.$$

Chaining these expressions we get:

$$\forall \psi \in \mathcal{H}_\eta\,,\ \forall j \in G^{N+1},\ \left[\Phi_\eta \circ \iota_{\eta\leftarrow\eta_2}\left(\Phi_{\eta_2}^{-1} A_2\, \Phi_{\eta_2}\right) \circ \iota_{\eta\leftarrow\eta_1}\left(\Phi_{\eta_1}^{-1} A_1\, \Phi_{\eta_1}\right)(\psi)\right](j) =$$
$$= \left(A_1 \otimes \left(\mathsf{L}_{(j_1)}^{-1} A_2\, \mathsf{L}_{(j_1)}\right)\right)\left(\Phi_\eta(\psi)\right)(j)\,. \tag{3.3.2}$$

For any bounded operator $A_1$ on $\mathcal{H}$, resp. $A_2$ on $\mathcal{H}^{\otimes N}$, we define a bounded operator $I(A_1\,,\,A_2)$ on $\mathcal{H}^{\otimes(N+1)}$ via:

$$\forall \psi \in \mathcal{H}^{\otimes(N+1)},\ \forall j \in G^{N+1},\ \left(I(A_1\,,\,A_2)\,\psi\right)(j) := \left[\left(A_1 \otimes \left(\mathsf{L}_{(j_1)}^{-1} A_2\, \mathsf{L}_{(j_1)}\right)\right)(\psi)\right](j)$$

(note that it follows in particular from the previous expression that this indeed defines a bounded operator on $\mathcal{H}^{\otimes(N+1)}$). We denote by $\mathcal{I}$ the vector subspace spanned by these operators in the space of bounded operators on $\mathcal{H}^{\otimes(N+1)}$:

$$\mathcal{I} := \mathrm{Vect}\left\{I(A_1\,,\,A_2)\ \middle|\ A_1 \text{ bounded operator on } \mathcal{H},\ A_2 \text{ on } \mathcal{H}^{\otimes N}\right\}$$

(*without* any completion, ie. considering only *finite* linear combinations). The recursion hypothesis, together with eq. (3.3.2) and the fact that both $\alpha$ and $\widetilde{\alpha}$ are $C^*$-algebra isomorphisms, then ensures:

$$\forall A \in \mathcal{I},\ \alpha\left(\left[\Phi_\eta^{-1} \circ A \circ \Phi_\eta\right]_{\sim_{\mathcal{L}^{[\lambda]}}}\right) = \widetilde{\alpha}\left(\left[\Phi_\eta^{-1} \circ A \circ \Phi_\eta\right]_{\sim_{\mathcal{L}^{[\lambda]}}}\right).$$

Let and $\epsilon > 0$. For any $u \in G$, $(u\,\cdot)_*\,\mu$ is a right-invariant Haar measure, so there exists $\Delta_u < \infty$

such that $(u\,.\,\cdot)_* \mu = \Delta_u \mu$. Defining, for any $\psi \in \mathcal{H}$ and any $u \in G$, $\psi^{(u)} \in \mathcal{H}$ via:

$$\forall j \in G,\ \psi^{(u)}(j) := \psi(u\,.\,j),$$

we then have $\|\psi^{(u)}\|_{\mathcal{H}} = \sqrt{\Delta_u}\ \|\psi\|_{\mathcal{H}}$ (which in particular ensures that $\psi^{(u)}$ is indeed in $\mathcal{H}$). Moreover, $u \mapsto \Delta_u$ is a smooth character on $G$ (aka. modular function of $G$, see [5, appendix C.4]). Hence, there exists an open neighborhood $D$ of $\mathbf{1}$ in $G$ such that:

$$\forall u \in D,\ |1 - \Delta_u| \leqslant \frac{\min(1,\ \epsilon)}{2}.$$

We now choose an orthonormal basis $(\phi_i)_{i\in\mathbb{N}}$ in $\mathcal{H}$ and an integer $M \in \mathbb{N}$. For any $i \leqslant M$, there exists a smooth, compactly supported function $\widetilde{\phi}_i$ on $G$ (with support $C_i \subset G$) such that:

$$\left\|\phi_i - \widetilde{\phi}_i\right\|_{\mathcal{H}} \leqslant \frac{\epsilon}{2}.$$

We define the compact $C := \bigcup_{i\leqslant M} C_i \subset G$. For any $i \leqslant M$, $\widetilde{\phi}_i$ is in particular uniformly continuous, so there exists an open neighborhood $V_i$ of $\mathbf{1}$ in $G$ such that:

$$\forall u \in V_i\,,\ \ \|\widetilde{\phi}_i - \widetilde{\phi}_i^{(u)}\|_\infty \leqslant \frac{\epsilon}{4\sqrt{\mu(C)}}.$$

In particular, we thus have:

$$\begin{aligned}
\forall u \in V_i\,,\ \ \|\phi_i - \phi_i^{(u)}\|_{\mathcal{H}} &\leqslant \|\phi_i - \widetilde{\phi}_i\|_{\mathcal{H}} + \|\widetilde{\phi}_i - \widetilde{\phi}_i^{(u)}\|_{\mathcal{H}} + \|\widetilde{\phi}_i^{(u)} - \phi_i^{(u)}\|_{\mathcal{H}} \\
&\leqslant \frac{\epsilon}{2}\left(1 + \sqrt{\Delta_u}\right) + \frac{\epsilon}{4}\sqrt{\frac{\mu(C_i\ \cup\ u^{-1}\,.\,C_i)}{\mu(C)}} \\
&\leqslant \frac{\epsilon}{2}\left(1 + \sqrt{\Delta_u} + \frac{\sqrt{1+\Delta_u}}{2}\right).
\end{aligned}$$

We define $V := D \cap \bigcap_{i\leqslant M} V_i$, so that $V$ is an open neighborhood of $\mathbf{1}$ in $G$. Let $(u_k)_{k\leqslant r}$ with $r < \infty$ be a *finite* family of points in $C$ such that $C \subset \bigcup_{k\leqslant r} V_{[u_k]}$, where $\forall u \in G,\ \ V_{[u]} := \{v\,.\,u \mid v \in V\} \cap \{u\,.\,v' \mid v' \in V\}$. For any $k \leqslant r$, we define:

$$F_k := \left(V_{[u_k]} \cap C\right) \setminus \left(\bigcup_{l<k} V_{[u_k]}\right).$$

Since $F_k \subset C$, $\mu(F_k) < \infty$, and we define $R := \{k \mid k \leqslant r\ \ \&\ \ \mu(F_k) > 0\}$. For any $k \in R$, we define $\chi_k \in \mathcal{H}$ as $\chi_k := {}^1\!/\!\sqrt{\mu(F_k)}\ \mathbb{1}_{F_k}$, where $\mathbb{1}_{F_k}$ denotes the indicator function of $F_k$. $(\chi_k)_{k\in R}$ is then an orthonormal family and we have, for any $i \leqslant M$:

$$\begin{aligned}
\left\|\widetilde{\phi}_i - \sum_{k\in R}\left\langle \chi_k \,\middle|\, \widetilde{\phi}_i\right\rangle \chi_k\right\|^2_{\mathcal{H}} &= \sum_{l\in R}\int_{F_l} d\mu(u)\, \left|\widetilde{\phi}_i(u) - \frac{1}{\mu(F_l)}\int_{F_l} d\mu(v)\, \widetilde{\phi}_i(v)\right|^2 \\
&\leqslant \sum_{l\in R}\int_{F_l} d\mu(u)\, \left(\left|\widetilde{\phi}_i(u) - \widetilde{\phi}_i(u_l)\right| + \frac{1}{\mu(F_l)}\int_{F_l} d\mu(v)\, \left|\widetilde{\phi}_i(u_l) - \widetilde{\phi}_i(v)\right|\right)^2 \leqslant \frac{\epsilon^2}{4},
\end{aligned}$$

so that $\left\|\phi_i - \sum_{k\in R}\left\langle \chi_k \,\middle|\, \widetilde{\phi}_i\right\rangle \chi_k\right\|_{\mathcal{H}} \leqslant \epsilon$. Since $\sum_{k\in R} |\chi_k\,\rangle\,\langle\,\chi_k|$ is an orthogonal projection, this implies $\|\phi_i - \sum_{k\in R}\langle\chi_k \mid \phi_i\rangle\,\chi_k\|_{\mathcal{H}} \leqslant \epsilon$.

Let $(i_p)_p\,,\,(j_p)_p \in \{0,\dots,M\}^{N+1}$. For any $k \in R$, we define:

$$A_1^k := \langle \chi_k \mid \phi_{i_1} \rangle \;\left|\chi_k \right\rangle \left\langle \phi_{j_1} \right|$$

$$\& \quad A_2^k := \Delta_{u_k} \left|\phi_{i_2}^{(u_k^{-1})} \otimes \phi_{i_3} \otimes \dots \otimes \phi_{i_{N+1}} \right\rangle \left\langle \phi_{j_2}^{(u_k^{-1})} \otimes \phi_{j_3} \otimes \dots \otimes \phi_{j_{N+1}} \right|,$$

so that $A^{(i_p)_p\,(j_p)_p} := \sum_{k \in R} I(A_1^k\,,\,A_2^k) \in \mathcal{I}$. Then, for any $\psi \in \mathcal{H}^{\otimes(N+1)}$, we have:

$$\forall h \in G^{N+1},\; \left[A^{(i_p)_p\,(j_p)_p}\,\psi\right](h) = \sum_{k \in R} \langle \chi_k \mid \phi_{i_1} \rangle\; \chi_k(h_1)\, \phi_{i_2}^{(u_k^{-1}\,.\,h_1)}(h_2) \times$$

$$\times \phi_{(i_p)_p}^{(3\dots)}(h_3\,,\,\dots\,,\,h_{N+1})\,\Delta_{h_1^{-1}\,.\,u_k} \left\langle \phi_{j_1} \otimes \phi_{j_2}^{(u_k^{-1}\,.\,h_1)} \otimes \phi_{(j_p)_p}^{(3\dots)} \;\middle|\; \psi \right\rangle,$$

where, for any $(i'_p)_p \in \{0,\dots,M\}^{N+1}$, $\phi_{(i'_p)_p}^{(3\dots)} \in \mathcal{H}^{\otimes(N-1)}$ is defined by:

$$\phi_{(i'_p)_p}^{(3\dots)} := \phi_{i'_3} \otimes \dots \otimes \phi_{i'_{N+1}}\,.$$

Using $\left\|\phi_{i_1} - \sum_{k \in R} \langle \chi_k \mid \phi_{i_1} \rangle\, \chi_k \right\|_{\mathcal{H}} \leqslant \epsilon$, this yields:

$$\left\| A^{(i_p)_p\,(j_p)_p}\,\psi - \left|\phi_{i_1} \otimes \phi_{i_2} \otimes \phi_{(i_p)_p}^{(3\dots)} \right\rangle \left\langle \phi_{j_1} \otimes \phi_{j_2} \otimes \phi_{(j_p)_p}^{(3\dots)} \;\middle|\; \psi \right\rangle \right\|_{\mathcal{H}^{\otimes(N+1)}}$$

$$\leqslant \epsilon\, \|\psi\|_{\mathcal{H}^{\otimes(N+1)}} + \left\| \sum_{k \in R} \langle \chi_k \mid \phi_{i_1} \rangle\, \beta_k \otimes \phi_{(i_p)_p}^{(3\dots)} \right\|_{\mathcal{H}^{\otimes(N+1)}},$$

where, for any $k \in R$, $\beta_k \in \mathcal{H}^{\otimes 2}$ is defined by:

$$\forall h_1\,,\,h_2 \in G,\;\; \beta_k(h_1,h_2) := \chi_k(h_1) \left[ \phi_{i_2}^{(u_k^{-1}\,.\,h_1)}(h_2)\,\Delta_{h_1^{-1}\,.\,u_k} \left\langle \phi_{j_1} \otimes \phi_{j_2}^{(u_k^{-1}\,.\,h_1)} \otimes \phi_{(j_p)_p}^{(3\dots)} \;\middle|\; \psi \right\rangle - \right.$$

$$\left. - \phi_{i_2}(h_2) \left\langle \phi_{j_1} \otimes \phi_{j_2} \otimes \phi_{(j_p)_p}^{(3\dots)} \;\middle|\; \psi \right\rangle \right].$$

Next, for any $k \in R$, and any $h_1 \in F_k$, we have $u_k^{-1}\,.\,h_1 \in V$, so that:

$$\|\beta_k\|_{\mathcal{H}^{\otimes 2}}^2 = \int_{F_k} \frac{d\mu(h_1)}{\mu(F_k)}\; \left\| \phi_{i_2}^{(u_k^{-1}\,.\,h_1)}\,\Delta_{h_1^{-1}\,.\,u_k} \left\langle \phi_{j_1} \otimes \phi_{j_2}^{(u_k^{-1}\,.\,h_1)} \otimes \phi_{(j_p)_p}^{(3\dots)} \;\middle|\; \psi \right\rangle - \right.$$

$$\left. - \phi_{i_2} \left\langle \phi_{j_1} \otimes \phi_{j_2} \otimes \phi_{(j_p)_p}^{(3\dots)} \;\middle|\; \psi \right\rangle \right\|_{\mathcal{H}}^2$$

$$\leqslant \int_{F_k} \frac{d\mu(h_1)}{\mu(F_k)} \left( \Delta_{u_k^{-1}\,.\,h_1} \left| \frac{1}{\Delta_{u_k^{-1}\,.\,h_1}} - 1 \right| + \sqrt{\Delta_{u_k^{-1}\,.\,h_1}}\, \left\| \phi_{i_2}^{(u_k^{-1}\,.\,h_1)} - \phi_{i_2} \right\|_{\mathcal{H}} + \right.$$

$$\left. + \left\| \phi_{j_2}^{(u_k^{-1}\,.\,h_1)} - \phi_{j_2} \right\|_{\mathcal{H}} \right)^2 \|\psi\|_{\mathcal{H}^{\otimes(N+1)}}^2$$

$$\leqslant \left( 5\,\epsilon\, \|\psi\|_{\mathcal{H}^{\otimes(N+1)}} \right)^2.$$

Moreover, for any $k \neq l$, we have $\langle \beta_k \mid \beta_l \rangle_{\mathcal{H}^{\otimes 2}} = 0$, so we get:

$$\left\| A^{(i_p)_p\,(j_p)_p} - \left|\phi_{i_1} \otimes \phi_{i_2} \otimes \phi_{(i_p)_p}^{(3\dots)} \right\rangle \left\langle \phi_{j_1} \otimes \phi_{j_2} \otimes \phi_{(j_p)_p}^{(3\dots)} \right| \right\|_{\mathcal{A}^{\otimes(N+1)}} \leqslant 6\,\epsilon\,,$$

where $\|\cdot\|_{\mathcal{A}^{\otimes(N+1)}}$ denotes the operator norm on $\mathcal{H}^{\otimes(N+1)}$.

Let $A_\eta \in \mathcal{A}_\eta$ and let $\eta' \in \mathcal{L}^{[\kappa]}$ such that $\alpha\left([A_\eta]_{\sim_{\mathcal{L}^{[\lambda]}}}\right) = \left[A'_{\eta'}\right]_{\sim_{\mathcal{L}^{[\kappa]}}}$ for some $A'_{\eta'} \in \mathcal{A}_{\eta'}$ (thanks to statement 2.8.2). Let $\eta'' := \eta' \cup T^{-1}\eta \in \mathcal{L}^{[\kappa]}$ and let $\rho_{\eta''}$ be a non-negative traceclass operator on $\mathcal{H}_{\eta''}$. Let $n'' \in \mathbb{N}$ such that $\eta'' \subset \kappa_{n''}$, choose $\tau_{n''} \in \mathcal{H}_{\kappa_{n''}\to\eta''}$ and, for any $n > n''$, choose $\tau_n \in \mathcal{H}_{\kappa_n\to\kappa_{n-1}}$. In a way similar to prop. 2.1, there exists $\rho' \in \mathcal{S}^{\otimes}_{(\mathcal{L}^{[\kappa]},\mathcal{H},\Phi)}$ such that:

$$\Phi_{\kappa_{n''}\to\eta''}\,\rho'_{\kappa_{n''}}\,\Phi^{-1}_{\kappa_{n''}\to\eta''} = |\tau_{n''}\rangle\langle\tau_{n''}|\otimes\rho_{\eta''} \quad \& \quad \forall n > n'',\ \Phi_{\kappa_n\to\kappa_{n-1}}\,\rho'_{\kappa_n}\,\Phi^{-1}_{\kappa_n\to\kappa_{n-1}} = |\tau_n\rangle\langle\tau_n|\otimes\rho'_{\kappa_{n-1}}.$$

In particular, we then have $\rho'_{\eta''} = \rho_{\eta''}$. Let $\rho := \left[\sigma(\rho')\right]_\eta$ and $\widetilde{\rho} := \mathrm{U}_{T^{-1}\eta}(T)\left(\mathrm{Tr}_{\eta''\to T^{-1}\eta}\,\rho_{\eta''}\right)\mathrm{U}_\eta(T^{-1})$. Next, for any $M \in \mathbb{N}$, we define a finite-dimensional vector subspace $\mathcal{J}_M \subset \mathcal{H}_\eta$ as:

$$\mathcal{J}_M := \mathrm{Vect}\left\{\Phi_\eta^{-1}\,|\phi_{i_1}\otimes\ldots\otimes\phi_{i_{N+1}}\rangle \;\middle|\; \left(i_p\right)_p \in \{0,\ldots,M\}^{N+1}\right\}.$$

Let $\epsilon > 0$. Applying [9, lemma 2.10.1] to the family $(\mathcal{J}_M)_{M\in\mathbb{N}}$ and the non-negative traceclass operators $\rho$ and $\widetilde{\rho}$, there exists $M \in \mathbb{N}$ such that:

$$\left\|\rho - \Pi_M\,\rho\,\Pi_M\right\|_1 \leqslant \frac{\epsilon}{4} \quad \& \quad \left\|\widetilde{\rho} - \Pi_M\,\widetilde{\rho}\,\Pi_M\right\|_1 \leqslant \frac{\epsilon}{4},$$

where $\|\cdot\|_1$ denotes the trace-norm and $\Pi_M$ is the orthogonal projection on $\mathcal{J}_M$. Now, from the previous step (with $M \in \mathbb{N}$ and $\epsilon/24\,(M+1)^2 > 0$), there exists, for any $\left(i_p\right)_p, \left(j_p\right)_p \in \{0,\ldots,M\}^{N+1}$, $A^{\left(i_p\right)_p\left(j_p\right)_p} \in \mathfrak{I}$ such that:

$$\left\|\Phi_\eta^{-1}\,\left|\phi_{i_1}\otimes\ldots\otimes\phi_{i_{N+1}}\right\rangle\left\langle\phi_{j_1}\otimes\ldots\otimes\phi_{j_{N+1}}\right|\,\Phi_\eta - \Phi_\eta^{-1}\,A^{\left(i_p\right)_p\left(j_p\right)_p}\,\Phi_\eta\right\|_{\mathcal{A}_\eta} \leqslant \frac{\epsilon}{4\,(M+1)^2},$$

where $\|\cdot\|_{\mathcal{A}_\eta}$ denotes the operator norm on $\mathcal{H}_\eta$. Thus, defining a bounded operator $\widetilde{A}$ on $\mathcal{H}_\eta$ as:

$$\widetilde{A} := \sum_{\left(i_p\right)_p,\left(j_p\right)_p\in\{0,\ldots,M\}^{N+1}} \left\langle\phi_{i_1}\otimes\ldots\otimes\phi_{i_{N+1}}\,\middle|\,\Phi_\eta\,A_\eta\,\Phi_\eta^{-1}\,\phi_{j_1}\otimes\ldots\otimes\phi_{j_{N+1}}\right\rangle\,\Phi_\eta^{-1}\,A^{\left(i_p\right)_p\left(j_p\right)_p}\,\Phi_\eta,$$

we have $\Phi_\eta\,\widetilde{A}\,\Phi_\eta^{-1} \in \mathfrak{I}$ and $\left\|\widetilde{A} - \Pi_M\,A_\eta\,\Pi_M\right\| \leqslant \frac{\epsilon}{4}\,\|A_\eta\|_{\mathcal{A}_\eta}$. Putting everything together, we obtain:

$$\begin{aligned}\left|\mathrm{Tr}_{\mathcal{H}_{\eta''}}\,\rho_{\eta''}\left(\iota_{\eta''\leftarrow\eta'}(A'_{\eta'}) - \iota_{\eta''\leftarrow T^{-1}\eta}(T^{-1}\triangleright A_\eta)\right)\right| &= \left|\mathrm{Tr}_{\mathcal{H}_\eta}\left(\rho-\widetilde{\rho}\right)A_\eta\right| \\ &\leqslant \epsilon\,\|A_\eta\|_{\mathcal{A}_\eta} + \left|\mathrm{Tr}_{\mathcal{H}_\eta}\left(\rho-\widetilde{\rho}\right)\widetilde{A}\right| \\ &= \epsilon\,\|A_\eta\|_{\mathcal{A}_\eta} + \left|\mathrm{Tr}\,\rho'\left(\alpha\left([\widetilde{A}]_{\sim_{\mathcal{L}^{[\lambda]}}}\right) - \widetilde{\alpha}\left([\widetilde{A}]_{\sim_{\mathcal{L}^{[\lambda]}}}\right)\right)\right| = \epsilon\,\|A_\eta\|_{\mathcal{A}_\eta}.\end{aligned}$$

Since this holds for any density matrix $\rho_{\eta''}$ on $\mathcal{H}_{\eta''}$ and any $\epsilon > 0$, it follows that $\iota_{\eta''\leftarrow\eta'}(A'_{\eta'}) = \iota_{\eta''\leftarrow T^{-1}\eta}(T^{-1}\triangleright A_\eta)$, ie. $\alpha\left([A_\eta]_{\sim_{\mathcal{L}^{[\lambda]}}}\right) = \widetilde{\alpha}\left([A_\eta]_{\sim_{\mathcal{L}^{[\lambda]}}}\right)$, which concludes the recursive proof of eq. (3.3.1).

From eq. (3.3.1), we have:

$$\forall A \in \mathcal{A}^{\otimes}_{(\mathcal{L}^{[\lambda]},\mathcal{H},\Phi)},\quad \alpha(A) = \widetilde{\alpha}(A),$$

hence, by continuity, $\alpha = \widetilde{\alpha}$. Finally, like in the proof of prop. 2.9, we can then define a bijective map $\widetilde{\sigma} : \overline{\mathcal{S}}^{\otimes}_{(\mathcal{L}^{[\kappa]},\mathcal{H},\Phi)} \to \overline{\mathcal{S}}^{\otimes}_{(\mathcal{L}^{[\underline{\kappa}]},\mathcal{H},\Phi)}$ via:

$$\begin{array}{rccl}\widetilde{\sigma} : & \overline{\mathcal{S}}^{\otimes}_{(\mathcal{L}^{[\kappa]},\mathcal{H},\Phi)} & \to & \overline{\mathcal{S}}^{\otimes}_{(\mathcal{L}^{[\lambda]},\mathcal{H},\Phi)} \\ & \left(\rho_\eta\right)_{\eta\in\mathcal{L}^{[\kappa]}} & \mapsto & \left(T\triangleright\rho_{T^{-1}\underline{\eta}}\right)_{\underline{\eta}\in\mathcal{L}^{[\lambda]}}\end{array},$$

$\widetilde{\sigma}$ and $\widetilde{\alpha} = \alpha$ are Tr-intertwined by construction, as are $\sigma$ and $\alpha$. Thus, for any $\rho \in \overline{\mathcal{S}}^{\otimes}_{(\mathcal{L}^{[\kappa]},\mathcal{H},\Phi)}$, any $\underline{\eta} \in \mathcal{L}^{[\lambda]}$ and any $A_{\underline{\eta}} \in \mathcal{A}_{\underline{\eta}}$, we have:

$$\mathrm{Tr}_{\mathcal{H}_{\underline{\eta}}}\left[\sigma(\rho)\right]_{\underline{\eta}} A_{\underline{\eta}} = \mathrm{Tr}\,\rho\,\alpha\left(\left[A_{\underline{\eta}}\right]_{\sim_{\mathcal{L}^{[\lambda]}}}\right) = \mathrm{Tr}_{\mathcal{H}_{\underline{\eta}}}\left[\widetilde{\sigma}(\rho)\right]_{\underline{\eta}} A_{\underline{\eta}},$$

which ensures that $\sigma = \widetilde{\sigma}$.

*Approximation of* $\mathcal{T}$ *by* $\mathcal{T}^{[\kappa]}$. $\mathcal{T}^{[\kappa]}$ is stable under composition and inverse, hence it forms a subgroup of $\mathcal{T}$. Moreover, using the characterization $\mathcal{L}^{[\kappa]} = \left\{\eta \in \mathcal{L}^{(\mathrm{aux})} \,\middle|\, \eta \subset K\right\}$, we have, for any $T \in \mathcal{T}^{[\kappa]}$ and any $\eta \in \mathcal{L}^{(\mathrm{aux})}$, $\eta \in \mathcal{L}^{[\kappa]} \Leftrightarrow T\eta \in \mathcal{L}^{[\kappa]}$. Thus, the group action of $\mathcal{T}$ on the projective system $\left(\mathcal{L}^{(\mathrm{aux})}, \mathcal{H}, \Phi\right)^{\otimes}$ induces a group action of $\mathcal{T}^{[\kappa]}$ on the projective system $\left(\mathcal{L}^{[\kappa]}, \mathcal{H}, \Phi\right)^{\otimes}$.

Next, let $T \in \mathcal{T}$ and let $V$ be an open neighborhood of $T$ in $\mathcal{T}$. Let $\epsilon > 0$ such that $B_\epsilon\,.\,T := \{T'.T \mid T' \in B_\epsilon\} \subset V$. Let $\sigma : \overline{\mathcal{S}}^{\otimes}_{(\mathcal{L}^{[\kappa]},\mathcal{H},\Phi)} \to \overline{\mathcal{S}}^{\otimes}_{(\mathcal{L}^{[\kappa]},\mathcal{H},\Phi)}$ and $\alpha : \overline{\mathcal{A}}^{\otimes}_{(\mathcal{L}^{[\kappa]},\mathcal{H},\Phi)} \to \overline{\mathcal{A}}^{\otimes}_{(\mathcal{L}^{[\kappa]},\mathcal{H},\Phi)}$ be as in prop. 2.9 with respect to the transformation $T \in \mathcal{T}$ and the open neighborhood $B_\epsilon$ of $\mathrm{id}_{]0,1]}$ in $\mathcal{T}$. Define $\underline{\sigma} : \overline{\mathcal{S}}^{\otimes}_{(\mathcal{L}^{[\kappa]},\mathcal{H},\Phi)} \to \overline{\mathcal{S}}^{\otimes}_{(\mathcal{L}^{[\underline{\kappa}]},\mathcal{H},\Phi)}$ and $\underline{\alpha} : \mathcal{A}^{\otimes}_{(\mathcal{L}^{[\underline{\kappa}]},\mathcal{H},\Phi)} \to \mathcal{A}^{\otimes}_{(\mathcal{L}^{[\kappa]},\mathcal{H},\Phi)}$ like in the proof of prop. 2.9, with $\underline{\kappa}$ the quasi-cofinal sequence $\underline{\kappa} := (T\kappa_n)_{n\in\mathbb{N}}$. Then, $\underline{\sigma} \circ \sigma^{-1} : \overline{\mathcal{S}}^{\otimes}_{(\mathcal{L}^{[\kappa]},\mathcal{H},\Phi)} \to \overline{\mathcal{S}}^{\otimes}_{(\mathcal{L}^{[\underline{\kappa}]},\mathcal{H},\Phi)}$ and $\alpha \circ \underline{\alpha} : \mathcal{A}^{\otimes}_{(\mathcal{L}^{[\underline{\kappa}]},\mathcal{H},\Phi)} \to \overline{\mathcal{A}}^{\otimes}_{(\mathcal{L}^{[\kappa]},\mathcal{H},\Phi)}$ fulfills the hypotheses of the first part of the present proof, hence there exists $T' \in B_\epsilon$ such that:

$$\left\{T\eta \,\middle|\, \eta \in \mathcal{L}^{[\kappa]}\right\} = \mathcal{L}^{[\underline{\kappa}]} = \left\{T'\eta \,\middle|\, \eta \in \mathcal{L}^{[\kappa]}\right\}$$
$$\&\quad \forall A \in \mathcal{A}^{\otimes}_{(\mathcal{L}^{[\kappa]},\mathcal{H},\Phi)},\ \ \alpha(A) = T'^{-1}\triangleright\underline{\alpha}^{-1}(A) = (T'^{-1}\,.\,T)\triangleright A\,.$$

In particular, $\widetilde{T} := T'^{-1}\,.\,T \in \mathcal{T}^{\kappa} \cap V$. Therefore, $\mathcal{T}^{\kappa}$ is dense in $\mathcal{T}$. □

As announced in the discussion preceding prop. 2.2, neither the universality of the restricted projective systems built from quasi-cofinal sequences, nor the possibility of approximating transformations (that directly follows from it), extend to the infinite tensor products that one can assemble from these sequences. In prop. 3.4 below, we construct an example of a deformation that provides an identification at the level of the projective systems, but fails to provide a unitary mapping of the corresponding ITP's. The proof is somewhat similar to the one of [9, theorem 2.11], and relies on the fact that grouping the tensor product factors pairwise yields an *inequivalent* ITP [27, section 4.2].

**Proposition 3.4** We consider the same objects as in prop. 3.1 and we assume that $G$ is non-trivial. For any quasi-cofinal sequence $\kappa = (\kappa_n)_{n\in\mathbb{N}}$ in $\mathcal{L}^{(\mathrm{aux})}$, we define $\mathcal{H}^{[\kappa]}_{\mathrm{seq}}$ and $\sigma^{[\kappa]}_{\mathrm{seq}}$ as in prop. 2.2 with respect to the projective system $\left(\mathcal{L}^{[\kappa]}, \mathcal{H}, \Phi\right)$ and the increasing sequence $(\kappa_n)_{n\in\mathbb{N}}$. There exists quasi-cofinal sequences $\kappa$, $\lambda$ and an homeomorphism $T \in \mathcal{T}$ such that:

$$\mathcal{L}^{[\lambda]} = \left\{ T\eta \;\middle|\; \eta \in \mathcal{L}^{[\kappa]} \right\} \quad \& \quad \sigma^{[\lambda]}_{\text{seq}} \left\langle \overline{\mathcal{S}}^{[\lambda]}_{\text{seq}} \right\rangle \neq \left\{ \left( T \triangleright \rho_{T^{-1}\underline{\eta}} \right)_{\underline{\eta} \in \mathcal{L}^{[\lambda]}} \;\middle|\; \rho \in \sigma^{[\kappa]}_{\text{seq}} \left\langle \overline{\mathcal{S}}^{[\kappa]}_{\text{seq}} \right\rangle \right\}$$

(where $\overline{\mathcal{S}}^{[\kappa]}_{\text{seq}}$, resp. $\overline{\mathcal{S}}^{[\lambda]}_{\text{seq}}$, denotes the space of non-negative traceclass operators on $\mathcal{H}^{[\kappa]}_{\text{seq}}$, resp. $\mathcal{H}^{[\lambda]}_{\text{seq}}$). In particular, this means that there does not exist any isomorphism of Hilbert spaces $\mathrm{U} : \mathcal{H}^{[\kappa]}_{\text{seq}} \to \mathcal{H}^{[\lambda]}_{\text{seq}}$ such that:

$$\forall \rho \in \overline{\mathcal{S}}^{[\kappa]}_{\text{seq}} ,\; \sigma^{[\lambda]}_{\text{seq}} \left( \mathrm{U}\, \rho\, \mathrm{U}^{-1} \right) = \left( T \triangleright \left[ \sigma^{[\kappa]}_{\text{seq}}(\rho) \right]_{T^{-1}\underline{\eta}} \right)_{\underline{\eta} \in \mathcal{L}^{[\lambda]}} . \tag{3.4.1}$$

**Proof** Let $\kappa = (\kappa_n)_{n\in\mathbb{N}}$ be a *strictly* increasing quasi-cofinal sequence (eg. the one we constructed in the proof of prop. 3.2) and let $\lambda := (\lambda_m)_{m\in\mathbb{N}}$, with $\forall m \in \mathbb{N}$, $\lambda_m := \kappa_{2m}$. Then, one can check that $\lambda$ is also a quasi-cofinal sequence and, setting $T := \mathrm{id}_{]0,1]} \in \mathcal{T}$, we have $\mathcal{L}^{[\lambda]} = \mathcal{L}^{[\kappa]} = \left\{ T\eta \;\middle|\; \eta \in \mathcal{L}^{[\kappa]} \right\}$.

We define $\mathcal{J}_o = \mathcal{J}'_o := \mathcal{H}_{\kappa_o} = \mathcal{H}_{\lambda_o} = \mathcal{H}_{\varnothing}$, as well as, for any $n > 0$, $\mathcal{J}_n := \mathcal{H}_{\kappa_n \to \kappa_{n-1}}$, and for any $m > 0$, $\mathcal{J}'_m := \mathcal{H}_{\lambda_m \to \lambda_{m-1}}$. Then, for any $m > 0$, $\Phi_{\kappa_{2m} \to \kappa_{2m-1} \to \kappa_{2m-2}}$ is an isomorphism $\mathcal{J}'_m \to \mathcal{J}_{2m} \otimes \mathcal{J}_{2m-1}$. We choose a normalized vector $\psi_o \in \mathcal{J}_o$. For any $n > 0$, we have $\kappa_{n-1} \subsetneq \kappa_n$, hence, $G$ being non-trivial, $\dim \mathcal{J}_n \geqslant 2$, so we can choose two normalized, mutually orthogonal vectors $\psi^{(1)}_n$, $\psi^{(2)}_n \in \mathcal{J}_n$. Then, we define:

$$\phi_o := \psi_o \in \mathcal{J}'_o \quad \& \quad \forall m > 0,\; \phi_m := \Phi^{-1}_{\kappa_{2m} \to \kappa_{2m-1} \to \kappa_{2m-2}} \left( \frac{\psi^{(1)}_{2m} \otimes \psi^{(1)}_{2m-1} + \psi^{(2)}_{2m} \otimes \psi^{(2)}_{2m-1}}{\sqrt{2}} \right) \in \mathcal{J}'_m ,$$

and, by prop. 2.1, we construct $\rho[\phi] \in \mathcal{S}^{\otimes}_{(\mathcal{L}^{[\lambda]}, \mathcal{H}, \Phi)}$ such that:

$$\rho_{\lambda_o}[\phi] = |\phi_o \rangle \langle \phi_o| \quad \& \quad \forall m > 0,\; \rho_{\lambda_m}[\phi] = \Phi^{-1}_{\lambda_m \to \lambda_{m-1}} \circ \left( |\phi_m \rangle \langle \phi_m| \otimes \rho_{\lambda_{m-1}}[\phi] \right) \circ \Phi_{\lambda_m \to \lambda_{m-1}} .$$

From prop. 2.2, we have $\rho[\phi] \in \sigma^{[\lambda]}_{\text{seq}} \left\langle \mathcal{S}^{[\lambda]}_{\text{seq}} \right\rangle \subset \sigma^{[\lambda]}_{\text{seq}} \left\langle \overline{\mathcal{S}}^{[\lambda]}_{\text{seq}} \right\rangle$.

Reasoning by contradiction we suppose that there exists a non-negative traceclass operator $\widetilde{\rho}$ on $\mathcal{H}^{[\kappa]}_{\text{seq}}$ such that:

$$\forall \eta \in \mathcal{L}^{[\lambda]} = \mathcal{L}^{[\kappa]},\; \rho_\eta[\phi] = T \triangleright \left[ \sigma^{[\kappa]}_{\text{seq}} \left( \widetilde{\rho} \right) \right]_{T^{-1}\eta} ,$$

ie. $\rho[\phi] = \sigma^{[\kappa]}_{\text{seq}} \left( \widetilde{\rho} \right)$. Using the same notations as in the proofs of props. 2.1 and 2.2, we have:

$$\rho[\phi] = \sigma^{[\kappa]}_{\text{seq}} \left( \widetilde{\rho} \right) = \sum_{[\![ \psi' ]\!]} \sigma^{[\kappa]}_{[\psi']} \left( \Pi_{[\![ \psi' ]\!]}\, \widetilde{\rho}\, \Pi_{[\![ \psi' ]\!]} \right) ,$$

as well as:

$$1 = \operatorname{Tr} \rho[\phi] = \operatorname{Tr}_{\mathcal{H}^{[\kappa]}_{\text{seq}}} \left[ \widetilde{\rho} \right] = \sum_{[\![ \psi' ]\!]} \operatorname{Tr}_{\mathcal{H}^{[\kappa]}_{[\psi']}} \left[ \Pi_{[\![ \psi' ]\!]}\, \widetilde{\rho}\, \Pi_{[\![ \psi' ]\!]} \right] .$$

Hence, there should exist $\psi' = \left( \psi'_n \right)_{n\in\mathbb{N}} \in \mathcal{Z}^{\otimes}_{(\mathbb{N}, \mathcal{J})}$ such that $\operatorname{Tr}_{\mathcal{H}^{[\kappa]}_{[\psi']}} \left[ \Pi_{[\![ \psi' ]\!]}\, \widetilde{\rho}\, \Pi_{[\![ \psi' ]\!]} \right] > 0$, and eq. (2.2.1) then yields:

$$\sup_{n\in\mathbb{N}} \inf_{n'>n} \left\langle \zeta'_{n' \to n} \;\middle|\; \left( \operatorname{Tr}_{\mathcal{K}_n} \widetilde{\Phi}_{n'}\, \rho[\phi]_{\kappa_{n'}}\, \widetilde{\Phi}^{-1}_{n'} \right) \zeta'_{n' \to n} \right\rangle =$$

$$= \sum_{[\![\psi'']\!]} \sup_{n\in\mathbb{N}} \inf_{n'>n} \left\langle \zeta'_{n'\to n} \;\middle|\; \left(\mathrm{Tr}_{\mathcal{K}_n} \widetilde{\Phi}_{n'} \left[\sigma^{[\kappa]}_{[\psi'']}\big(\Pi_{[\![\psi'']\!]} \widetilde{\rho}\, \Pi_{[\![\psi'']\!]}\big)\right]_{\kappa_{n'}} \widetilde{\Phi}^{-1}_{n'}\right) \zeta'_{n'\to n} \right\rangle$$

$$\geqslant \sup_{n\in\mathbb{N}} \inf_{n'>n} \left\langle \zeta'_{n'\to n} \;\middle|\; \left(\mathrm{Tr}_{\mathcal{K}_n} \widetilde{\Phi}_{n'} \left[\sigma^{[\kappa]}_{[\psi']}\big(\Pi_{[\![\psi']\!]} \widetilde{\rho}\, \Pi_{[\![\psi']\!]}\big)\right]_{\kappa_{n'}} \widetilde{\Phi}^{-1}_{n'}\right) \zeta'_{n'\to n} \right\rangle$$

$$= \mathrm{Tr}_{\mathcal{H}^{[\kappa]}_{[\psi']}} \left[\Pi_{[\![\psi']\!]} \widetilde{\rho}\, \Pi_{[\![\psi']\!]}\right] > 0 \,,$$

with $\forall n < n' \in \mathbb{N}$, $\zeta'_{n'\to n} := \psi'_{n'} \otimes \ldots \otimes \psi'_{n+1} \in \mathcal{K}_{n'\to n}$ (we have used that the sum $\sum_{[\![\psi'']\!]}$ is absolutely convergent in trace norm, and that the argument of $\inf_{n'>n}$, resp. of $\sup_{n\in\mathbb{N}}$, is positive and decreasing with $n'$, resp. increasing with $n$).

From the definition of $\widetilde{\Phi}_n$ and $\rho[\phi]$, we have $\widetilde{\Phi}_o\, \rho[\phi]_{\kappa_o}\, \widetilde{\Phi}^{-1}_o = |\phi_o\,\rangle\,\langle\,\phi_o|$, as well as, for any $m > 0$:

$$\widetilde{\Phi}_{2m}\, \rho[\phi]_{\kappa_{2m}}\, \widetilde{\Phi}^{-1}_{2m} = \frac{1}{2} \left|\psi^{(1)}_{2m} \otimes \psi^{(1)}_{2m-1} + \psi^{(2)}_{2m} \otimes \psi^{(2)}_{2m-1} \right\rangle \left\langle \psi^{(1)}_{2m} \otimes \psi^{(1)}_{2m-1} + \psi^{(2)}_{2m} \otimes \psi^{(2)}_{2m-1}\right| \otimes$$
$$\otimes \left(\widetilde{\Phi}_{2m-2}\, \rho_{\kappa_{2m-2}}[\phi]\, \widetilde{\Phi}^{-1}_{2m-2}\right).$$

Hence, we get, for any $m > 0$:

$$\widetilde{\Phi}_{2m}\, \rho[\phi]_{\kappa_{2m}}\, \widetilde{\Phi}^{-1}_{2m} = \frac{1}{2^m} \left|\psi^{(1)}_{2m} \otimes \psi^{(1)}_{2m-1} + \psi^{(2)}_{2m} \otimes \psi^{(2)}_{2m-1} \right\rangle \left\langle \psi^{(1)}_{2m} \otimes \psi^{(1)}_{2m-1} + \psi^{(2)}_{2m} \otimes \psi^{(2)}_{2m-1}\right| \otimes$$
$$\otimes \ldots \otimes \left|\psi^{(1)}_{2} \otimes \psi^{(1)}_{1} + \psi^{(2)}_{2} \otimes \psi^{(2)}_{1} \right\rangle \left\langle \psi^{(1)}_{2} \otimes \psi^{(1)}_{1} + \psi^{(2)}_{2} \otimes \psi^{(2)}_{1}\right| \otimes |\phi_o\,\rangle\,\langle\,\phi_o|\,.$$

Let $n \in \mathbb{N}$ and let $m > n/2$. If $n = 2p$, we have $p < m$ and:

$$\left\langle \zeta'_{2m\to n} \;\middle|\; \left(\mathrm{Tr}_{\mathcal{K}_n} \widetilde{\Phi}_{2m}\, \rho[\phi]_{\kappa_{2m}}\, \widetilde{\Phi}^{-1}_{2m}\right) \zeta'_{2m\to n} \right\rangle = \prod_{k=p+1}^{m} \xi_k\,,$$

where, for any $k > 0$, $\xi_k := \frac{1}{2} \left|\left\langle \psi'_{2k} \otimes \psi'_{2k-1} \;\middle|\; \psi^{(1)}_{2k} \otimes \psi^{(1)}_{2k-1} + \psi^{(2)}_{2k} \otimes \psi^{(2)}_{2k-1} \right\rangle\right|^2$. If $n = 2p+1$, we also have $p < m$ and:

$$\left\langle \zeta'_{2m\to n} \;\middle|\; \left(\mathrm{Tr}_{\mathcal{K}_n} \widetilde{\Phi}_{2m}\, \rho[\phi]_{\kappa_{2m}}\, \widetilde{\Phi}^{-1}_{2m}\right) \zeta'_{2m\to n} \right\rangle = \widetilde{\xi}_{p+1} \prod_{k=p+2}^{m} \xi_k\,,$$

where $\widetilde{\xi}_{p+1} := \frac{1}{2} \left|\left\langle \psi'_{2p+2} \;\middle|\; \psi^{(1)}_{2p+2} \right\rangle\right|^2 + \frac{1}{2} \left|\left\langle \psi'_{2p+2} \;\middle|\; \psi^{(2)}_{2p+2} \right\rangle\right|^2$. Now, for any $k > 0$, we have:

$$\xi_k = \frac{1}{2} \left|\sum_{\epsilon=1}^{2} \left\langle \psi'_{2k} \;\middle|\; \psi^{(\epsilon)}_{2k} \right\rangle \left\langle \psi'_{2k-1} \;\middle|\; \psi^{(\epsilon)}_{2k-1} \right\rangle\right|^2$$
$$\leqslant \frac{1}{2} \left(\sum_{\epsilon=1}^{2} \left|\left\langle \psi'_{2k} \;\middle|\; \psi^{(\epsilon)}_{2k} \right\rangle\right|^2\right) \left(\sum_{\epsilon=1}^{2} \left|\left\langle \psi'_{2k-1} \;\middle|\; \psi^{(\epsilon)}_{2k-1} \right\rangle\right|^2\right)$$
$$\leqslant \frac{1}{2}\, \|\psi'_{2k}\|^2\, \|\psi'_{2k-1}\|^2 = \frac{1}{2}.$$

Thus, we get:

$$\left\langle \zeta'_{2m\to n} \;\middle|\; \left(\mathrm{Tr}_{\mathcal{K}_n} \widetilde{\Phi}_{2m}\, \rho[\phi]_{\kappa_{2m}}\, \widetilde{\Phi}^{-1}_{2m}\right) \zeta'_{2m\to n} \right\rangle \leqslant \frac{1}{2^m - \lfloor (n+1)/2 \rfloor}\,,$$

where $\lfloor \cdot \rfloor$ denotes the floor function. This yields, for any $n \in \mathbb{N}$:

$$\inf_{n'>n} \left\langle \zeta'_{n'\to n} \,\middle|\, \left(\mathrm{Tr}_{\mathcal{K}_n} \widetilde{\Phi}_{n'}\, \rho[\phi]_{\kappa_{n'}} \widetilde{\Phi}_{n'}^{-1}\right) \zeta'_{n'\to n} \right\rangle = 0,$$

and therefore:

$$\sup_{n\in\mathbb{N}} \inf_{n'>n} \left\langle \zeta'_{n'\to n} \,\middle|\, \left(\mathrm{Tr}_{\mathcal{K}_n} \widetilde{\Phi}_{n'}\, \rho[\phi]_{\kappa_{n'}} \widetilde{\Phi}_{n'}^{-1}\right) \zeta'_{n'\to n} \right\rangle = 0,$$

which provides the desired contradiction. Hence, $\rho[\phi] \notin \left\{ \left(T \triangleright \rho_{T^{-1}\eta}\right)_{\eta\in\mathcal{L}^{[\lambda]}} \,\middle|\, \rho \in \sigma^{[\kappa]}_{\mathrm{seq}} \left\langle \overline{\mathcal{S}}^{[\kappa]}_{\mathrm{seq}} \right\rangle \right\}$, so:

$$\sigma^{[\lambda]}_{\mathrm{seq}} \left\langle \overline{\mathcal{S}}^{[\lambda]}_{\mathrm{seq}} \right\rangle \neq \left\{ \left(T \triangleright \rho_{T^{-1}\eta}\right)_{\eta\in\mathcal{L}^{[\lambda]}} \,\middle|\, \rho \in \sigma^{[\kappa]}_{\mathrm{seq}} \left\langle \overline{\mathcal{S}}^{[\kappa]}_{\mathrm{seq}} \right\rangle \right\}.$$

□

# 4 Outlook

It is easy to check that projective state spaces are not affected if their underlying label set is restricted to a *cofinal* part [9, prop. 2.6]. In the present work, we have displayed a strengthening of this result: sequences of labels satisfying a rather weak condition of 'quasi-cofinality' (def. 2.7) turn out to be sufficient to capture a physically accurate picture of the algebra of observables. This makes the benefits of working with a *countable* label subset (subsection 2.1) available even for label sets that *do not* admit cofinal sequences, like the label set $\mathcal{L}_{\mathrm{HF}}$ introduced in [7]. As a proof of principle for this approach, the existence of such quasi-cofinal sequences was proved in a simple toy-model built on a one-dimensional version of $\mathcal{L}_{\mathrm{HF}}$.

To exhibit a quasi-cofinal sequence for $\mathcal{L}_{\mathrm{HF}}$ in the physically more interesting $d = 3$ case, the idea would be to construct a discrete but dense, *fractal-like* structure made of edges and surfaces: note that the sequence constructed above for our one-dimensional toy-model can also be seen as a fractal in $]0, 1]$. The proof that such a structure can be designed that satisfies the quasi-cofinality property 2.7.3 is however significantly more involved than in the one-dimensional case, and is not yet finished.

An interesting use of the possibility to start from a *non-directed*, extended label set $\mathcal{L}^{(\mathrm{ext})}$ would be that we could drop the somewhat ad-oc (semi-)analyticity requirement for edges and surfaces: recall that analyticity was used solely in [7, lemmas 2.6 and 2.10], with the aim of proving the directedness of $\mathcal{L}_{\mathrm{HF}}$. Also, we could take advantage of this possibility to eliminate the unphysical (in the sense of having no equivalent in the classical continuum phase space described eg. in [25, sections I.4 and IV.33]), *degenerated* fluxes. These fluxes supported on geometrical objects of dimension strictly less that $d - 1$ had to be included in $\mathcal{L}_{\mathrm{HF}}$ because they arose from the commutators of non-degenerate, $(d - 1)$-supported fluxes: we could not exclude *a priori* that an edge, hitting the intersection, would see the non-vanishing commutator. In contrast, in the context of a fractal-structure, where a discrete set of admissible edges and surfaces is chosen *beforehand*, we can effectively forbid that an edge will ever go through the intersection of any two surfaces, and simply *set* the commutator of *degenerately* intersecting fluxes to zero (this would not, however completely solve the problem of the 'non-commuting fluxes' [25, subsection II.6.1], since fluxes may intersect *non-degenerately*).

In addition, the simplification of the algebra of observables achieved by going over to such a fractal setup could help *solving the constraints* of LQG in the projective setting. Recall that we put forward in the outlook of [7] that solving the Gauss constraints will require to 'anchor' the fluxes [23, def. 3.5], in order to improve their transformation properties under gauge transformations. The problem was then to build a *directed* label set while keeping track of the auxiliary systems of paths entering the definition of these anchored fluxes. Remarkably, a fractal-like structure as mentioned above could provide these anchoring paths automatically. Indeed, considering a given face, with a conjugate edge starting from it, the refinement taking place as we go deeper and deeper in the quasi-cofinal sequence would simultaneously subdivide the original face and ramify the original edge. In this way, we would recursively construct a dense tree reaching this face, that one could use for anchoring the corresponding flux.

Finally, it would also be natural to perform the regularization of the Hamiltonian constraints *along* the quasi-cofinal sequence, eg. by adapting the approximation scheme developed in [22], and doing so could potentially help the dynamical *stability* of the factorized semi-classical states constructed along the lines of subsection 2.1 (by contrast, an important limitation of the fixed-graph coherent states usually used in LQG is that they are not well adapted to graph-changing Hamiltonian constraints, see the discussion in [4, subsection 1.1]). Interestingly, the need for fractal constructions also emerges from this perspective [25, subsection II.12.2.5].

## Acknowledgements

This work has been financially supported by the Université François Rabelais, Tours, France (via a 3-years doctoral stipend from the French Ministry of Education – Contrat Doctoral Normalien), and by the Friedrich-Alexander-Universität Erlangen-Nürnberg, Germany (via the Bavarian Equal Opportunities Sponsorship – Förderung von Frauen in Forschung und Lehre (FFL) – Promoting Equal Opportunities for Women in Research and Teaching).

This research project has also been supported by funds to Emerging Field Project "Quantum Geometry" from the FAU Erlangen-Nürnberg within its Emerging Fields Initiative.

# A References